\documentclass[a4paper,11pt]{article}
\usepackage{Mypos}
\usepackage{hyperref}
\usepackage{mathtools}
\usepackage{physics}
\usepackage{float}
\usepackage[version=4]{mhchem}
\usepackage{cellspace}
\usepackage{setspace}
\makeatletter
\renewcommand{\tableofcontents}{%
  \@starttoc{toc}
}
\makeatother

\newcommand{\Tc}{T_{\rm c}}
\newcommand{\Te}{T_{\rm e}}

\title{Astrophysical Axion Bounds: The 2024 Edition}
\ShortTitle{Astrophysical Axion Bounds}

\author*[a]{Andrea Caputo}
\author*[b]{Georg Raffelt}

\affiliation[a]{Theoretical Physics Department, CERN, 1211 Geneva 23, Switzerland}

\affiliation[b]{Max-Planck-Institut für Physik (Werner-Heisenberg-Institut),
Boltzmannstr.~8, 85748 Garching, Germany}

\abstract{We review the current status of astrophysical bounds on QCD axions, primarily based on the observational effects of nonstandard energy losses on stars, including black-hole superradiance. Over the past few years, many of the traditional arguments have been reexamined both theoretically and using modern data and new ideas have been put forth. This compact review updates similar Lecture Notes written by one of us in 2006 [\href{https://arxiv.org/abs/hep-ph/0611350}{Lect.\ Notes Phys.\ 741 (2008) 51--71}].
}

\FullConference{%
  First Training School COST Action {\em COSMIC WISPers} (CA21106)\\
  11--14 September 2023 \\
  Rettorato Universit\'a del Salento, Lecce, Italy
}

\begin{document}

\maketitle

{\setstretch{1.71}
\setlength{\parskip}{-11.2pt}\tableofcontents}

\section{Introduction}
\label{sec:introduction}

Stars are powerful factories for feebly interacting particles such as neutrinos, gravitons, hypothetical axions, and other new particles that are light enough to be produced by nuclear reactions or by thermal processes in the stellar interior. Even when this particle flux cannot be measured, the backreaction on stars can lead to observable consequences.  This ``energy-loss argument'' has been widely used to constrain new particle properties \hbox{\cite{Gamow:1940eny, Gamow:1941gis, Bernstein:1963qh, Stothers:1970ap, Ruderman:1964era, Sato:1975vy, Dicus:1978fp, Vysotsky:1978dc, Mikaelian:1978jg, Sato:1978vy, Turner:1989vc, Raffelt:1990yz, Raffelt:1996wa, Raffelt:2006cw, Giannotti:2017hny, Agrawal:2021dbo, Antel:2023hkf}} and has enjoyed a lively recent resurrection driven by heightened interest in low-mass particles, notably axions \& friends, often called ALPs,\footnote{The term {\em axion-like particle\/} was long used for a pseudoscalar Higgs-like boson. The acronym ALP, specifically denoting a particle with a generic two-photon vertex, gained circulation since 2006 with Joerg Jaeckel's talk at Moriond in March 2006 \cite{Jaeckel:2006id} and in an earlier seminar by Andreas Ringwald at the University of Zurich on 11~January 2006.} WISPs,\footnote{WISP was invented by Andreas Ringwald in his talk {\em Low-Energy Photons as a Probe of Weakly Interacting Sub-eV Particles\/} at the 3rd ILIAS-CERN-DESY Axion-WIMP Workshop, Patras, Greece, on 21~June 2017 (\href{https://axion-wimp2007.desy.de/e30/e122/talk_Ringwald1.pdf}{https://axion-wimp2007.desy.de/e30/e122/talk\_Ringwald1.pdf}). Occasionally, the acronym is also interpreted as Weakly Interacting Slim Particle. It also suggests something a bit eerie, like a ghostly Will-o'-Wisp.} and FIPs (Feebly Interacting Particles), but also keV-range dark matter, and generally low-mass dark sectors. 

We summarize here the main arguments, observational evidence, and resulting constraints on QCD axions as our main example. While a broader contemporary review is sorely missing, it would be too ambitious here to summarize the explosive recent development. In the First Training School of the COSMIC WISPers network in Sept.~2023 in Lecce (\href{https://agenda.infn.it/event/34190/}{https://agenda.infn.it/event/34190/}), one of us (GR) gave lectures on the subject {\em Axions and the Stars\/} and the other (AC) related student tutorials. Many years ago, GR gave similar lectures at an axion school (Nov.~2005) at CERN \cite{Raffelt:2006cw}. We use this opportunity to update the old Lecture Notes to contemporary times.

Broadly following the earlier structure, we review in Sect.~\ref{sec:Interactions} the properties of QCD axions. In Sect.~\ref{sec:Sun} we discuss the Sun as an axion source, notably by the Primakoff process, and review limits on the axion-photon interaction strength by the measured neutrino flux and the CAST experiment before mentioning in Sect.~\ref{sec:Magnetospheres} axion-photon conversion in astrophysical magnetic fields. In \hbox{Sects.~\ref{sec:GC}--\ref{sec:SN}} we review axion limits from globular-cluster stars, white-dwarf and neutron-star cooling, and SN~1987A, and in Sect.~\ref{sec:BH} from black-hole superradiance. In Sect.~\ref{sec:EoS} we mention the backreaction of the axion field on the equation of state in compact stars, in Sect.~\ref{sec:HDM} cosmological hot-dark-matter bounds that complement stellar arguments, and conclude in Sect.~\ref{sec:Conclusions}.

\section{Axion properties}
\label{sec:Interactions}

We briefly review the phenomenological properties of QCD axions, the pseudo Nambu-Goldstone bosons of the Peccei-Quinn (PQ) symmetry \cite{Peccei:1977ur, Peccei:1977hh}. All low-energy properties relevant for stellar physics depend on one energy scale $f_a$, the axion decay constant or PQ scale. The axion mass and interaction strength with standard particles scale with $f_a^{-1}$, meaning that for axions, feeble interactions correlate with small mass. The original Weinberg-Wilczek axion \cite{Weinberg:1977ma, Wilczek:1977pj} assumed $f_a$ at the weak-interaction scale of 246~GeV, implying an axion mass in the 100~keV range, the small $m_a$ allowing for first astrophysical bounds \cite{Dicus:1978fp, Vysotsky:1978dc, Mikaelian:1978jg, Sato:1978vy}. Here we consider invisible axions with much larger $f_a$ and much smaller $m_a$, effectively eliminating any production threshold.

\subsection{Defining properties}

The quantum theory of strong interactions contains a CP-violating term
$\overline\Theta ( g_{\rm s}^2/32\,\pi^2)\,G_{\mu\nu}^b\widetilde G^{b\mu\nu}$,
where $g_{\rm s}$ is the strong gauge coupling constant, $G$ the gluon field-strength tensor, $\widetilde G$ its dual, and $b$ a color index. Among other consequences, it induces a much larger neutron electric dipole moment (EDM) than is experimentally allowed. The PQ solution of this {\em strong CP problem\/} consists of dynamical symmetry restoration by adding the axion field $\overline\Theta\to\overline\Theta+a/f_a$: the potential provided by the CP-violating term drives $a$ to the CP-conserving minimum at $a=-\overline\Theta/f_a$. By construction, at energies far below $f_a$ and the electroweak scale, the most general Lagrangian containing axions~is
\begin{equation}\label{Eq:General}
    \mathcal{L}_a = \frac{1}{2}(\partial_\mu a)^2 
    + \frac{g_{\rm s}^2}{32\,\pi^2}\,\frac{a}{f_{a}}\, G_{\mu\nu}^b\widetilde G^{b\mu\nu}
    + \ldots
\end{equation}
This interaction with gluons is the defining property of QCD axions. The ellipsis stand for additional model-dependent terms such as tree-level couplings to quarks of the type $\frac{\partial_\mu a}{2 f_a}\sum_q C_{aqq}^0 \overline{q}\gamma^\mu \gamma_5 q$. As we will see, often an axion-photon coupling obtains on top of the generic one implied by Eq.~\eqref{Eq:General}, involving the ratio $E/N$ of the electromagnetic and color anomalies.

At leading order in $1/f_a$, the axion can be treated as an external source and the low-energy behavior of the theory is well described by the chiral Lagrangian, which elegantly provides some of the most relevant axion properties. In a two-flavor approximation, with the quark mass matrix $M_q = \big(\begin{smallmatrix}m_u & 0\\ 0 & m_d \end{smallmatrix}\big)$, and defining the matrix $Q_a = M_q^{-1}/\rm Tr(M_q^{-1})$, chosen so that no tree-level mass mixing between the axion and the neutral pion arises, the crucial terms defining some of the basic axion properties are
\begin{equation}\label{Eq:ChiralLagrangianAxionPions}
\mathcal{L}_{a\pi} = \underbrace{\frac{2m_\pi^2}{m_u+m_d} \frac{f_\pi^2}{4} \, 
{\rm Tr}\,\Big(U M_a^\dag + M_a U^\dag\Big)}_\text{from here we read the axion mass}
~+~ \underbrace{\frac{m_d-m_u}{m_d + m_u}\, 
\frac{i f_\pi^2}{4}\, {\rm Tr} \, \Big(\sigma_3 \left\{ U, (\partial_\mu U)^\dag\right\}\Big)\,\frac{\partial_\mu a}{2 f_a}}_\text{from here we read the axion-pion couplings},
\end{equation}
where $U = e^{i \Pi/f_\pi}$, with $\Pi = \big(\begin{smallmatrix}
  \pi^0 & \sqrt{2}\pi^+\\
  \sqrt{2}\pi^- & -\pi^0
\end{smallmatrix}\big)$ the pion matrix, $\sigma_3$ the third Pauli matrix, and $M_a = e^{i \, a Q_a/2 \, f_a} \, M_q \, e^{i \, a Q_a/2 \, f_a}$.

\subsection{Axion mass}

Expanding the chiral Lagrangian in Eq.~\eqref{Eq:ChiralLagrangianAxionPions} to second order in the fields, it is straightforward to derive the axion mass from the quadratic axion term, leading to the traditional result \smash{$m_a^2 = \frac{m_u m_d}{(m_u+m_d)^2} \frac{m_\pi^2 f_\pi^2}{f_a^2}$}. Including next-to-leading-order (NLO) corrections,
the mass is~\cite{GrillidiCortona:2015jxo}
\begin{equation}\label{eq:axmass}
  m_{a}=C_{m_a}\,\frac{f_\pi m_\pi}{f_{a}}
  =\frac{5.69(5)\,{\rm meV}}{f_{a}/10^9\,{\rm GeV}}
\quad\hbox{where}\quad
C_{m_a}=\frac{z^{1/2}}{1+z}+{\rm NLO}=0.457(4),
\end{equation}
where $z=m_{u}/m_{d}=0.472(11)$ and $z^{1/2}/(1+z)=0.467$. The accuracy and very good convergence of the chiral expansion was confirmed~\cite{Gorghetto:2018ocs}, where the authors included QED and NNLO corrections, both of which were found to be an order of magnitude below the NLO ones. 

\clearpage

\subsection{Electric dipole portal}

\begin{figure}[ht]
    \centering
    \vskip-4pt
    \includegraphics[width=0.58\textwidth]{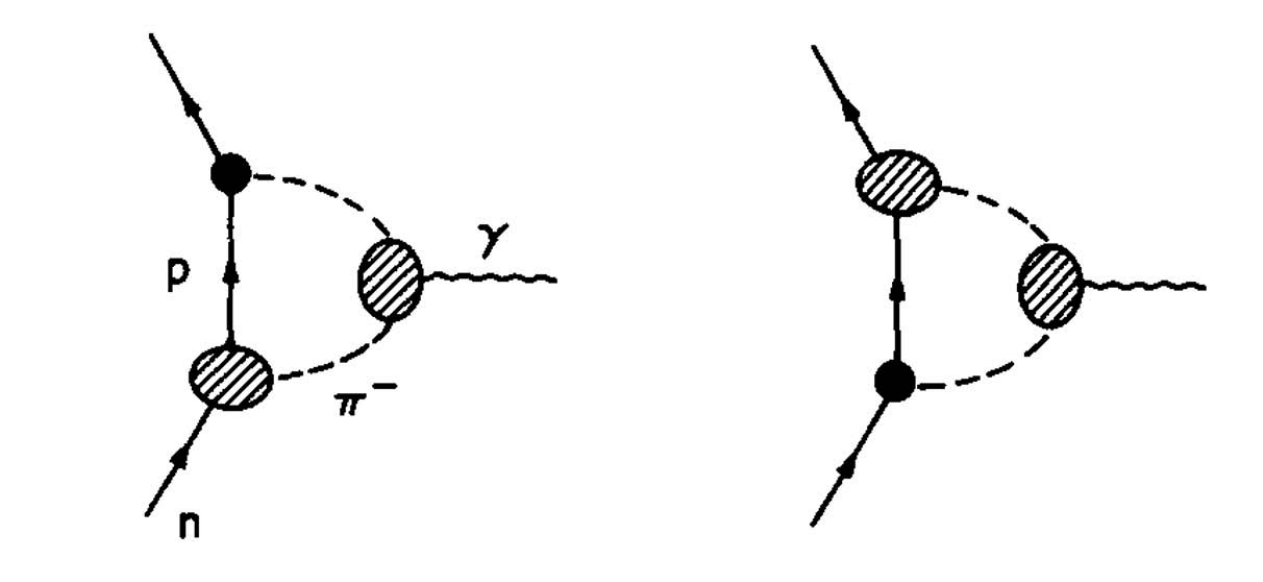}
    \caption{Diagrams for the nucleon EDM. The solid dark blobs indicate the CP-violating pion-nucleon vertex. Switching the role of neutron and proton, one also finds an analogous proton EDM of nearly opposite-equal strength. (Figure from the classic work~\cite{Crewther:1979pi}.)
    }
    \label{fig:Dipole}
    \vskip-6pt
\end{figure}

The defining axion-gluon interaction of Eq.~\eqref{Eq:General} implies several of the generic hadronic axion interactions, notably with nucleons and pions. The most fundamental is a model-independent axion-nucleon-photon vertex, also called nucleon electric-dipole portal, represented by~\cite{Hong:1990yp, Hong:1991fp, Graham:2013gfa}
\begin{equation}\label{eq:EDMportal}
    {\cal L}_{aNN\gamma}^{\rm EDM}=-\frac{i}{2}\,C_{aNN\gamma}\,\mu_N 
    \,\frac{a}{f_a}\,
    \overline{N}\gamma_5\sigma_{\mu\nu}N\,F^{\mu\nu},
\end{equation}
where $C_{aNN\gamma}$ is a numerical coefficient, $\mu_N=e/2m_p = 1.05 \times 10^{-14} \, e\,{\rm cm}$ the nuclear magneton, $N=n$ or $p$ is a nucleon field, and $F^{\mu\nu}$ the EM field-strength tensor. With $a=\overline\Theta f_a$, this vertex reproduces the static nucleon EDM implied by the Theta term, in particular $d_n=\overline\Theta\, C_{ann\gamma} \mu_N$. The origin of the EDM can be understood in that the $G\widetilde{G}$ vertex implies a CP-violating pion-nucleon coupling, which then leads to a nucleon EDM as shown in Fig.~\ref{fig:Dipole}, where the solid dark blobs indicate these CP-violating pion-nucleon vertices.

QCD sum rules imply $C_{ann\gamma}=0.022$ to approximately 40\% precision~\cite{Pospelov:1999ha, Pospelov:1999mv}, while previous simpler estimates based on the effective chiral Lagrangian predicted larger values, either using the MIT quark bag model~\cite{Baluni:1978rf}, or the more sophisticated chiral loop~\cite{Crewther:1979pi} as reported in Table~\ref{tab:DipoleNeutrons}. The best measurement $d_n=(0.0\pm1.1_{\rm stat}\pm0.2_{\rm sys})\times10^{-26}e\,{\rm cm}$ implies $d_n<1.8\times10^{-26}\,e\,{\rm cm}$ at 90\%~C.L.~\cite{Workman:2022ynf} and thus $\overline\Theta\lesssim 1.3\times10^{-10}$ within the error of the theory prediction~\cite{Pospelov:1999ha, Pospelov:1999mv}. Its smallness is the original strong CP problem. Other baryons acquire similar EDMs and especially $C_{app\gamma}\simeq -C_{ann\gamma}$ up to small corrections \cite{Seng:2014pba}. A future proton storage ring could advance the $\overline\Theta$ sensitivity by 3~orders of magnitude \cite{Omarov:2020kws}. 

\begin{table}[b]
    \centering
    \begin{tabular*}{\columnwidth}{@{\extracolsep{\fill}}lllll}
    \hline
     \hline
       &QCD Sum Rules& Quark Bag Model & Chiral Loop  & Lattice QCD\\
       \hline
       $d_n \bigl[10^{-16}\,e\,{\rm cm}\,\overline\Theta\bigr]$\vphantom{$\Big|$} & $2.4 \pm 1.0$~~\cite{Pospelov:1999ha}   & 8.2~~\cite{Baluni:1978rf} & 12~~\cite{Crewther:1979pi} & $1.52 \pm 0.71$~~\cite{Dragos:2019oxn}\\
\hline
\end{tabular*}
    \caption{Neutron EDM computed with different techniques.  
    The quoted result \cite{Pospelov:1999ha} comes from v3 (2005) of the arXiv posting (see also v4 (2005) of the arXiv posting for Ref.~\cite{Pospelov:1999mv}), replacing the published result (1999) that was erroneously a factor of 2 smaller.}    \label{tab:DipoleNeutrons}
\end{table}

Very recently, substantial progress has been made also in the calculation of the Theta term in lattice QCD. One can compute the nucleon EDM directly in terms of CP-violating operators at the quark level. Assuming the Theta term to be the only source of CP violation, Ref.~\cite{Dragos:2019oxn} found $\overline\Theta\lesssim 1.98\times10^{-10}$\hbox{~(90\% C.L.)}, a result later confirmed by other groups. For a recent review of the latest lattice results see Ref.~\cite{Acharya:2023swl}, in particular the right panel of Fig.~4 for the neutron EDM induced by the QCD Theta term.

Intriguingly, the EDM coupling provides a new channel to search for axion dark matter because the classical axion field oscillations spawn time-varying nucleon EDMs~\cite{Graham:2013gfa, JacksonKimball:2017elr}. The vertex in Eq.~\eqref{eq:EDMportal} also implies that stars can emit axions by $\gamma+N\to N+a$ \cite{Lucente:2022vuo}. However, this process dominates in a SN core only if both the axion-neutron and proton couplings vanish with high precision everywhere. If such a situation is conceivable deserves further studies along the lines of what was done for renormalization group evolution effects \cite{DiLuzio:2022tyc}.

\subsection{Interaction with nucleons and pions}

\label{Sect:NucleonsPions}

The couplings to nucleons require some extra care, as they cannot be read directly from the Lagrangians in Eq.~\eqref{Eq:ChiralLagrangianAxionPions}, but involve matching hadronic physics to that of elementary quarks. The generic interaction structure for the CP-conserving axion interaction with a fermion $\psi$ has derivative structure so that it is invariant under the shift $a\to a+a_0$ as behooves a Nambu-Goldstone boson,
\begin{equation}\label{eq:axionfermioncoupling}
  {\cal L}_{a\psi\psi}=\frac{C_{a\psi\psi}}{2f_{a}}\,
  \overline\psi\gamma^\mu\gamma_5\psi\partial_\mu a.
\end{equation}
Here, $\psi$ is the fermion field, $m$ its mass, and $C_{a\psi\psi}$ a model-dependent numerical coefficient. 

This structure often provides the same results as the pseudoscalar form $-i\,(C_{a\psi\psi} m/f_{a})\,\overline\psi\gamma_5\psi a$ and one defines the combination $g_{a\psi\psi}\equiv C_{a\psi\psi} m/f_{a}$ as a Yukawa coupling and \smash{$\alpha_{a\psi\psi}\equiv g_{a\psi\psi}^2/4\pi$} as a fine-structure constant. But one has to be careful because the two interactions are physically distinct theories. This is seen, for example, in the scattering process $a\psi \to a\psi$, where the derivative interaction has vanishing amplitude in the soft limit (as it should by the Adler zero condition), which implies among others that the forward-scattering amplitude and thus possible in-medium refraction vanishes, in contrast to the pseudoscalar interaction. In other words, the two theories differ at the nonlinear axion level and to recover the same soft limit, one needs to add the quadratic term $2m\,\overline{\psi}\psi(a/f_a)^2$ to the Yukawa Lagrangian. It stems from the Lagrangians \smash{$\overline{\psi}(i \gamma^\mu \partial_\mu - m \, e^{i \, 2 \gamma_5 \, a/f_a} )\psi + \frac{1}{2}(\partial a)^2$} and \smash{$\overline{\psi}(i \gamma^\mu \partial_\mu - m)\psi + \frac{1}{f_a} \partial_\mu a \overline{\psi} \gamma^5 \gamma^\mu \psi + \frac{1}{2}(\partial a)^2$} being equivalent up to a field redefinition \smash{$\psi \rightarrow e^{i \gamma_5a/f_a}\psi$}, although this equivalence strictly applies only on the classical level and one has to worry about the well-known chiral anomaly. It is then not surprising that some important axion rates, such as nucleon bremsstrahlung, give different results for the derivative and pseudoscalar case~\cite{Raffelt:1987yt, Carena:1988kr}.

In models without direct axion-quark couplings, the leading-order coupling to nucleons is roughly estimated to be, neglecting sea-quark contributions, \cite{GrillidiCortona:2015jxo}
\begin{equation}\label{Eq:ExpressionQCDNucleonCoupling}
    C_{app} \simeq - \frac{m_d \Delta u + m_u \Delta d}{m_u+m_d}
    \quad\hbox{and}\quad
    C_{ann} \simeq - \frac{m_d \Delta d + m_u \Delta u}{m_u+m_d}.
\end{equation}
Here we introduced the matrix elements $\Delta q$ ($q = u$ or $d$) defined by $s^\mu \Delta q = \bra{p} \overline{q}\gamma^\mu \gamma^5 q \ket{p}$, where $\ket{p}$ is the proton state at rest and $s^\mu$ its spin. With $\Delta u=0.897(27)$ and $\Delta d=-0.376(27)$ and $m_u/m_d=0.467$ \cite{GrillidiCortona:2015jxo} one finds $C_{app}=0.49$ and $C_{ann}=0.03$, compatible with zero within uncertainties. This phenomenologically important fact derives from an accidental cancellation between $m_u/m_d=0.47\simeq 1/2$ and $\Delta_u/\Delta_d=-2.4\simeq -2$, i.e.,  between the $u/d$ mass ratio 1/2 and the $u/d$ number ratio 2 of valence quarks in the neutron. A more precise analysis reveals~\cite{GrillidiCortona:2015jxo}
\begin{equation}\label{eq:KSVZ-NucleonCouplings}
    C_{app}=-0.47(3)
    \quad\hbox{and}\quad
    C_{ann}=-0.02(3).
\end{equation}
These couplings arise from the model-independent axion coupling to gluons and apply, in particular, to the KSVZ model that does not have any direct axion couplings to ordinary quarks or leptons.

For the DFSZ model, for which there are explicit axion couplings to quarks axial currents, the nucleon interactions read \cite{GrillidiCortona:2015jxo}
\begin{equation}\label{eq:DFSZ-NucleonCouplings}
    C_{app}=-0.182(25)-0.435\,\sin^2\beta
    \quad\hbox{and}\quad
    C_{ann}=-0.160(25)+0.414\,\sin^2\beta.
\end{equation}
Here, $\tan\beta=v_u/v_d$ is the ratio of two Higgs vacuum expectation values of this model.\footnote{In the older axion literature, including the standard reference \cite{GrillidiCortona:2015jxo}, $\cot\beta=v_u/v_d$ was used instead. Therefore, what was $\cos^2\beta$ in the axion-electron coupling is now $\sin^2\beta$. This change of convention was introduced in the 2018 edition of the Review of Particle Physics \cite{ParticleDataGroup:2018ovx} to homogenize notation with the Higgs literature. It is also used in the recent and often-cited review \cite{DiLuzio:2020wdo}. Still, for each newer paper one needs to check which convention is followed. Notice that $\tan\beta=1$ implies $\sin^2\beta=1/2$. It was argued that the possible range is $\tan\beta=0.25$--170 \cite{DiLuzio:2020wdo}, translating to $\sin^2\beta=0.06$--1.\label{fn:beta}} $C_{app}$ never becomes very small as a function of $\beta$, whereas $C_{ann}$ vanishes for $\sin^2\beta=0.39(6)$, corresponding to $\tan\beta=0.80(10)$. In this case $C_{app}=0.346$ 

However, the main interest is for processes in a nuclear medium (SN core, neutron star), where nucleon properties significantly change. These finite-density modifications \cite{Balkin:2020dsr, Stelzl:2023} imply that $C_{ann}$ never vanishes everywhere in a compact star, so neutrons always produce some axions.

In analogy to axions, for completeness, we also report here the standard-model nucleon-pion interactions \cite{Carena:1988kr, Choi:2021ign}
\begin{equation}
    {\cal L}_{NN\pi}=\frac{g_A}{2f_\pi}\left[
    \bigl(\overline p\gamma^\mu\gamma_5 p-\overline n\gamma^\mu\gamma_5 n\bigr)\partial_\mu\pi^0
    +\sqrt{2}\,\overline p\gamma^\mu\gamma_5 n\,\partial_\mu\pi^+
    +\sqrt{2}\,\overline n\gamma^\mu\gamma_5 p\,\partial_\mu\pi^-\right],
\end{equation}
with the axial-current coupling constant being $g_A=1.2723(23)$. The coupling constant is often expressed as $g_A/2f_\pi\to f/m_\pi$ with $f$ the ``pion-nucleon coupling constant'' and $m_\pi=139.6$~MeV the charged pion mass. With $f_\pi=92.4$~MeV, this relation implies $f=g_A m_\pi/2 f_\pi=0.961$. The measured value is $f=0.967$--1.007 \cite{Matsinos:2019kqi}, compatible with $f=1$ often used in the older axion literature, but essentially also compatible with $g_A m_\pi/2 f_\pi$. Of course, none of the calculated processes in a nuclear medium is so precise that these few-percent differences would matter.

Passing to the axion couplings to pions, they can be derived from the second term in Eq.~\eqref{Eq:ChiralLagrangianAxionPions}. Expanding the chiral Lagrangian, the leading order terms read~\cite{Chang:1993gm, Choi:2021ign}
\begin{eqnarray}\label{Eq:PionsCouplings}
{\cal L}_{a\pi}&=&\Big(\frac{m_u - m_d}{m_u+m_d} \, + \, C_{auu}^0 - C_{add}^0\Big)\,\frac{\partial_\mu a}{3 \, f_\pi f_a} \Bigl(\pi^0\pi^+\partial^\mu\pi^-
    +\pi^0\pi^-\partial^\mu\pi^+
    -2\pi^+\pi^-\partial^\mu\pi^0\Bigr) 
    \nonumber\\[1.5ex]
    &+& \frac{1}{2} \Big(\frac{m_u - m_d}{m_u+m_d} \, + \, C_{auu}^0 - C_{add}^0\Big)\, \frac{f_\pi}{f_a} \partial_\mu a \partial^\mu \pi^0,
\end{eqnarray}
where for completeness we added \textit{model-dependent} terms, proportional to $C_{auu}^0 - C_{add}^0$, which arise from tree-level coupling between axion and quarks, the ellipsis in our Eq.~\eqref{Eq:General}. Axion-pion processes appear in the early universe just after the QCD confinement epoch \cite{Chang:1993gm}.

The nucleon-pion and nucleon-axion couplings allow for processes such as $\pi+N\to N+a$. In addition, there is also a direct pion-nucleon-axion vertex \cite{Carena:1988kr, Chang:1993gm, Choi:2021ign}
\begin{equation}\label{Eq:AxionNucleonPion}
    {\cal L}_{aNN\pi}=i\frac{C_{aNN\pi}}{f_\pi}\,\frac{\partial_\mu a}{2 f_a}
    \Bigl(\overline p\gamma^\mu\gamma_5 n\,\pi^+
    -\overline n\gamma^\mu\gamma_5 p\,\pi^-\Bigr),
    \quad\hbox{where}\quad
    C_{aNN\pi}=\frac{C_{app}-C_{ann}}{\sqrt{2}\,g_A}.
\end{equation}
In nonrelativistic $NN$ bremsstrahlung, this direct coupling is subdominant, but matters in $\pi N\to N a$. Such processes can be particularly relevant for supernova axions, both for their production and for their detection on Earth \cite{Choi:2021ign, Ho:2022oaw, Lella:2022uwi, Li:2023thv}.

\subsection{Interaction with electrons}

The DFSZ model is a popular axion model that couples both to ordinary quarks and leptons. The derivative interaction structure is analogous to that with nucleons, and the relation to the often-used pseudoscalar structure is as discussed in Sect.~\ref{Sect:NucleonsPions}. The DFSZ electron coupling is~\cite{Dine:1981rt,Zhitnitsky:1980tq}
\begin{equation}\label{Eq:CouplingElectrondDFSZ}
  C_{aee}=\frac{\sin^2\beta}{3}\;
  \quad\hbox{and}\quad
  g_{aee}=\frac{\sin^2\beta}{3}\,\frac{m_e}{f_a}=
  1.50\times10^{-14}\,\frac{m_a}{\rm meV}\,2\sin^2\beta.
\end{equation}
Here, $\tan\beta$ is the ratio of two Higgs vacuum expectation values of this model (see footnote \ref{fn:beta}).

Sometimes a second variant of the DFSZ model is discussed (DFSZ-II), where the assignment of PQ charges to electrons is opposite relative to quarks, meaning that nothing changes for the hadronic interactions, whereas $C_e=-\frac{1}{3}\,\cos^2\beta$ \cite{DiLuzio:2020wdo}. This case should not be confused with the older $\beta$ convention (footnote \ref{fn:beta}), where $C_e=\frac{1}{3}\,\cos^2\beta_{\rm old}$ in the DFSZ-I model. For the anomaly ratio, we have $E/N=8/3$ for DFSZ-I and $E/N=2/3$ for DFSZ-II.

\begin{figure}[ht]
\vskip-4pt
    \centering
\includegraphics[width=0.40\textwidth]{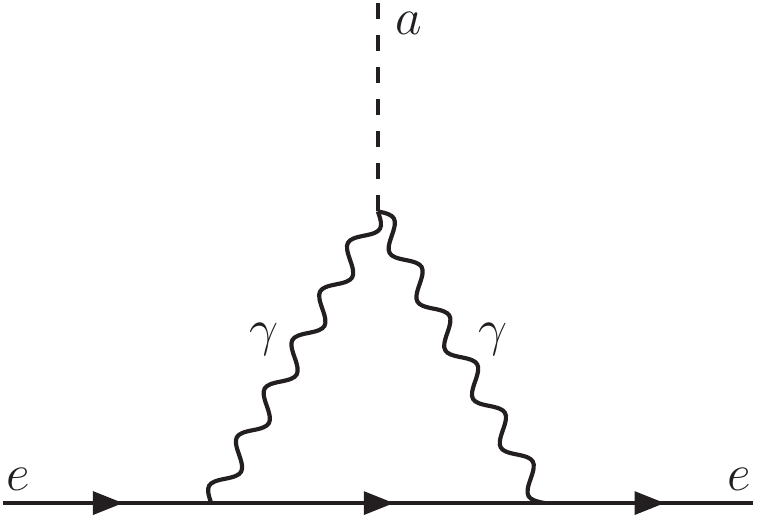}
    \caption{Feynman diagram contributing to the renormalization of the derivative coupling of the axion to electrons and thus to a small effective axion-electron coupling even in the absence of a tree-level coupling.
    }
    \label{fig:KSVZelectrons}
    \vskip-4pt
\end{figure}

In hadronic models such as KSVZ~\cite{Kim:1979if, Shifman:1979if}, by definition axions do not couple to ordinary quarks and leptons at tree level. Nevertheless, the axion-photon interaction implies a loop-induced axion-electron interaction~\cite{Srednicki:1985xd} (see Fig.~\ref{fig:KSVZelectrons}), later corrected by a factor $1/2\pi$ \cite{Chang:1993gm}, providing\footnote{In these Lectures Notes we use {\bf ln} for the Napierian logarithm, and {\bf log} for the logarithm with base 10.}
\begin{equation}
    C_{aee}^{\rm loop}=\frac{3\alpha_{EM}^2}{8\pi^2}\left[\frac{E}{N}\,\ln\left(\frac{f_a}{\mu}\right)-\frac{2(4+z)}{3(1+z)}\ln\left(\frac{\Lambda}{\mu}\right)\right],
\end{equation}
where $\Lambda$ is the QCD scale and $\mu<\Lambda$ a renormalization scale.
In the KSVZ model, $E/N=0$, and the first term drops out.

\subsection{Interaction with photons}

Another generic property of axions is their two-photon interaction that plays a key role for most searches,\footnote{We use the notation that a coupling constant in capital letters is dimensionful, $G_{a\gamma\gamma}$ with dimension of inverse energy, whereas the dimensionless Yukawa couplings to fermions are denoted, for example, by $g_{aee}$.}
\begin{equation}
  {\cal L}_{a\gamma\gamma}=
  \frac{G_{a\gamma\gamma}}{4}\,F_{\mu\nu}\widetilde F^{\mu\nu}a
  =-G_{a\gamma\gamma}\,{\bf E}\cdot{\bf B}\,a .
\end{equation}
Here, $F$ is the electromagnetic field-strength tensor, $\widetilde F$ its dual, and {\bf E} and ${\bf B}$ the electric and magnetic
fields, respectively. The coupling constant is \cite{GrillidiCortona:2015jxo}
\begin{equation}\label{eq:axionphoton}
  G_{a\gamma\gamma}=\frac{\alpha}{2\pi f_{a}}\,C_{a\gamma\gamma}
  \quad\hbox{and}\quad
  C_{a\gamma\gamma}
  =-\frac{2}{3}\,\frac{4+z}{1+z}+{\rm NLO}\,+ \frac{E}{N}
  =-1.92(4) \, + \frac{E}{N},
\end{equation}
where the piece proportional to $E/N$ is \textit{model dependent}. $E$ and $N$ are the electromagnetic and color anomaly of the axial current associated with the axion field. The other piece in $ C_{a\gamma\gamma}$ is instead model \textit{independent} and arises from the axion coupling to gluons. The simplest case is the KSVZ model~\cite{Kim:1979if, Shifman:1979if}, that involves a new Higgs singlet with large vacuum expectation value (vev) and the axion-gluon coupling arises through a triangle graph of a new heavy electrically neutral quark. This implies $E/N=0$ so that the axion-photon interaction arises solely through its mixing with the $\pi^0$ and $\eta$ mesons caused by the axion-gluon interaction. In the DFSZ model~\cite{Zhitnitsky:1980tq, Dine:1981rt}, a grand unified model in which axions also couple to charged leptons, $E/N=8/3$. In the \hbox{DFSZ-II} model mentioned in the previous section, $E/N=2/3$, but with DFSZ we will always mean the usual \hbox{DFSZ-I} model. The KSVZ and DFSZ models are often used as reference cases with $C_{a\gamma\gamma}^{\rm KSVZ}=-1.92(4)$ and $C_{a\gamma\gamma}^{\rm DFSZ}=0.75(4)$, although, in principle, models with practically any value for $E/N$ can be constructed \cite{DiLuzio:2016sbl, Diehl:2023uui}.

Axions or similar pseudoscalars with a two-photon vertex (axion-like particles or ALPs) decay into two photons with the rate
\begin{equation}\label{eq:ALP-decay}
  \Gamma_{a\to\gamma\gamma}
  =\frac{G_{a\gamma\gamma}^2m_{a}^3}{64\,\pi}
  =\frac{C_{a\gamma\gamma}^2}{4C_{m_a}^2}\,\frac{\alpha^2}{64\,\pi^3}
  \,\frac{m_{a}^5}{m_\pi^2 f_\pi^2}
  =3.14\times10^{-25}\,{\rm s}^{-1}\,
  \left(\frac{m_{a}}{\rm eV}\right)^5\,C_{a\gamma\gamma}^2,
\end{equation}
where we used Eq.~\eqref{eq:axmass} in the second step. The age of the universe of $13.8\,{\rm Gyr}=4.35\times10^{17}\,{\rm s}$ reveals that axions decay on a cosmic time scale if $m_{a}\gtrsim24\,{\rm eV}$ for $|C_{a\gamma\gamma}|\simeq1$.

In the $G_{a\gamma\gamma}$--$m_a$ parameter plane, the ``ALP-scape'' often shown for experimental searches and bounds~\cite{cajohare}, QCD axions lie on a line, or rather, within a usually shown golden band that reflects the model uncertainty of $C_{a\gamma\gamma}$.  A plausible range over a broad model range is $|C_{a\gamma\gamma}|=0.074$--17.3 \cite{DiLuzio:2016sbl, Diehl:2023uui}, meaning $|C_{a\gamma\gamma}|/2 C_{m_a}=0.081$--19. We mention in passing that among standard-model particles, the $\pi^0$ and $\eta$ mesons lie in that band and are standard-model ALPs. Specifically, the pion decay rate from the chiral anomaly is $\Gamma_{\pi^0\to\gamma\gamma}=(\alpha^2/64\,\pi^3)\,(m_{\pi}^3/f_\pi^2)=7.75\,{\rm eV}$, very close to the experimental world average of $7.71(11)\,{\rm eV}$ \cite{Workman:2022ynf}, i.e., for $C_{a\gamma\gamma}/2 C_{m_a}=1$, it agrees with the axion for $m_a\to m_\pi$.  Likewise, the $\eta$ meson with $m_\eta=548\,{\rm MeV}$ has $\Gamma_{\eta\to\gamma\gamma}=516\,{\rm eV}$, which is within errors $\Gamma_{\pi^0\to\gamma\gamma}(m_\eta/m_\pi)^3$ is also on that line. In other words, $\pi^0$ and $\eta$ are ALPs that lie on the line defined by Eq.~\eqref{eq:ALP-decay} with $C_{\pi\gamma\gamma}=C_{\eta\gamma\gamma}=0.914$.

\section{The Sun as an axion source}
\label{sec:Sun}

\subsection{How well do we know the inner Sun?}

The Sun at an average distance of $149.6\times10^6$~km is the closest stellar factory for low-mass particles, producing a neutrino flux at Earth of approximately $6.6\times10^{10}~{\rm cm}^{-2}~{\rm s}^{-1}$ with energies up to 18~MeV. The overall process $2e+4p\to{}^4{\rm He}+2\nu_e$ proceeds through various nuclear reactions of the PP chains and CNO cycle. In addition, thermal plasma reactions such as electron-nucleus bremsstrahlung or plasmon decay $\gamma\to\nu\overline\nu$ produce keV-range pairs, in total $3.1\times 10^6~{\rm cm}^{-2}~{\rm s}^{-1}$ of $\nu$ at Earth and an equal $\overline\nu$ flux \cite{Haxton:2000xb, Vitagliano:2017odj, Vitagliano:2019yzm}. Analogous processes produce keV-range axions that can be constrained by the properties of the Sun, and more importantly, potentially detected by helioscopes \cite{Sikivie:1983ip} such as the now-decommissioned CAST \cite{SolarAxionTelescopicAntenna:1999gln, CAST:2017uph}, the future IAXO \cite{IAXO:2019mpb}, and BabyIAXO that is currently under construction \cite{IAXO:2020wwp, Schneekloth:2023}, providing a realistic axion detection opportunity.

How well do we understand the Sun as a particle source? The first solar neutrino observations beginning in 1968 \cite{Davis:1968cp} inaugurated the solar neutrino problem, but also its solution in the form of flavor conversion \cite{Gribov:1968kq}. After decades of neutrino-oscillation research using solar, atmospheric, and laboratory neutrinos, determining the mixing parameters has become precision science \cite{Esteban:2020cvm, Capozzi:2021fjo}. After including the effects of flavor conversion, the predicted and measured solar neutrino flux spectra agree extremely well \cite{Bergstrom:2016cbh, BOREXINO:2018ohr, Gonzalez-Garcia:2023kva}. It is intriguing that very recently, low-energy solar pp neutrinos were measured for the first time in a dark-matter detector, PandaX-4T \cite{Lu:2024ilt}.

Another way to probe the inner Sun is helioseismology, where one uses the measured surface motion from seismic waves (p modes) to infer the sound-speed profile, originally yielding impressive agreement with standard solar models (SSMs). Since the early 2000s, however, agreement turned to tension when modern determinations of the photospheric element abundances appeared. The exact structure of the Sun depends on radiative energy transfer, which in turn depends on the opacity, corresponding to the average photon mean free path. This depends on processes such as Compton scattering, inverse bremsstrahlung (free-free transitions) as well as free-bound and bound-bound atomic transitions---recall that heavier elements in the Sun are not fully ionized. The mass fraction $Z$ of metals (everything other than hydrogen and helium, mass fractions $X$ and $Y$ with $X+Y+Z=1$) is therefore a crucial SSM ingredient~\cite{Vinyoles:2016djt}. For the older GS98 composition \cite{Grevesse:1998bj}, the photospheric metal fraction is $Z/X=0.02292$ and in a SMM corresponds to $Z_{\rm ini}=0.0187(13)$ and also implies $Y_{\rm ini}=0.2718(56)$. (Gravitational settling reduces $Z$ somewhat from $Z_{\rm ini}$.) The modern AGSS09 composition \cite{Asplund:2009fu}, on the other hand, yields $Z_{\rm ini}=0.0149(9)$ and $Y_{\rm ini}=0.2613(55)$, leading to strong tension with the seismic sound-speed profile and depth of the convective zone. It reaches from the surface down to $0.713(1)$ of $R_\odot$ (seismic determination) in agreement with 0.712(5) for the SMM with GS98, but in tension with 0.722(5) for the one using AGSS09 \cite{Vinyoles:2016djt}. This issue is called the solar abundance, metallicity or opacity problem. Very recently (2022), however, yet a new photospheric abundance determination was performed, using new observational material for the Sun, updated atomic data, and up-to-date NLTE (non local thermal equilibrium) model atoms~\cite{Magg:2022rxb}. The resulting $Z$ is similar to GS98, although the element distribution differs significantly. The agreement with helioseismic information is again excellent and the authors think that the 20-year solar abundance problem may be largely resolved.

Neutrinos from the sub-dominant CNO cycle are sensitive to the metal abundance, coming primarily from ${}^{13}{\rm N}\to{}^{13}{\rm C}+e^++\nu_e$ ($E_{\rm max}=1.2$~MeV) and  ${}^{15}{\rm O}\to{}^{15}{\rm N}+e^++\nu_e$ ($E_{\rm max}=1.7$~MeV), with a combined flux at Earth of $4.8\times10^8~{\rm cm}^{-2}~{\rm s}^{-1}$ for a GS98 model (high $Z$), uncertainties $\pm 15\%$ and $\pm 17\%$ for the two components, whereas AGSS09 (low $Z$) yields 3.5 in these units \cite{Vinyoles:2016djt}. Since 2020, the now-decommissioned Borexino experiment has succeeded in measuring this combined flux against large backgrounds in the energy range 0.8--1.2~MeV and found $\Phi_{\rm CNO}= 6.7^{+1.2}_{-0.8}\times10^8~{\rm cm}^{-2}~{\rm s}^{-1}$, strongly favoring the high-$Z$ case \cite{BOREXINO:2020aww, BOREXINO:2023ygs}.

It remains to be seen if these latest developments put to rest the solar opacity uncertainty, but even if it is not completely settled, we know so much detail about the Sun that the remaining uncertainties are not a major concern for its role as an axion source.

\subsection{How much energy loss is allowed by Primakoff emission?}
\label{sec:SolarPrimakoff}

If the Sun were to radiate energy in the form of axions or other new particles, it would need to burn faster to supply energy for both the measured photon flux and the new channel. At its age of 4.57~Gy, the Sun is roughly halfway to exhausting hydrogen in its center, suggesting that its axion luminosity $L_a$ must not exceed $L_\odot=3.85\times10^{33}~{\rm erg}~{\rm s}^{-1}$ for it to live as long, a rough estimate supported by self-consistent solar models evolved with axion losses included \cite{Raffelt:1987yu}.

As an often-used reference case, we consider hadronic axions or general ALPs that are primarily produced by the Primakoff process $\gamma+Ze\to Ze+a$ through their two-photon coupling $G_{a\gamma\gamma}$ in the fluctuating electric fields of the stellar plasma \cite{Dicus:1978fp, Raffelt:1985nk, Raffelt:1987np, Hoof:2021mld}. Besides temperature $T$, two scales of the plasma are important, the plasma frequency $\omega_{\rm p}$ and Debye-H\"uckel screening scale $k_{\rm s}$ given~by
\begin{equation}\label{eq:plasma}
    \omega_{\rm p}^2=4\pi\alpha \sum_i\frac{Z_i^2 n_i}{m_i}\simeq
    \frac{4\pi\alpha\,n_e}{m_e}
\quad\hbox{and}\quad
    k_{\rm s}^2=4\pi\alpha \sum_i\frac{Z_i^2 n_i}{T}.
\end{equation}
The sum is over all charged particles, notably electrons and ions, masses $m_i$ and number densities~$n_i$. Near the solar center, $T=1.3$~keV, $\omega_{\rm p}=0.3$~keV, and $k_{\rm s}=9$~keV, and in our case an overall hierarchy $m_a\ll\omega_{\rm p}\ll T\ll k_{\rm s}$. The plasma frequency provides us with the photon dispersion relation $k^2=\omega^2-\omega_{\rm p}^2$, i.e., photons propagate as if they had a mass $\omega_{\rm p}$, although in general media, the dispersion relation is more complicated. The screening scale, on the other hand, tells us the exponential decline of the electric field of a test charge in plasma and in this sense is also an effective photon mass. The large hierarchy between plasma mass and screening scale is characteristic for a nonrelativistic plasma and depends on $m_e\gg T$ as we can see in Eq.~\eqref{eq:plasma}.

For Primakoff emission, one may safely neglect $m_a$ and $\omega_{\rm p}$ relative to $\omega$, but with average photon energies $\langle\omega\rangle\simeq3T$, there is no strong hierarchy of a typical $\omega$ relative to $k_{\rm s}$.  Ignoring recoil and degeneracy effects one finds \cite{Raffelt:1985nk}
\begin{equation}\label{eq:Gamma-ag}
  \Gamma_{\gamma\to a}= \frac{G_{a\gamma\gamma}^2T k_{\rm s}^2}{32\pi}
  \bigg[\bigg(1+\frac{k_{\rm s}^2}{4\omega^2}\bigg)
  \ln\bigg(1+\frac{4\omega^2}{k_{\rm s}^2}\bigg)-1\bigg].
\end{equation}
Integrating over a thermal photon distribution, the energy-loss rate per unit volume is
\begin{equation}\label{eq:primakofflossrate}
  Q=\frac{G_{a\gamma\gamma}^2T^7}{4\pi}\,F(\kappa^2)\;,
\end{equation}
where $F(\kappa^2)$ is a numerical factor with $\kappa=k_{\rm s}/2T$. In the Sun, $\kappa^2\simeq12\pm15\%$ throughout and for these conditions $F(\kappa^2)\simeq 1.842\,(\kappa^2/12)^{0.31}$ \cite{Schlattl:1998fz}. Integrating over a SMM, using $G_{10}=G_{a\gamma\gamma}/(10^{-10}\,{\rm GeV}^{-1})$, one finds an axion flux at Earth of
\begin{equation}\label{eq:Primakoff-Sun-Flux}
  \Phi_{a}=G_{10}^2\,3.75\times10^{11}\,{\rm cm}^{-2}\,{\rm s}^{-1}
\quad\hbox{and}\quad
  L_{a}=G_{10}^2\,1.85\times 10^{-3} L_\odot.
\end{equation}
The maximum of the distribution is at $3.0\,{\rm keV}$, the average energy is $4.2\,{\rm keV}$. A detailed recent error analysis reveals that these traditional results are surprisingly accurate, the main uncertainty of around $5\%$ derives from differences between SSMs~\cite{Hoof:2021mld}. Conversely, after a future solar axion detection by IAXO, the measured flux could probe details about the inner Sun~\cite{Hoof:2023jol}.

Another recent reexamination of Primakoff emission goes beyond the static limit, allowing for an energy shift of the final-state axion \cite{Liang:2023jlz}. In a dynamical treatment, including electron collective motion, Primakoff emission can be interpreted as transverse to longitudinal plasmon decay $\gamma_{\rm T}\to\gamma_{\rm L}+a$ or coalescence $\gamma_{\rm T}+\gamma_{\rm L}\to a$ \cite{Raffelt:1987np}. The solar axion spectrum is slightly broadened and the overall emission rate reduced by 1--2\%  \cite{Liang:2023jlz}, a small modification, but perhaps interesting if axions were detected and used to probe the Sun.

\begin{figure}[b]
    \centering
    \includegraphics[width=0.6\textwidth]{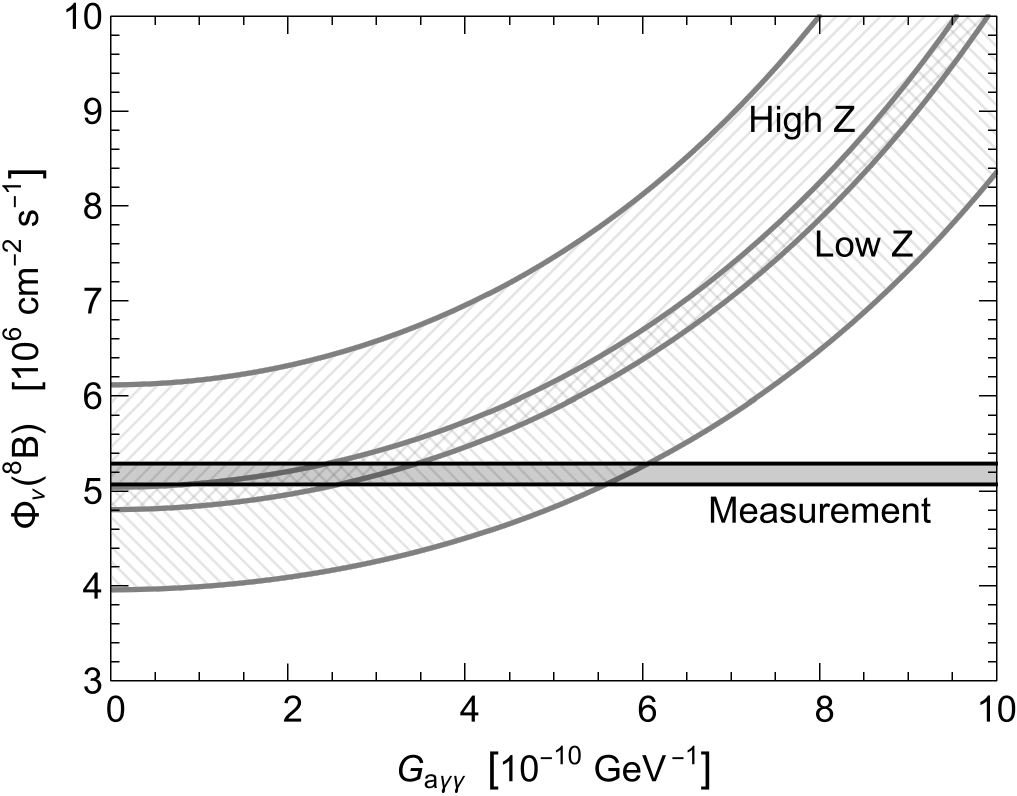}
    \caption{Flux of solar $^8$B neutrinos as a function of assumed Primakoff axion losses. The predicted range from a high-$Z$ and low-$Z$ SSM \cite{Vinyoles:2015aba} as well as the measured value \cite{Gonzalez-Garcia:2023kva} are shown.
    }
    \label{fig:NeutrinoFlux}
\end{figure}

The solar-age requirement of $L_a\lesssim L_\odot$ discussed earlier implies $G_{10}\lesssim23$. However, the increased nuclear burning required to provide $L_a$ implies an increased inner $T$ and thus increased neutrino fluxes, notably from ${}^8{\rm B}\to{}^8{\rm Be}^*+e^++\nu_e$ ($E_{\rm max}\simeq15$~MeV). The flux at Earth is $5.46(67)\times10^6~{\rm cm}^{-2}~{\rm s}^{-1}$ (GS98) and 4.50(54) in these units (AGSS09), to be compared with the measured value of $5.20(10)$ \cite{Gonzalez-Garcia:2023kva}. The $^8$B flux roughly scales with $T_{\rm c}^{18}$ of the solar central temperature and self-consistent solar models imply \cite{Schlattl:1998fz, Gondolo:2008dd, Vinyoles:2015aba}
\begin{equation}\label{eq:SolarNeutrinoFlux}
    \frac{\Phi({}^8{\rm B})}{\Phi_{\rm SMM}({}^8{\rm B})}
    \simeq \left(1+\frac{L_a}{L_\odot}\right)^{4.4}.
\end{equation}
In Fig.~\ref{fig:NeutrinoFlux} we show the range thus predicted for a high-$Z$ and low-$Z$ SSM \cite{Vinyoles:2015aba} as a function of axion losses as well as the measured range, suggesting a constraint
\begin{equation}\label{eq:SolarLimit-Photon}
    G_{a\gamma\gamma}\lesssim 6\times10^{-10}~{\rm GeV}^{-1},
    \quad\hbox{KSVZ:}\quad
    f_a\gtrsim0.4\times10^7~{\rm GeV},\quad
    m_a\lesssim1.5~{\rm eV}.    
\end{equation}
The modified sound-speed profile provides additional constraints \cite{Schlattl:1998fz} and imply $G_{10}<4.1$ from a recent global analysis \cite{Vinyoles:2015aba}. In view of the solar opacity problem, such limits seem a bit more questionable, but should be reconsidered in view of the possible resolution discussed earlier.

\subsection{Trapping limit}

Is there a ceiling to this constraint, i.e., a value of $G_{a\gamma\gamma}$ so large that axions are again allowed? The Primakoff mean free path (MFP) is the inverse of 
Eq.~\eqref{eq:Gamma-ag}. For $4~{\rm keV}$ axions and with $T\simeq 1.3~{\rm keV}$ and $k_{\rm s}\simeq9~{\rm keV}$ at the solar center we find $\lambda_{a}\simeq
G_{10}^{-2}\,6\times10^{24}\,{\rm cm}\simeq G_{10}^{-2}\,8\times10^{13}\,R_{\odot}$ and so one would need something like $G_{a\gamma\gamma}\gtrsim 10^{-3}~{\rm GeV}^{-1}$ for axions to be reabsorbed before leaving the Sun. However, in this case they would contribute to radiative energy transfer much more than photons. To avoid a strong modification of the Sun, their MFP should be less than that of photons, about 10~cm near the solar center~\cite{Raffelt:1988rx}.

However, for QCD axions rather than generic ALPs, large $G_{a\gamma\gamma}$ also implies large $m_a$, specifically $G_{a\gamma\gamma}=C_{a\gamma\gamma}\,2.04\times10^{-7}~{\rm GeV}^{-1}(m_a/{\rm keV})$. Therefore, the energy-loss argument can be applied until the flux is Boltzmann suppressed when $m_a\gg T\simeq 1$~keV. In this regime, axions are primarily produced by photon coalescence (inverse decay) $2\gamma\to a$ \cite{DiLella:2000dn}, not Primakoff production. Coalescence has been invoked for heavy-ALPs production in the Sun \cite{DiLella:2000dn, Beaufort:2023zuj}, horizontal-branch stars~\cite{Carenza:2020zil}, SN cores \cite{Lucente:2020whw, Caputo:2022mah, Ferreira:2022xlw, Muller:2023vjm}, or all main-sequence stars~\cite{Nguyen:2023czp}. For photon energies $\omega> m_a\gg T$ one may use Maxwell-Boltzmann statistics, providing an energy-loss rate per unit volume of
\begin{eqnarray}\label{eq:photoncoalescence}
    Q_a=\frac{G_{a\gamma\gamma}^2m_a^4}{128\pi^3}
    \int_{m_a}^\infty d\omega\,\omega\sqrt{\omega^2-m_a^2}\,e^{-\omega/T}
    =\frac{g_{a\gamma\gamma}^2m_a^6 T}{128\pi^3}K_2\left(\frac{m_a}{T}\right)
    \simeq \frac{g_{a\gamma\gamma}^2m_a^{11/2}\,T^{3/2}}{128\sqrt2\,\pi^{5/2}}e^{-m_a/T},
\end{eqnarray}
where $K_2(x)$ is a modified Bessel function of the second kind, which in {\sc Mathematica} notation is {\tt BesselK[2,x]}, and the last expression obtains for $m_a\gg T$. (For the full expression including quantum statistics see the Supplemental Material of Ref.~\cite{Caputo:2022mah}.) Integrating over a SMM with the KSVZ value $C_{a\gamma\gamma}=-1.92$ we find  $L_a<0.04\,L_\odot$ if $m_a\gtrsim 41$~keV. 
For this mass, $\tau_{a\to\gamma\gamma}\simeq8$~s, to be compared with $R_\odot=6.957\times10^{10}~{\rm cm}=2.32~{\rm s}$, so most of the produced axions escape.

Subsequent decays $a\to2\gamma$ between Sun and Earth would produce hard solar X-rays (HXRs). In the 30--100~keV range, such a flux from the quiet Sun was constrained by the RHESSI satellite to below around $10^{-4}~{\rm cm}^{-2}~{\rm s}^{-1}~{\rm keV}^{-1}$ \cite{Hannah:2010kh}, approximately corresponding to an energy flux below $10^{-15}\,L_\odot$. Axions so heavy would mostly decay before reaching the solar surface, roughly implying $m_a\gtrsim 80$~keV to avoid excessive HXRs, the mass range of the original Weinberg-Wilczek axion.

\subsection{Helioscope searches}

The solar axion flux can be searched with the inverse Primakoff process where axions convert to photons in a macroscopic $B$ field, the ``axion helioscope'' technique~\cite{Sikivie:1983ip}. One would look at the Sun through a ``magnetic telescope'' and place an X-ray detector at the far end. The conversion can be coherent over a large propagation distance and can then be pictured as a particle oscillation phenomenon \cite{Raffelt:1987im}. If an axion with energy $\omega$ travels a distance $L$ in a transverse magnetic field $B$, the conversion probability is
\begin{equation}\label{eq:axion-photon-conversion}
    P_{a\to\gamma}=\left(\frac{G_{a\gamma\gamma}B}{q}\right)^2\sin^2\left(\frac{qL}{2}\right)
    \simeq\left(\frac{G_{a\gamma\gamma}B L}{2}\right)^2~\hbox{for}~|qL|\ll 1,
\end{equation}
where $q=k_a-k_\gamma\simeq (m_a^2-m_\gamma^2)/2\omega$ is the axion-photon momentum difference and the second expression is for the ultrarelativistic limit.

\begin{figure}[ht]
    \centering
    \includegraphics[width=0.6\textwidth]{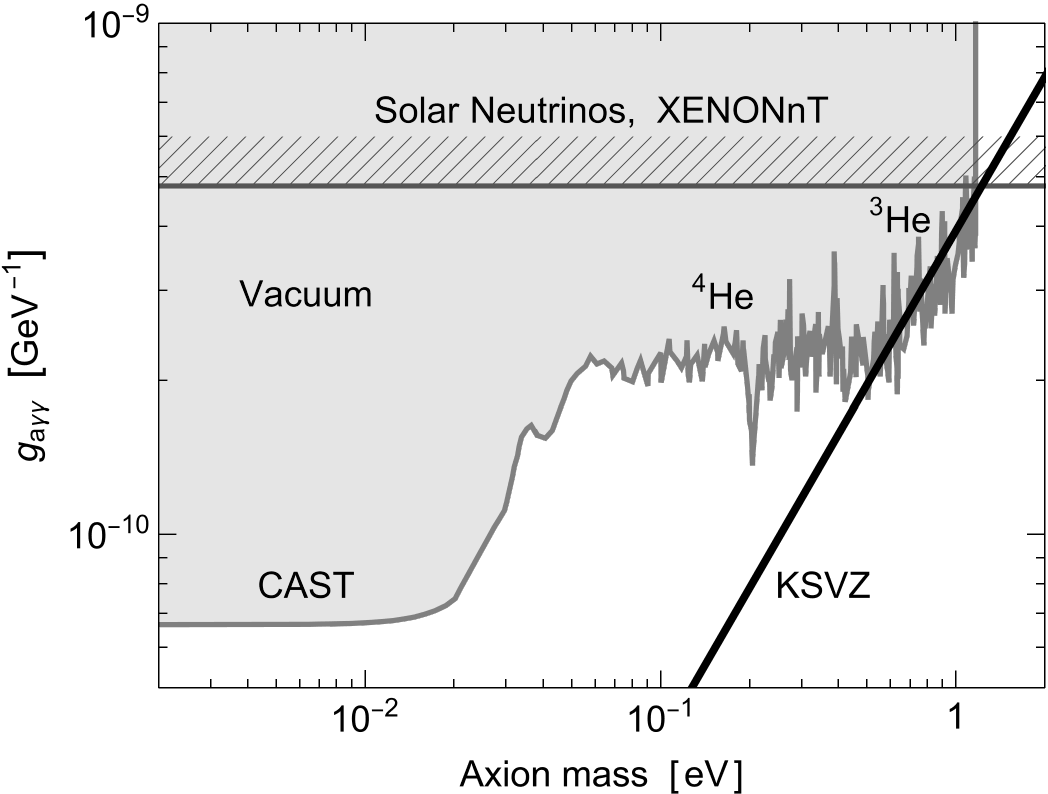}
    \caption{Solar bounds on the axion-photon coupling. The limit from the measured solar neutrino flux was given in Eq.~\eqref{eq:SolarLimit-Photon}, the nearly identical one from the XENONnT experiment in Eq.~\eqref{eq:XENON-photon}. For small search masses, CAST \cite{CAST:2017uph} used evacuated pipes, whereas for larger masses, $^4$He \cite{CAST:2008ixs, CAST:2015qbl} and $^3$He \cite{CAST:2011rjr, CAST:2013bqn} as buffer gas provides photons with a pressure-dependent refractive mass to achieve resonance $m_a\simeq m_\gamma$.
    }
    \label{fig:CAST}
\end{figure}

Early helioscope searches were performed in Brookhaven~\cite{Lazarus:1992ry} and Tokyo \cite{Moriyama:1998kd, Inoue:2002qy, Inoue:2008zp}, whereas the largest one was the now-decommissioned CAST experiment at CERN \cite{CAST:2008ixs, CAST:2015qbl, CAST:2011rjr, CAST:2013bqn, CAST:2017uph}. It used an LHC test magnet ($L=9.26$~m, $B\simeq9$~T) that was mounted such that it could follow the Sun for 1.5~h at dawn and dusk ($\pm8^\circ$ vertical motion). The CAST operation can be seen in an impressive YouTube movie (\href{https://youtu.be/XY2lFDXz8aQ}{https://youtu.be/XY2lFDXz8aQ}). For $m_a$ so small that the inverse axion-photon momentum transfer is small compared with the magnet length, CAST found a limit~\cite{CAST:2017uph}
\begin{equation}\label{eq:castlimit}
  G_{a\gamma\gamma}<0.66\times 10^{-10}\,{\rm GeV}^{-1}
  \hbox{~(95\% C.L.)\qquad for\quad}m_{a}\lesssim0.02\,{\rm eV}.
\end{equation}
For larger masses, many axion-photon oscillation wiggles fit within $L$ and as a function of energy, the oscillatory piece in Eq.~\eqref{eq:axion-photon-conversion} averages to $1/2$ so that $\langle P_{a\to\gamma}\rangle\simeq2 G_{a\gamma\gamma}^2B^2\omega^2/m_a^4$. The solar axion flux scales with $G_{a\gamma\gamma}^2$, so in this case the overall sensitivity scales with $G_{a\gamma\gamma}^4/m_a^4$ and thus degrades with mass as $G_{a\gamma\gamma}\propto m_a$, a line parallel to the axion line in the $G_{a\gamma\gamma}$--$m_a$ parameter space (see Fig.~\ref{fig:CAST}). It is this effect that makes it difficult to probe realistic QCD axion parameters in any macroscopic axion-photon conversion scenario in the laboratory or in astrophysics. In this regime, the CAST constraint is $C_{a\gamma\gamma}\lesssim20$, an order of magnitude short of the KSVZ value. As the overall sensitivity scales with $C_{a\gamma\gamma}^4$ means that four orders are missing in counting rate. 

The suppression by axion-photon momentum mismatch can be removed in a narrow mass range by providing photons with a refractive mass in the presence of a low-$Z$ gas~\cite{vanBibber:1988ge}. This method was first used in the Tokyo experiment~\cite{Inoue:2002qy} and later in CAST with the results shown in Fig.~\ref{fig:CAST}. For a plasma frequency $\omega_{\rm p}=m_\gamma< 0.4$~eV, $^4$He was used as a buffer gas with variable pressure \cite{CAST:2008ixs, CAST:2015qbl}. At the cryogenic $T=1.8$~K of the magnet bore, no larger vapor pressure can be achieved. (Indeed, the superconducting magnet is cooled by liquid helium.) In this range, a typical limit of $G_{a\gamma\gamma}<2.2\times10^{-10}~{\rm GeV}^{-1}$ was achieved (see Fig.~\ref{fig:CAST}). The vapor pressure of $^3$He is larger and the search range up to 0.64~eV was covered with a typical limit of $G_{a\gamma\gamma}<2.3\times10^{-10}~{\rm GeV}^{-1}$  \cite{CAST:2011rjr}. Finally, the range up to 1.17~eV was covered \cite{CAST:2013bqn} with pressure settings and exposure times such as to follow the KSVZ axion line (see Fig.~\ref{fig:CAST}). While the actual limit fluctuates with mass, one can say that for KSVZ axions,
\begin{equation}\label{eq:CAST-mass}
    0.64~{\rm eV}<m_a<1.17~{\rm eV}
\end{equation}
is excluded by CAST.

For DFSZ axions, $C_{a\gamma\gamma}^{\rm DFSZ}=0.75$ compared with  $C_{a\gamma\gamma}^{\rm KSVZ}=-1.92$, so the detection rate is down by a factor $(0.75/1.92)^2=0.15$. On the other hand, the main solar flux now derives from the axion-electron coupling through the ABC processes (see Sect.~\ref{sec:solar-axion-electron}) with a number flux 50 times larger, but with a softer spectrum. While CAST has not analyzed their data for this case, we suspect that DFSZ axions are excluded for the same mass range if $\sin^2\beta$ is not too small.

While the CAST sensitivity range for QCD axions is excluded by other astrophysical bounds to be discussed later, the helioscope technique is one of the few realistic approaches that can turn up QCD axions. A much larger instrument IAXO (International Axion Observatory) has been conceived \cite{IAXO:2019mpb} and as an intermediate step, the smaller BabyIAXO \cite{IAXO:2020wwp} is under construction at DESY in Hamburg. The BabyIAXO sensitivity forecast is down to $G_{a\gamma\gamma}\simeq1.5\times10^{-11}~{\rm GeV}^{-1}$ for $m_a\lesssim0.25$~eV. For KSVZ axions, BabyIAXO has detection potential for $m_a\gtrsim60$~meV, the full IAXO for $m_a\gtrsim 6.8$~meV. For axion models with nonvanishing electron coupling, the sensitivity range is much larger, but other astrophysical constraints are more restrictive, so the open parameter domain with realistic detection opportunity is more complicated.

\subsection{Axion-electron interaction}
\label{sec:solar-axion-electron}

When axions have a direct interaction with electrons as in the DFSZ model, the most efficient emission processes are photo production (Compton scattering) $\gamma+e\to e+a$, bremsstrahlung $e+I\to I+e+a$ (free-free transition), electron-electron bremsstrahlung, as well as free-bound $e+I\to I^-+a$ or bound-bound $I^*\to I+a$ transitions of an excited ion, also called ABC processes for Atomic, Bremsstrahlung, and Compton \cite{Redondo:2013wwa}. The atomic processes contribute around 1/3 of the total flux so that a detailed treatment is necessary. Redondo observed that the axion processes can be scaled to those involving photons and the latter have been studied in detail for the solar opacity, i.e., essentially the photon mean-free path as a function of energy. Based on existing opacity tables, he produced the solar axion flux from the ABC processes that shows many discrete lines \cite{Redondo:2013wwa}. This flux uses an AGSS09 (low $Z$) SSM and thus underestimates the atomic contributions compared to the newly favored high-$Z$ models that we discussed earlier. 

\eject

Integrating Redondo's tabulated ABC flux and using $g_{11}=g_{aee}/10^{-11}$, one finds an axion flux at Earth of
\begin{equation}
  \Phi_{a}=g_{11}^2\,2.65\times10^{13}\,{\rm cm}^{-2}\,{\rm s}^{-1}
\quad\hbox{and}\quad
  L_{a}=g_{11}^2\,7.15\times 10^{-2} L_\odot.
\end{equation}
The average energy is $2.3\,{\rm keV}$, much smaller than for Primakoff production. With the nominal constraint $L_a<0.04\,L_\odot$ as in the Primakoff case one finds
\begin{equation}\label{eq:SolarLimit-electroncoupling}
    g_{aee}< 0.75\times10^{-11},
    \quad\hbox{DFSZ:}\quad
    f_a>1.14\times10^7~{\rm GeV}\,(2\sin^2\beta)
    \quad
    m_a<\frac{0.50~{\rm eV}}{2\sin^2\beta}.    
\end{equation}
For $\sin^2\beta=1/2$, this limit corresponds to $G_{a\gamma\gamma}<0.76\times10^{-10}~{\rm GeV}^{-1}$ and a Primakoff number flux at Earth 1.9\% that of the ABC flux.

The sensitivity of direct dark-matter search experiments has improved to the point where solar neutrinos begin forming the main background, the neutrino floor or neutrino fog for WIMP searches~\cite{OHare:2021utq, Lu:2024ilt}. The low detection threshold for nuclear recoils implies good sensitivity to keV-energy particles from the Sun and notably axions \cite{LUX:2017glr, PandaX:2017ock, XENON:2020rca, XENON:2022ltv}. The XENONnT experiment currently reports the best bound on the axion-electron interaction, based on the solar ABC flux, of \cite{XENON:2022ltv}
\begin{equation}\label{eq:XENONnT-electroncoupling}
    g_{aee}< 1.9\times10^{-12}~\hbox{(90\% C.L.)}
    \quad\hbox{DFSZ:}\quad
    f_a>4.5\times10^7~{\rm GeV}\,(2\sin^2\beta)
    \quad
    m_a<\frac{0.13~{\rm eV}}{2\sin^2\beta}.    
\end{equation}
An earlier signal reported by XENON1T \cite{XENON:2020rca} is now attributed to background.

Even if axions do not couple to electrons, the solar Primakoff flux can still trigger a signal through keV-photons from inverse Primakoff production \cite{Gao:2020wer, Dent:2020jhf, Abe:2020mcs}. XENONnT reports a limit on the {\em photon\/} coupling of
\begin{equation}\label{eq:XENON-photon}
    G_{a\gamma\gamma}\lesssim 4.8\times10^{-10}~{\rm GeV}^{-1},
    \quad\hbox{KSVZ:}\quad
    f_a\gtrsim0.46\times10^7~{\rm GeV}\quad
    m_a\lesssim1.2~{\rm eV}.
\end{equation}
This result is practically identical with the solar energy-loss bound of Eq.~\eqref{eq:SolarLimit-Photon}. 

Inverse Primakoff conversion in a crystal lattice was previously used, notably in the context of experiments otherwise searching for neutrinoless double-$\beta$ decay. The most recent such bound of $G_{a\gamma\gamma}<1.45\times10^{-9}~{\rm GeV}^{-1}$ (95\% C.L.) comes from the Majorana Demonstrator \cite{Majorana:2022bse}. According to Eq.~\eqref{eq:Primakoff-Sun-Flux}, this limiting coupling strength would imply $L_a\simeq 0.4\,L_\odot$ and thus is barely compatible with the properties of the Sun and would imply a huge solar neutrino flux.

\subsection{Axions from nuclear reactions in the Sun}

The Sun would emit QCD axions also by their nuclear interaction. One frequent process is $p+d\to{}^3{\rm He}+\gamma$(5.5~MeV), the second step in the solar PP Chains, where occasionally an axion can substitute for the photon, but there are several other similar reactions. With a larger Lorentz factor for the same mass, the escaping axion flux is less suppressed by decays within the Sun and the limit on solar $\gamma$-rays is more restrictive than that on hard X-rays, allowing one to exclude yet larger axion masses or other types of new particles that decay between the Sun and Earth \cite{Raffelt:1982dr, Massarczyk:2021dje, Gustafson:2023hvm}. The Sun does not seem to leave a ceiling to the energy-loss limit of Eq.~\eqref{eq:SolarLimit-Photon} for QCD axions.

More interesting would be the direct detection of such mono-energetic axions or other particles through their photon, electron, or nucleon interaction \cite{Krcmar:2001si, Belli:2012zz, CAST:2009klq, Borexino:2012guz, Bhusal:2020bvx, Derbin:2013zba, Lucente:2022esm, DEramo:2023buu}, but the sensitivity is not competitive with the solar energy-loss limit. On the other hand, for special model parameters, the axion-photon or axion-electron coupling might be strongly suppressed and one may wish to rely on the axion-nucleon coupling alone. One analysis uses SNO data, based on the solar production of $p+d\to{}^3{\rm He}+a$(5.5~MeV) and the detection by $a+d\to p+n$ with a threshold of 2.2~MeV, providing a constraint on the isovector axion-nucleon coupling of \cite{Bhusal:2020bvx}
\begin{equation}\label{eq:SNOlimit}
    \left|\frac{g_{app}-g_{ann}}{2}\right|< 2\times 10^{-5}~\hbox{(95\% C.L.),}
    \quad\hbox{KSVZ:}
    \quad
    f_a>1.1\times10^4~{\rm GeV}, 
    \quad
    m_a<540~{\rm eV}.    
\end{equation}
These constraints on $m_a$ and $f_a$ follow from the KSVZ nuclear couplings of Eq.~\eqref{eq:KSVZ-NucleonCouplings}.

A similar case is the solar process ${}^7{\rm Be}+e\to{}^7{\rm Li}^*+\nu_e$ followed by ${}^7{\rm Li}^*\to{}^7{\rm Li}+\gamma$(477.6~keV), with the axion substituting for the photon, and then detection by the inverse resonant process $a+{}^7{\rm Li}\to{}^7{\rm Li}^*\to{}^7{\rm Li}+\gamma$(477.6~keV) in the laboratory~\cite{Krcmar:2001si}. The most restrictive limit from this channel of $m_a<8.6$~keV \cite{Belli:2012zz} is not competitive with Eq.~\eqref{eq:SNOlimit}.

More restrictive limits arise from low-lying nuclear levels that are thermally excited and occasionally produce axions in their deexcitation, similar to the atomic bound-bound process to be discussed in Sect.~\ref{sec:solar-axion-electron}. The first example is the M1 transition in $^{57}$Fe (natural abundance 2.2\%) with energy 14.4~keV, not too high relative to $T\simeq1.3$~keV in the inner Sun \cite{Haxton:1991pu, Moriyama:1995bz}. Detection by the inverse process $a+{}^{57}{\rm Fe}\to{}^{57}{\rm Fe}^*\to{}^{57}{\rm Fe}+\gamma$ provides a constraint \cite{Derbin:2011zz} 
\begin{equation}\label{eq:solar-Fe}
    |\beta_N g_{aNN}^0+g_{aNN}^1|<3.0\times 10^{-6}~\hbox{(95\% C.L.)}
    \quad\hbox{KSVZ:}
    \quad
    f_a>2.1\times10^4~{\rm GeV}, 
    \quad
    m_a<273~{\rm eV},
\end{equation}
where $\beta_N=-1.19$ derives from the nuclear matrix element \cite{Haxton:1991pu}. For $\beta_N=-1$, the constraint would be purely on the neutron coupling and in any case strongly depends on $C_n$. With older couplings, Ref.~\cite{Derbin:2011zz} found a stronger constraint of $m_a<145$~eV. Other authors found $\beta_N=-1.31$ \cite{Avignone:2017ylv}, implying  $m_a<191$~eV. A new dedicated project ISAI (Investigating Solar Axion by Iron-57) is being commissioned \cite{Onuki:2023zhq} and, after a few years of data taking, aims at a sensitivity to $m_a\simeq3$~eV, while the future EISAI could push it below 2~eV.

An analogous search used the 9.4~keV first excited level of $^{83}$Kr \cite{Gavrilyuk:2015aea, Akhmatov:2018kjv}, essentially a neutron M1 transition, similar to the case of $^{57}{\rm Fe}$. Assuming it is a pure neutron transition, the relevant coupling is $g_{ann}=g_{aNN}^0-g_{aNN}^1$ and the experimental limit is $|g_{ann}|<0.84\times10^{-6}$. While this is more restrictive than Eq.~\eqref{eq:solar-Fe}, it is purely on the neutron coupling which in the KSVZ model is compatible with zero. So it makes little sense to express this bound in terms of the axion mass.

The most recent case uses the 8.4~keV first excited level of $^{169}$Tm (Thulium), with the great advantage of this being a nearly pure proton M1 transition with $\beta_N\simeq+1$ \cite{Derbin:2023yrn}. They find a bound
\begin{equation}\label{eq:Thulium}
    |g_{app}|< 8.89\times 10^{-6}~\hbox{(90\% C.L.),}
    \quad\hbox{KSVZ:}
    \quad
    f_a>4.9\times10^4~{\rm GeV}, 
    \quad
    m_a<115~{\rm eV}.    
\end{equation}
These authors instead state the mass limit as 141~eV,
based on older assignments of $C_{ann}$ and $C_{app}$ used in their Eq.~(10). For KSVZ axions, this is currently the most restrictive limit by this method. Of course, the cooling speed of neutron stars (Sect.~\ref{sec:NS}) and the neutrino signal of SN 1987A (Sect.~\ref{sec:SN}) provide far more restrictive limits on the axion nuclear couplings.

\clearpage

\subsection{Hidden photons}

We have seen that the Sun is most useful as a particle source for direct searches and still holds the promise of future QCD axion detection with (Baby)IAXO, whereas the impact of energy loss on the Sun provides less restrictive limits. However, beyond axions, we mention one exception to this rule, the case of hidden photon (HP) emission. They are hypothetical photon-like particles that have a vacuum mass and mix weakly with ordinary photons. If $m_{\rm HP}<\omega_{\rm p}$ of the ambient plasma, there is always a momentum $k$ where their $(\omega,k)$ with $\omega^2=m_{\rm HP}^2+k^2$ matches that of longitudinal plasmons $(\omega_{\rm p},k)$ which have the dispersion relation $\omega_{\rm p}\simeq k$. One then obtains efficient resonant emission \cite{An:2013yfc, Redondo:2013lna, An:2020bxd}, constrained by the condition $m_{\rm HP}<\omega_{\rm p}$ which is different for different types of stars. Solar constraints apply for $m_{\rm HP}\lesssim 0.3$~keV, a typical solar plasma frequency. Notice that the variation of the solar neutrino flux with HP luminosity is slightly different than that of axions: the exponent of 4.4 in Eq.~\eqref{eq:SolarNeutrinoFlux} is found to be 5.7 instead \cite{Vinyoles:2015aba}.

\subsection{Solar basin}

Some of the particles produced in stars will be gravitationally retained \cite{Hannestad:2001xi, DiLella:2002ea}, i.e., those produced with such small energies that they are slower than the stellar escape velocity. This effect will be especially significant for particles with masses of the order of the interior stellar $T$, the keV range for the Sun. The original discussions envisioned Kaluza-Klein excitations, i.e., a tower of particles with a large range of masses, allowing for a large population of trapped particles that individually interact very feebly.

An intriguing fresh incarnation of this idea under the name of {\em solar basin\/} finds that the gravitationally trapped particles can build up to a density at Earth exceeding that of  dark matter~\cite{VanTilburg:2020jvl}. Therefore, direct-detection experiments, such as XENON1T, LUX, and PandaX-II, strongly constrain the axion-electron coupling for keV-mass ALPs. Further limits, both on electron and photon couplings, also derive from the recent X-ray measurements of the quiescent Sun taken by the NuSTAR satellite to look for the radiative decay of the basin axions~\cite{DeRocco:2022jyq, Beaufort:2023zuj}.

For our main particles of interest in this review, QCD axions, solar or stellar basins to not seem to provide new insights, simply because of their small mass, but of course for more general FIPs and WISPs, it is a fantastic laboratory.

\section{Axion-photon conversion in neutron-star magnetospheres}
\label{sec:Magnetospheres}

In a magnetic field, axions and photons mix \cite{Sikivie:1983ip, Raffelt:1987im}, leading to various potentially observable effects in astrophysics and cosmology, such as changing photon polarization or very-high energy $\gamma$-rays reaching to larger distances than they otherwise would. Many papers have thus derived constraints in the ALP-scape spanned by $G_{a\gamma\gamma}$ and $m_a$, of which the CAST exclusion plot Fig.~\ref{fig:CAST} shows a small section. The full panorama and concomitant references are found on Ciaran O'Hare's github page \cite{cajohare}, whereas a comprehensive formal review seems to be missing.

For a perspective on the conceivable sensitivity of such methods, we glean from Eq.~\eqref{eq:axion-photon-conversion} that the conversion is large if $G_{a\gamma\gamma}BL\simeq1$. The same combination $BL$ for a typical field strength and spatial extent determines the energy that can be achieved for cosmic-ray acceleration. A collection of different astrophysical environments in the parameter space of $B$ and $L$ is known as a Hillas plot \cite{Hillas:1984ijl}, ranging from pulsars to the intergalactic medium. To achieve proton acceleration up to $10^{20}$~eV requires $BL\simeq3\times10^{11}$~GeV, with many environments near or below this line. Considering efficient ALP-photon conversion, this line corresponds to $G_{a\gamma\gamma}\simeq3\times10^{-12}~{\rm GeV}^{-1}$. Indeed, the best bounds on $G_{a\gamma\gamma}$ from very different astrophysical systems are not far from this estimate which looks hard to surpass. Of course, the larger is $L$ the smaller is $m_a$ before the conversion is suppressed by momentum mismatch. None of the many arguments in the ALP-scape come close to reaching the axion line, defined by the $G_{a\gamma\gamma}$--$m_a$ relation for typical models. Except in CAST and the future (Baby)IAXO, axion-photon conversion limits do not come close to QCD axions.

This limitation is overcome in the search for dark-matter axions \cite{Sikivie:1983ip, Irastorza:2018dyq}, where typically one engineers a resonance between a chosen search mass $m_a$ and the produced photon that could be, for example, an excitation of a microwave cavity. It is intriguing that a similar effect obtains in pulsar magnetospheres, where $B \gtrsim 10^{12}~{\rm G}$ is a typical scale. Dark-matter axions falling towards a pulsar will typically encounter a resonance between $m_a$ and the plasma frequency $\omega_{\rm p}$, potentially leading to a sharp radio line, an early idea~\cite{Pshirkov:2007st} that was recently developed further by several groups \cite{Huang:2018lxq, Hook:2018iia, Leroy:2019ghm,  Battye:2019aco, Battye:2021xvt, Witte:2021arp, Millar:2021gzs, McDonald_2023, McDonald:2023shx} and has already led to nontrivial constraints using radio-telescope data \cite{Safdi:2018oeu, Foster:2020pgt, Battye:2021yue, Foster:2022fxn, Battye:2023oac}. 

Independently of axion dark matter, the pulsar itself can produce dense axion clouds in its polar cap regions, where large values of ${\bf E}\cdot{\bf B}$ obtain that act as a source. Many of these axions will stay gravitationally trapped. These bound clouds are difficult to dissipate and can thus grow to enormous densities over the lifetime of the star, possibly producing spectacular signals~\cite{Prabhu:2021zve, Noordhuis:2022ljw, Noordhuis:2023wid, Caputo:2023cpv}.

\section{Globular-cluster stars}
\label{sec:GC}

\begin{figure}[ht]
  \centering
  \includegraphics[width=0.65\textwidth]{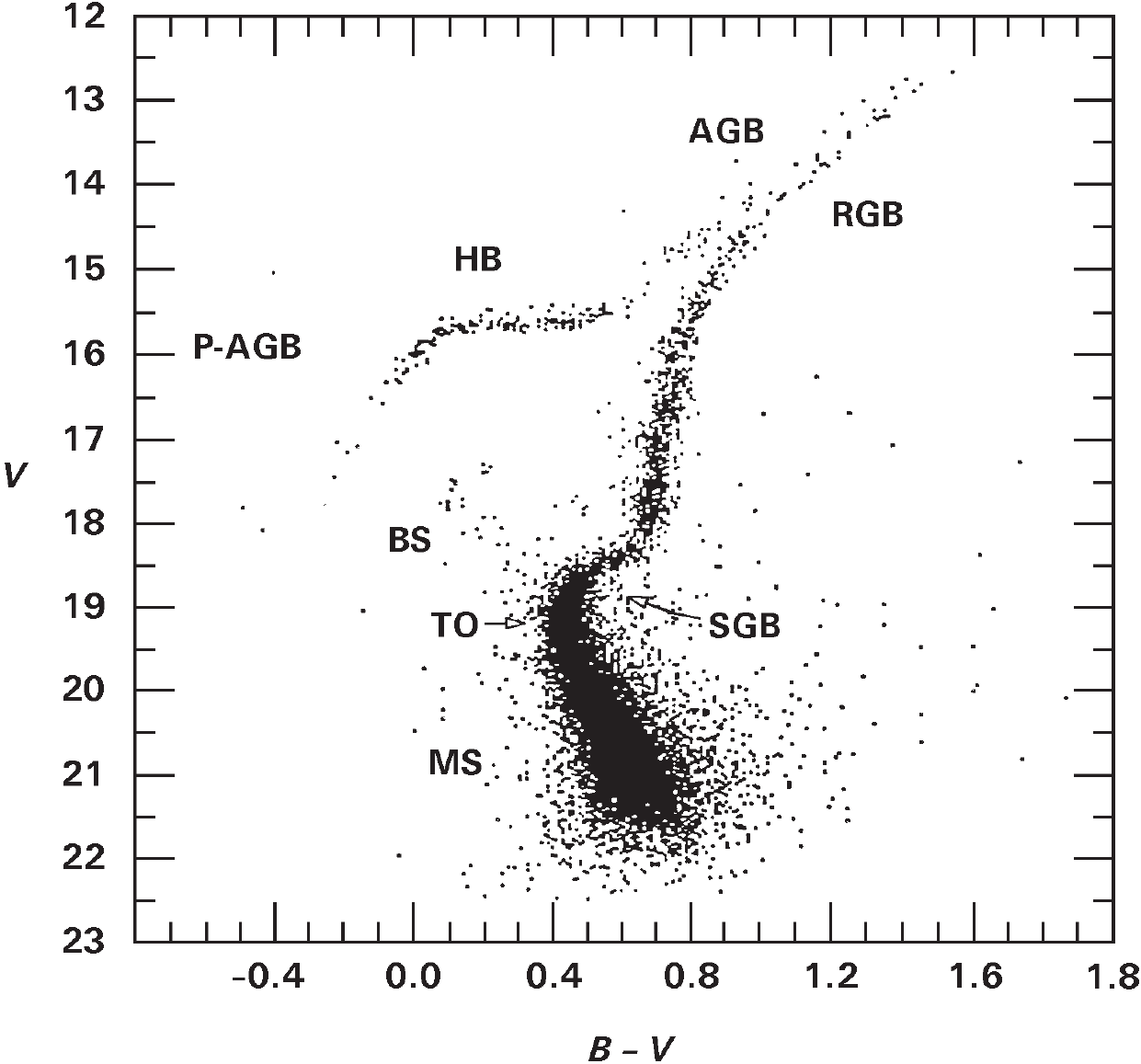}
  \caption{Color-magnitude (CM) diagram for the globular cluster M3, based on 10,637 stars~\cite{Buonanno:1986}. Vertically is the brightness in the visual ($V$) spectral band, horizontally the difference between $B$ (blue) and $V$ brightness, i.e., a~measure of color and thus surface temperature, where blue (hot) stars lie toward the left.  The classification for the evolutionary phases is as follows~\cite{Renzini:1988cs}.  MS (main sequence): core hydrogen burning. BS (blue stragglers). TO (main-sequence turnoff): central hydrogen is exhausted.  SGB (subgiant branch): hydrogen burning in a thick shell. RGB (red-giant branch): hydrogen burning in a thin shell with a growing core until helium ignites. HB (horizontal branch): helium burning in the core and hydrogen burning in a shell. AGB (asymptotic giant branch): helium and hydrogen shell burning. P-AGB (post-asymptotic giant branch): final evolution from the AGB to the white-dwarf stage.}
  \label{fig:colmag}
\end{figure}

Restrictive limits on axion couplings arise from globular-cluster (GC) stars. A GC is a gravitationally bound system of up to a few million stars that formed at the same time and thus differ primarily in their mass. Star formation ended when the first SNe blew out the gas. Our own galaxy hosts around 150 GCs \cite{Harris:1996kt} in a spherical halo, i.e., GCs are usually not in the galactic disk and represent an early generation of stars. A GC provides a rather homogeneous population, allowing for detailed tests of stellar-evolution theory~\cite{Renzini:1988cs}. The stars surviving today since formation have masses somewhat below $0.85\,M_\odot$.  In a color-magnitude (CM) diagram (Fig.~\ref{fig:colmag}), where one plots essentially the surface brightness vs.\ the surface temperature, stars appear in characteristic loci, allowing one to identify their state of evolution as indicated in the figure. Any modification of stellar evolution will distort the entire CM diagram and the distribution of stars along the different branches, but the effect of axion emission can have a rather specific effect. We will consider the number of HB vs.\ RGB stars, the so-called $R$-parameter and the number of AGB vs.\ HB stars ($R_2$-parameter) as a good criterion to constrain $G_{a\gamma\gamma}$ and the brightness of the tip of the red-giant branch (TRGB), providing the best sensitivity to $g_{aee}$.

\subsection{Tip of the red giant branch (TRGB) and axion-electron interaction}

\label{sec:TRGB}

\begin{figure}[ht]
  \centering
  \includegraphics[width=0.55\textwidth]{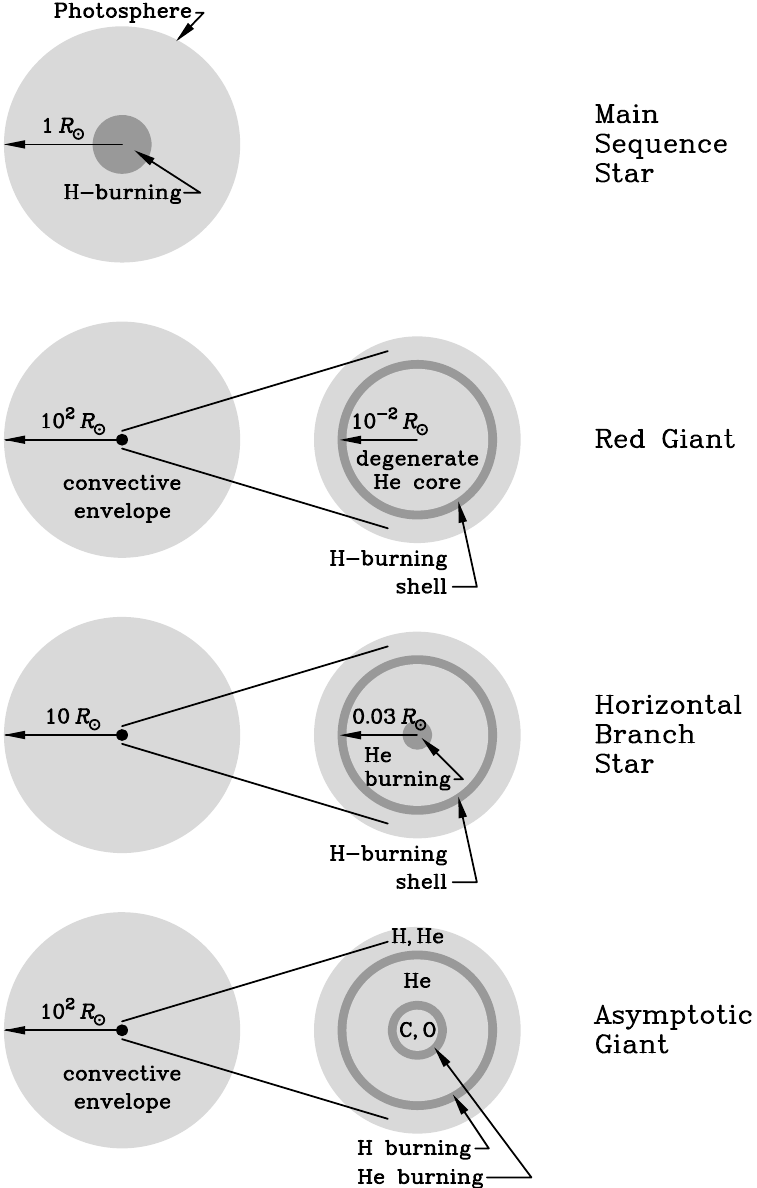}
  \caption{Evolutionary phases of low-mass stars. This figure only gives a rough impression of the dimensions of the core and the envelope, as well as the various elements that burn in the different stellar phases. RGB, HB and AGB stars are all crucial laboratories for axion physics as we discuss in this and next sections.}
  \label{fig:PhasesStellar}
\end{figure}

Stars on the main sequence (MS), like our Sun, burn hydrogen to helium in their central regions. In typical GCs, stars with $M\gtrsim 0.85\,M_\odot$ have exhausted central hydrogen and turn off (TO) from the~MS. Before the advent of precision cosmology, the MS--TO in GCs provided a critical lower limit to the age of the universe. No fresh stars form because the required gas was blown out by~SNe. After thick-shell H burning during the sub-giant (SGB) phase, a degenerate He core forms that shrinks in size with increasing mass ($R\propto M^{-1/3}$) as behooves a system supported by electron degeneracy pressure. As the He core becomes more massive and smaller, its surface gravitational potential increases, driving up $T$ in the H-burning shell, and thus driving up the luminosity. Stars along the MS have increasing total mass, stars along the RGB have a nearly fixed total mass, but increasing He-core mass. The star ascends the RGB until the core is so dense and hot that He ignites, roughly at $\rho \simeq 10^6~{\rm g/cm}^3$ and $T \simeq 10^8~{\rm K}= 8.6~{\rm keV}$. Notice that degenerate He ignition happens only for low-mass stars with $M\lesssim2$--$3\,M_\odot$, heavier stars ignite it under nondegenerate conditions. The degenerate RG core before He ignition is essentially a helium WD, except that it carries an H-burning shell. After He ignition, a quick run-away takes place, sometimes described as the He flash, the core expands to nondegenerate density of some $10^4~{\rm g}~{\rm cm}^{-3}$ with $T\simeq10^8$~K and quietly burns He to C and O, becoming a horizontal branch (HB) star. It burns He in the core and continues to burn H in a shell, but overall is much dimmer because the larger core radius dials down $T$ in the H shell and thus the dominant H luminosity. In the next sections we will turn to He-burning stars as particle-physics laboratories. Figure~\ref{fig:PhasesStellar} shows the main evolutionary phases of low mass stars, providing a schematic impression of the dimensions of the core and the envelope, and the different burning elements.

Helium ignition depends sensitively on $\rho$ and $T$ so that a novel cooling agent can force the core to grow more massive and luminous before it flashes: the tip of the red giant branch (TRGB) would become brighter. (The figure of speech that axions delay He ignition was critiqued because in a temporal sense, it is actually advanced because the star ascends the RGB a bit faster.) The idea that He ignition should not be prevented entirely led to the first bounds on axions, majorons, and familons in 1985 \cite{Dearborn:1985gp}. This argument was refined by considering the putative TRGB brightness increase relative to observations, leading to bounds on neutrino magnetic dipole moments \cite{Raffelt:1989xu, Raffelt:1990pj}. They would enhance the plasmon-decay emissivity $\gamma\to\nu\overline\nu$ which is the dominant standard energy-loss channel near the TRGB. The latest bound of this type of $\mu_\nu<1.5\times10^{-12}\mu_{\rm B}$ (95\% C.L.) \cite{Capozzi:2020cbu} remains the most restrictive limit on this quantity. Returning to axions, a refined treatment implies a bound $g_{aee}\lesssim2.5\times10^{-13}$ \cite{Raffelt:1994ry}, so henceforth we use the scaled variable $g_{13}=g_{aee}/10^{-13}$ in this context. One innovation was an interpolation formula for both photo production ($\gamma+e\to e+a$) and bremsstrahlung ($e + Z e \rightarrow e + Z e + a$), where the Compton term is needed in the nondegenerate parts of the star. Very recently, these approximate rates were shown to be uncannily accurate~\cite{Carenza:2021osu}. Together with another mid-1990s study \cite{Catelan:1995ba}, the picture emerged that a novel energy loss at $T\simeq10^8~{\rm K}$ and an average density $\langle\rho\rangle=2\times10^5~{\rm g}~{\rm cm}^{-3}$ should not exceed about $10~{\rm erg}~{\rm g}^{-1}~{\rm s}^{-1}$. At these conditions, the standard neutrino emission is about $4~{\rm erg}~{\rm g}^{-1}~{\rm s}^{-1}$. In other words, the bounds from that generation were still quite rough.

\begin{figure}[b]
  \centering
  \hbox to\hsize{\includegraphics[height=0.34\textwidth]{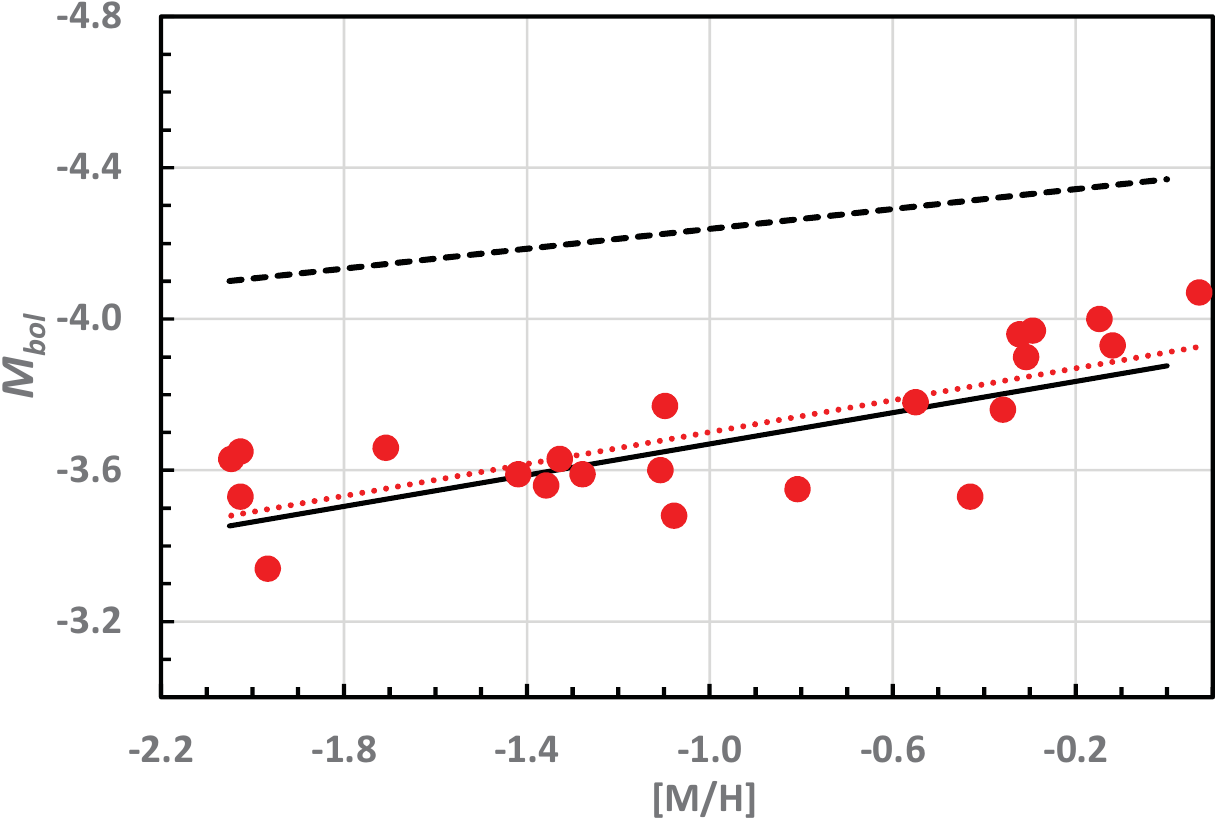}\hfill
  \includegraphics[height=0.34\textwidth]{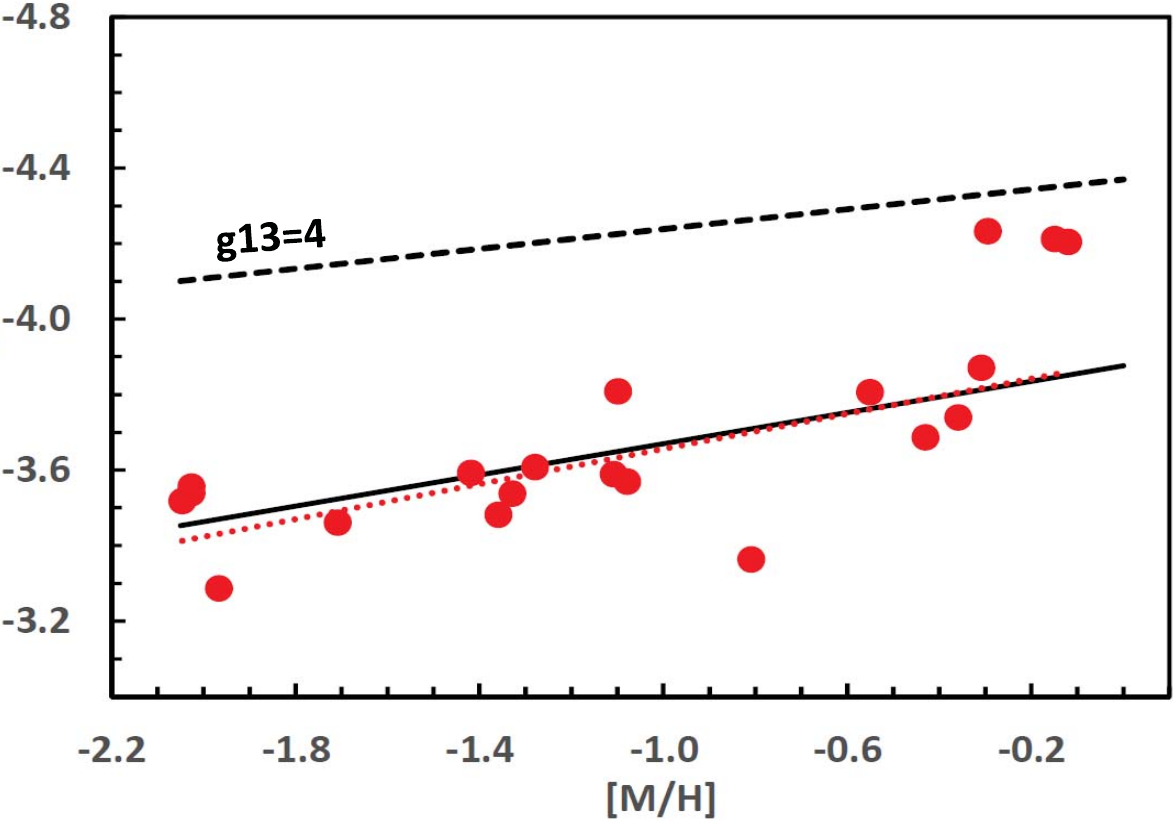}}
  \caption{TRGB in 22 GCs ({\em left}) \cite{Straniero:2020iyi} and 21 GCs ({\em right}) \cite{Straniero:2023} as a function of metallicity ${\rm [M/H]}=\log(Z/X)-\log(Z/X)_\odot$. {\em Left}: Distances determined by aligning the ZAHB, anchored to a parallax measurement of 47~Tuc, leading to the bound $g_{13}<1.48$. {\em Right}: Distances determined by GAIA EDR3 and other independent means, leading to $g_{13}<0.96$. {\em Dashed red line:} Best fit to the red circles. {\em Solid black line:} Standard prediction. {\em Dashed black line:} Prediction for $g_{13}=4$.}
  \label{fig:TRGB}
\end{figure}

Twenty years later, since 2013, the topic has been reexamined by different groups. The new series began with a detailed study of the globular cluster M5, for the first time estimating a realistic error budget \cite{Viaux:2013hca, Viaux:2013lha}. The bound $g_{13}<4.3$ (95\% C.L.) was poor because the data preferred a bit of extra cooling. However, at least half the effect crept in from an incorrect screening prescription in the nuclear reaction rates \cite{Serenelli_2017}. A different group later found $g_{13}<2.6$ (95\% C.L.) from the cluster M3 \cite{Straniero:2018fbv}. One eternal problem with such arguments is the true distance of the objects. Using new geometric distance indicators, a restrictive bound $g_{13}< 1.3$ (95\% C.L.) came from the large GC $\omega$~Centauri \cite{Capozzi:2020cbu}, although this result was later critiqued  because of $\omega$~Cen's ellipticity. At the same time, a sample of 22 GCs was used, with distances relatively aligned by their zero-age HBs, and commonly anchored to 47~Tuc with a parallax distance from the GAIA Satellite DR2  (Data Release~2). The estimated bolometric TRGB brightnesses are shown in Fig.~\ref{fig:TRGB} (left panel) as a function of metallicity ${\rm [M/H]}=\log(Z/X)-\log(Z/X)_\odot$ that affects the bolometric brightness. The linear regression of these points (red dashed line) nearly coincides with the standard prediction (solid black line). From a detailed error budget, the theoretical uncertainty is $\sigma_{\rm theory}=\pm0.04$~mag. For reference, this figure also shows as a dashed black line the extreme case of what one expects for axion cooling with $g_{13}=4$. A measurement of $g_{13}=0.60^{+0.32}_{-0.58}$ was described as a cooling hint, but is only a weak effect and so their bound $g_{13}<1.48$ (95\% C.L.) is more relevant \cite{Straniero:2020iyi}.

The latest improvement comes from using 21 GCs \cite{Straniero:2023}, where the highest-metallicity case of the previous sample was eliminated, with distances independently determined by (i) Parallaxes (GAIA EDR3 = Early Data Release 3). (ii)~Kinematic (GAIA EDR3, HST). (iii)~Subdwarf MS fitting (HST and ground-based). (iv)~ZAHB fitting (HST and ground-based). (v)~RR Lyrae P--L relation. (vi) Eclipsing binaries. (vii)~Sometimes other methods. The estimated TRGBs are shown in Fig.~\ref{fig:TRGB} (right panel) as a function of metallicity.  The reported bound is \cite{Straniero:2023}
\begin{equation}\label{eq:TRGB-gaee}
    g_{13}<0.96~\hbox{(95\% C.L.)},
\end{equation}
which for the moment has not yet been formally published. For DFSZ axions, it translates into the 95\% C.L.\ bounds
\begin{equation}\label{eq:TRGB-mass-bound}
    f_a> 0.88\times 10^9~{\rm GeV} (2\sin^2\beta)
    \quad\hbox{and}\quad
    m_a< \frac{\rm 6.4~meV}{2\sin^2\beta}.
\end{equation}
At the limit $g_{13}=1$, the core-averaged axion luminosity roughly corresponds to $1/5$ of the standard neutrino luminosity, see Fig.~3 of Ref.~\cite{Straniero:2020iyi}, from which we extract $\langle\epsilon_\nu\rangle\simeq 4.4~{\rm erg}~{\rm g}^{-1}~{\rm s}^{-1}$ and
$\langle\epsilon_a\rangle\simeq 0.89~{\rm erg}~{\rm g}^{-1}~{\rm s}^{-1}$. In other words, the latest bounds are rather ambitious indeed!

Recently, bounds based on the TRGB were heavily questioned on the basis of a large grid of numerical models (interpolated with machine learning) that varied stellar mass $M$, metal abundance $Z$, helium abundance $Y$, and axion-electron coupling $\alpha_{26}=10^{26} g_{aee}^2/4\pi$ in a wide range \cite{Dennis:2023kfe, Franz:2023gic}. All of these parameters were fitted simultaneously with the input of the measured TRGB absolute $I$-band brightness $M_I$. The posterior distribution for $\alpha_{26}$ showed no upper limit other than the imposed prior of $\alpha_{26}<2$. We do not believe that such an exercise makes sense because the task is not to fit the stellar parameters from the measured $M_I^{\rm TRGB}$, which is not very sensitive to them. The influence of uncertainties of input parameters, including global stellar parameters, was quantified in the critiqued studies. The dominant uncertainties do not come from stellar parameters, but from distance, microphysics, treatment of convection, bolometric corrections, and so forth. The initial stellar mass is less crucial than mass loss on the RGB that was not varied in this study. In an auxiliary analysis (Fig.~4 of \cite{Dennis:2023kfe}), sensible priors were put on $M$, $Y$, and $Z$, but of course, their approach remains to find posteriors for these parameters. In this case,  the MCMC returned posteriors for $Y$ and $Z$ nearly identical with the priors, which makes sense because the measured observable is not sensitive to them and thus cannot provide new information. On the other hand, the authors interpret this situation as meaning that the priors were chosen unphysically narrow. Confusingly, the posterior for the stellar mass $M$ was much broader than the prior and the mean shifted some $8\sigma$ away from the Gaussian prior, different from what the authors write and physically hard to understand. This effect is what seems to drive the broad posterior for $\alpha_{26}$, which still is not limited from above. In principle, however, a global study of this kind could be useful, in the sense of creating a stellar model space with many varied input parameters and use machine learning to interpolate between models, and include other measurements besides the TRGB brightness that are actually informative on the varied parameters.

We finally mention a TRGB bound without GCs. The free halo stars in a galaxy also derive from an early generation. The luminosity function of a field of stars at the outskirts of a galaxy shows a sharp break caused by the TRGB. This feature can be used to gauge the relative distances between galaxies and thus can be used as one rung in the cosmic distance ladder \cite{Freedman:2021ahq}. It plays an interesting role in the context of the Hubble tension, the difference between cosmological and astrophysical determinations of the cosmic expansion parameter $H_0$. Conversely, the galaxy NGC~4258 hosts a water mega maser, allowing for a quasi-geometric distance determination and thereby among the best absolute TRGB calibrations. This logic implies yet another limit $g_{13}<1.6$ at 95\% C.L.\ \cite{Capozzi:2020cbu}, no longer competitive with the latest GC bounds, but still of some interest because it depends on other systematics and thus provides an independent line of defense.

After a long hiatus, there has been a break-neck push over the past decade to advance this method to new levels of precision. Several groups have strongly improved the TRGB bounds both quantitatively and qualitatively. They have put them on a much sounder footing both observationally and theoretically, providing a much better sense of their error budget and credibility.

\subsection{Helium-burning lifetime and axion-photon interaction}

Stars on the horizontal branch (HB) have a nondegenerate core (about $0.5\,M_\odot$) that generates energy by fusing helium to carbon and oxygen with a core-averaged energy release of about $80~\rm erg~g^{-1}~s^{-1}$. At the edge, there is a hydrogen-burning shell, i.e., HB stars have two sources of energy. A typical core density is $10^4~{\rm g}~{\rm cm}^{-3}$ and a typical temperature $10^8\,{\rm K}$. The Primakoff energy loss rate~Eq.~\eqref{eq:primakofflossrate} implies that the energy-loss rate per unit mass, $\epsilon=Q/\rho$, is proportional to $T^7/\rho$. Averaged over a typical HB-star core one finds $\langle (T/10^8\,{\rm K})^7\,(10^4\,{\rm g}\,{\rm cm}^{-3}/\rho)\rangle\approx0.3$. Therefore, the core-averaged energy loss rate is about $G_{10}^2\,30~\rm erg~g^{-1}~s^{-1}$. The main effect would be accelerated consumption of helium and thus a reduction of the HB lifetime by a factor \hbox{$80/(80+30\,G_{10}^2)$}, i.e., by about 30\% for $G_{10}=1$.

The HB lifetime can be measured relative to the red-giant evolutionary time by comparing the number of HB stars with the number of RGB stars that are brighter than the HB, expressed as $R=N_{\rm HB}/N_{\rm RGB}=t_{\rm HB}/t_{\rm RGB}$. It can be translated to a bound on $G_{a\gamma\gamma}$ if axion losses have no impact on RGB evolution, so we assume that axions do not couple to electrons or else they would affect the TRGB (Sect.~\ref{sec:TRGB}). A long-standing limit $G_{a\gamma\gamma}<10^{-10}~{\rm GeV}^{-1}$ \cite{Raffelt:1996wa} was based on number counts in 15 GCs that provided an overall estimate of $R$ within some 10\% \cite{Buzzoni:1983}. The purpose of determining $R$, as originally advanced by Iben \cite{Iben:1968}, was to determine the helium abundance in these old systems; $Y_{\rm GGC}=0.23\pm0.02$ was found in 1983 \cite{Buzzoni:1983}, where GGC stands for Galactic Globular Clusters. Twenty years later, fresh number counts in 57 GCs, a third of the galactic population, provided  $Y_{\rm GGC}=0.250\pm0.006$ \cite{Cassisi:2003nm, Salaris:2004xd, Pietrinferni:2004vy}.

About ten years ago, 39~GCs from that sample were selected with metallicity $\log(Z/X)<-1.1+\log(Z/X)_\odot$, implying $R=1.39\pm0.04$ \cite{Ayala:2014pea}. As usual, $Z$ is the mass fraction of all elements except helium (Y) and hydrogen (X) with $X+Y+Z=1$. $R$ does not depend much on $Z$ or stellar mass; models that include axion cooling reveal $R=6.26\,Y-0.41\,G_{10}^2-0.12$ \cite{Ayala:2014pea}. In the absence of axion cooling, this result implies $Y_{\rm GGC}=0.241\pm0.006$. For comparison, the primordial helium abundance from standard big-bang nucleosynthesis is $Y_{\rm SBBN}=0.2467(6)$ or 0.2471(5) \cite{Planck:2018vyg}, and is a lower limit to $Y_{\rm GGC}$. Observational determinations of the primordial $Y$ use spectroscopic determinations in extragalactic H~II clouds \cite{Izotov:2013waa, Aver:2013wba} and extrapolate to zero metallicity. Instead, one may use these same measurements and select those objects with metallicities similar to GCs, leading to an assumed $Y_{\rm GGC}=0.2535(36)$ or 0.255(3) \cite{Ayala:2014pea}. In the presence of axion cooling, this implies $G_{10}=0.45^{+0.12}_{-0.16}$ (68\% C.L.) and $G_{10}<0.66$ (95\% C.L.) \cite{Ayala:2014pea}. Subsequently the analysis was improved \cite{Straniero:2015nvc}, notably by generating synthetic CM diagrams to calibrate theoretical expectations, leading to $R=1.39\pm0.03$, by including errors for key nuclear reaction rates, and using updated helium measurements \cite{Izotov:2014fga, Aver:2015iza} that implied $Y_{\rm GGC}=0.255\pm0.002$. 
With these inputs, $G_{10}=0.29\pm0.18$ (68\% C.L.) was found, implying a limit \cite{Straniero:2015nvc}
\begin{equation}\label{Eq:Bound_R_parameter}
     G_{a\gamma\gamma} < 0.65 \times 10^{-10}~{\rm GeV}^{-1}~\hbox{(95\% C.L.)},
\end{equation}
which for KSVZ axions implies
\begin{equation}\label{Eq:HB-MassBound}
   f_a>3.4\times10^7~{\rm GeV}
    \quad\hbox{and}\quad
    m_a<0.17~{\rm eV}.
\end{equation}
It is not selfconsistent to interpret Eq.~\eqref{Eq:Bound_R_parameter} for DFSZ axions because their interaction with electrons would strongly modify RGB evolution and the $R$-parameter no longer measures the axion-photon interaction in isolation. Instead, the TRGB constraint (Sect.~\ref{sec:TRGB}) implies a much more restrictive constraint on $f_a$ and $m_a$.

The restrictive limits from the $R$-method beg the question if there could be compensating effects that hide a large $G_{a\gamma\gamma}$ in violation of the nominal bound. The duration of He burning is determined by convection in the core that provides the central furnace with fuel from the entire convective region. (In contrast, the solar core is radiative and so H burning does not benefit from this enhanced supply.) The convective region is large, perhaps 50\% of the He core, but there is plenty more He if it could be brought in and prolong the HB phase. Indeed, shortly before central He exhaustion, the ``core breathing pulses'' (CBPs) found in numerical simulations \cite{Castellani:1985, Dorman:1993} were an early concern because they would do precisely this, increase the $R$-parameter, and modify the inferred $Y_{\rm GGC}$. However, this effect would also modify AGB evolution in contrast to observations (see Sect.~\ref{sec:AGB} below), leading to the perception that CBPs may be an artifact and must be suppressed in numerical simulations \cite{Caputo:1989, Straniero:2003, Cassisi:2003nm, Salaris:2004xd, Pietrinferni:2004vy}. The recent studies of axion bounds \cite{Ayala:2014pea, Straniero:2015nvc, Carenza:2020zil, Lucente:2022wai} do not mention this topic, but their stellar evolution code suppresses CBPs \cite{Straniero:2003}.

The bound Eq.~\eqref{Eq:Bound_R_parameter} is identical to the CAST limit Eq.~\eqref{eq:castlimit}, but extends to much larger masses of around 10~keV (the HB-star core temperature). For larger ALP masses, production is suppressed, but eventually photo-coalescence $\gamma\gamma\to a$ overtakes Primakoff production and ALP decays $a\to\gamma\gamma$ within the star become important. An exclusion range in the $G_{a\gamma\gamma}$--$m_a$ plane was presented that reaches to $m_a\lesssim400$~MeV \cite{Carenza:2020zil, Lucente:2022wai}, but such large masses are irrelevant for QCD axions.

The helium-burning lifetime can also be measured in massive stars, although the evolution is more complicated. It was argued \cite{Friedland:2012hj} that in 8--$12\,M_\odot$ stars, too much axion emission would shorten and eventually eliminate the blue-loop phase of the evolution. This would contradict observational data, notably the existence of Cepheid stars, that correspond to the blue loop crossing the instability strip in the CM diagram. This argument implies a conservative bound $G_{10}<0.8$, that was competitive at the time, but is now superseded by Eq.~\eqref{Eq:Bound_R_parameter}.

\subsection{New bounds from the asymptotic giant branch (AGB)}

\label{sec:AGB}

After helium has been exhausted in the core of an HB star, it develops a degenerate carbon-oxygen core with helium burning in a shell and hydrogen burning continuing in a second shell. As the C/O core grows more massive and thus smaller, He shell burning accelerates and the star ascends asymptotically the RGB. This occurs for stars with $M\lesssim 8\,M_\odot$ that never ignite the next burning phase, but instead shed their envelope and end as WDs. AGB evolution is interesting but complicated in that the burning shells can interact and the occurrence of thermal pulses. 

Of course, new channels of energy loss or transfer can influence this evolution. The mass of the resulting WD can be lower, implying a deficit of luminous AGB stars and massive WDs~\cite{Dominguez:1999gg}. Moreover, the amount of material processed by nuclear burning that arrives at the surface (3rd~dredge~up) can increase, modifying the chemical composition of the photosphere \cite{Dominguez:1999gg}. The limiting mass $M_{\rm up}$ that marks the transition between stars that will or will not end up as core-collapse SNe is modified \cite{Dominguez:2017yhy} as well as the initial mass--final luminosity relation of core-collapse SN progenitors~\cite{Straniero:2019dtm}. Some of these studies were driven by the idea that there might be a hint of axion cooling in some stellar systems, but in view of the latest bounds, all of the effects in intermediate-mass stars appear to be marginal for the allowed parameters of QCD axions. On the other hand, for general ALPs with larger masses for given coupling strength, hotter environments could be relevant for larger ALP masses. A recent study of  the WD initial--final mass relation excludes a vast range in the $G_{a\gamma\gamma}$--$m_a$ plane that bites into the ``cosmological triangle'' \cite{Dolan:2021rya}, but does not further constrain QCD axions. On the other hand, a study of nucleosynthesis in a $16\,M_\odot$ star reveals that even for $G_{a\gamma\gamma}=10^{-11}~{\rm GeV}^{-1}$, below any available limit, the yields are significantly changed, e.g., neon by a factor of~3 \cite{Aoyama:2015}. This intriguing finding does not represent a constraint, but suggests that strong effects can obtain in the otherwise allowed parameter range.

\begin{figure}[ht]
  \centering
  \includegraphics[width=0.6\textwidth]{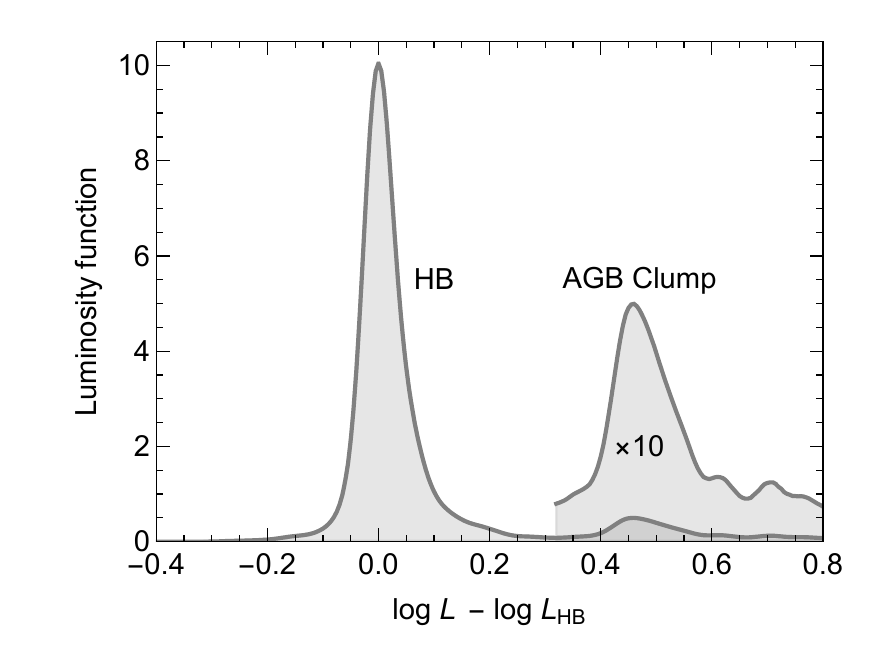}
  \caption{Empirical luminosity function, normalized to one, of the HB plus AGB stars from 14 GCs without a blue extension of the HB \cite{Constantino_2016}, using HST photometric data of Piotto et~al.~\cite{Piotto:2002pi} and Sarajedini et~al.~\cite{Sarajedini:2006vg}. We show a superposition of the two curves of Fig.~6 of \cite{Constantino_2016}, weighted with the number of stars (2414 for Sarajedini et al.\ and 4036 for Piotto et al.), although there is large overlap between the used clusters. Each one was aligned to its HB luminosity, defined as the maximum of the distribution. The AGB clump at $\log L-\log L_{\rm HB}=0.455$ sticks out as a narrow peak. The separation between HB and AGB is taken at the minimum between the peaks, and the AGB itself until $\log L-\log L_{\rm HB}=1$. A total of 725 AGB and 5725 HB stars went into the construction of this distribution, corresponding to $R_2=0.127$ for the clusters that went into this plot.
  }
  \label{fig:AGB}
\end{figure}

Returning to low-mass stars in GCs, recently it was shown that AGB evolution offers the potentially  best sensitivity yet on $G_{a\gamma\gamma}$ \cite{Dolan:2022kul}. HB stars begin with core He burning (CHeB) that proceeds at a nearly fixed HB luminosity. In terms of bolometric luminosity, without color filter effects, the HB is truly horizontal. When core He is exhausted, a transition to shell burning occurs, eventually the C/O core becomes degenerate, and the star ascends the AGB. However, this transition is slow while the star stays pinned to around three times the HB luminosity. This hesitation before ascending the AGB can be seen in the empirical luminosity function in Fig.~\ref{fig:AGB} that shows the distribution of HB and AGB stars as a function of bolometric luminosity derived from 14~GCs~\cite{Constantino_2016}. Low-mass AGB stars spend most of their time in the AGB clump, whereas the actual ascent on the AGB is fast and ends quickly when the star runs out of envelope and becomes a~WD. Helium burns now at higher $T$ than during CHeB so that neutrino or axion emission is more effective. Axion losses shorten both the HB and AGB-clump phase, but the latter more than the former, so the ratio $R_2=N_{\rm AGB}/N_{\rm HB}$ must decrease and provides a measure of new energy losses~\cite{Constantino_2016}. (Traditionally one defines three $R$-parameters: $R=N_{\rm HB}/N_{\rm RGB}$, $R_1=N_{\rm AGB}/N_{\rm RGB}$, and $R_2=N_{\rm AGB}/N_{\rm HB}=R\times R_1$.) Exploiting the HST photometry of 48 GCs, $R_2=0.117 \pm 0.005$ was recently found~\cite{Constantino_2016}, similar to several earlier studies. Moreover, the luminosity of the AGB clump was found at $\Delta \log L_{\rm HB}^{\rm AGB}=0.455\pm0.012$, corresponding to $\Delta M_{\rm bol}^{\rm HB-AGB}=1.14\pm0.03$~mag. 

The astrophysical motivation for this study was a renewed examination of how convection and mixing should be treated in a He-burning star, where in particular convective instabilities in the form of CBPs can develop as mentioned in the previous section. This study relates to an earlier one by the same authors concerning the asteroseismic determination of the core structure \cite{Constantino:2015kja}. Convection is, of course, a 3D phenomenon and is always implemented in 1D (spherically symmetric) stellar evolution codes in different physics-inspired parametrizations. The final conclusion, based on simulations with the MONash University STellAR evolution code (MONSTAR), is apparently the same as reached a generation earlier and discussed in the previous section, namely that CBPs extend the HB lifetime and thus predict an $R_2$ parameter smaller than observed. An interesting new argument derives from the luminosity and narrowness of the AGB clump shown in Fig.~\ref{fig:AGB}. The~scatter introduced by numerical instability of CBPs should strongly broaden this distribution and thus likely do not exist in nature.

The idea that the $R_2$ parameter would be reduced by Primakoff axion emission~\cite{Constantino_2016} motivated a recent study by a different group \cite{Dolan:2022kul}, who produced stellar models with the MESA code, once more adopting different recipes to deal with convection, and in addition implementing axion losses with different assumed values of $G_{10}$. CBP instabilities were not manually suppressed. For a fixed value of $G_{10}$, the predicted $R_2$ and $R$ values scatter over a broad range as seen in their Fig.~2, depending on the number and duration of CBPs that depend on the spatial and temporal numerical resolution, i.e., for each fixed set of input parameters, 20~HB and AGB simulations were performed with varying numerical resolution, and the bands in their Fig.~2 reflect this numerical scatter. Of~course, for a given run, the resulting $R_2$ and $R$ values are not independent, and so a scatter plot in the $R$--$R_2$--plane for fixed $G_{10}$ might have been more informative. 

Another significant input uncertainty is the rate for the $^{12}{\rm C}(\alpha,\gamma){}^{16}{\rm O}$ reaction that was varied over $\pm20$\% of the standard value. As a fiducial value \cite{Dolan:2022kul} uses the updated nuclear rate from \cite{Xu:2013fha}, while \cite{Constantino_2016} used the values reported in \cite{ANGULO19993}. For typical HB temperatures of $T\simeq 10^8~{\rm K}$, the  $^{12}{\rm C}(\alpha,\gamma){}^{16}{\rm O}$ rate from \cite{Xu:2013fha} is smaller than the one from \cite{ANGULO19993} by some 40\%. A smaller nuclear rate leads to a larger $R_2$, so at least some difference between \cite{Dolan:2022kul} and \cite{Constantino_2016} may be attributed to this difference.

Without suppressing CBPs, limits on $G_{10}$ for different values of this rate and four different convection schemes are shown in their Table~1, varying in the range $G_{10}<0.13$--0.47, nominally at 95\% C.L., referring to the C.L.\ of the measured $R_2$. Following these authors and using the most conservative case, one finds~\cite{Dolan:2022kul}
\begin{equation}\label{Eq:Bound_R2_parameter}
     G_{a\gamma\gamma} < 0.47 \times 10^{-10}~{\rm GeV}^{-1}~\hbox{(95\% C.L.)},
\end{equation}
which for KSVZ axions implies
\begin{equation}\label{eq:R2-KSVZ}
    f_a>4.7\times10^7~{\rm GeV}
    \quad\hbox{and}\quad
    m_a<0.12~{\rm eV}.
\end{equation}
Once more, it makes no sense to translate this constraint to the case of DFSZ axions because emission by their electron coupling would have to be included in the analysis. Probably this bound is conservative in the sense that CBPs reduce $R_2$ and axion cooling adds to this effect. In contrast to $R$, one cannot directly compensate one against the other.

\section{White-dwarf cooling}
\label{sec:WD}

Stars with $M \lesssim 8 \, M_\odot$ do not evolve beyond core helium burning and, after losing most of their envelopes, remain as carbon-oxygen white dwarfs (WDs) with masses in the approximate 0.5--$1.4\,M_\odot$ range. They are supported by electron degeneracy pressure, leading to the inverse mass-radius relation $R\simeq 10^4~{\rm km}\,(0.6 M_\odot/M)^{1/3}$, roughly the size of the Earth with a mass roughly that of the Sun. Their mass is bounded by the Chandrasekhar limit, originating in electrons becoming relativistic, and then can no longer prevent gravitational collapse. An axion-modified equation of state has been constrained because it would lead to a gap in the mass-radius relation (Sect.~\ref{sec:EoS}). WD~evolution simply consists of contraction and cooling, a gravothermal process \cite{mestel1952theory}, the speed being regulated by the surface layer of hydrogen (spectral type DA) or helium and heavier elements (non-DA). Axion cooling proceeds through electron bremsstrahlung and can show up in a modified WD luminosity function (WDLF), the distribution of galactic WDs per brightness interval, or in the cooling speed of individual WDs that can be measured for variable WDs by a drift of the pulsation period---for a recent review see Ref.~\cite{Isern:2022vdx} that also covers several non-axion cases and and see also a recent application to scalar interactions \cite{Bottaro:2023gep}.

\subsection{White-dwarf luminosity function (WDLF)}

The density of WDs in our galactic neighborhood can be plotted against their brightness, expressed in absolute bolometric magnitudes $M_{\rm bol}$, forming the WDLF \cite{2016NewAR..72....1G}. Assuming a constant birthrate $B$, the slope of this curve manifests the cooling speed---for slower cooling there are more WDs in a given $M_{\rm bol}$ interval. Initially, energy is lost by neutrino emission through plasmon decay $\gamma\to\nu\overline\nu$, whereas later photon surface emission takes over. For a simple model of photon transport through the surface layer, establishing a connection between surface and inner temperature, the WDLF is according to Eq.~(2.9) of Ref.~\cite{Raffelt:1996wa}
\begin{equation}\label{eq:almostmestel}
\frac{dN}{dM_{\rm bol}} = B_3 \, 2.2 \times 10^{-4}~{\rm pc}^{-3}~{\rm mag}^{-1}
\, \frac{10^{-4M_{\rm bol}/35}L_\odot}{78.7 \, L_\odot 10^{-2M_{\rm bol}/5}+L_\nu + L_a} \left(\frac{M}{M_\odot}\right)^{5/7}\sum_j \frac{X_j}{A_j}, 
\end{equation}
where $X_j$ is the mass fraction of element $j$ with atomic mass $A_j$, the chemical composition determining the WD heat capacity. $B_3$ is the birthrate normalised to $10^{-3}~{\rm pc}^{-3}~{\rm Gyr}^{-1}$, $L_\nu$ the neutrino luminosity, and $L_a$ the possible axion luminosity. Without axions and assuming an equal mixture of carbon and oxygen with $M = 0.6 \, M_{\odot}$, this is the famous Mestel cooling law~\cite{mestel1952theory}
\begin{equation}
 \log\,(dN/dM_{\rm bol}) = \frac{2}{7} \, M_{\rm bol} - 6.84 + \log(B_3)
\end{equation}
that provides a very good fit to data for intermediate luminosities. At the bright end, the WDLF is not represented by a power law and rather shows a depression caused by initial neutrino cooling. At the faint end, where cooling is slow and most WDs build up, the core begins to crystallize \cite{Isern:1997na, 2019Natur.565..202T}, defying a simple analytic description.
Moreover, the assumption of a constant birthrate is just a first approximation, and more realistic models will take into account the strong dependence of the stellar lifetime with mass and its impact on the WD birthrate \cite{Isern_2019}.

Axions can change both the amplitude and shape of the WDLF. Using analytic estimates, the first constraint on the axion-electron coupling of $g_{aee} < 4 \times 10^{-13}$ was derived by one of us a long time ago \cite{Raffelt:1985nj}. A few years later, it was relaxed by a factor of two based on WD evolution simulations and taking into account systematic uncertainties in the WD birthrate~\cite{Blinnikov:1994eoa}. More recently, another self-consistent evolutionary computation was performed~\cite{MillerBertolami:2014rka} and compared with the WDLF of the Galactic Disk, obtained using both SDSS~\cite{Harris:2005gd} and SuperCOSMOS~\cite{rowell2011white} sky surveys. 
The resulting bound is $g_{aee}<2.8\times10^{-13}$ at a nominal $99\%$ C.L.\ \cite{MillerBertolami:2014rka}. (In Table~\ref{tab:WD-Cooling} we list the more restrictive but lower-significance bound quoted in Ref.~\cite{Isern:2022vdx}.) For DFSZ axions, the original limit implies
\begin{equation}\label{Eq:BoundMillerBertolamiWDLF}
    f_a>3\times10^8~{\rm GeV}
    \quad\hbox{and}\quad
    m_a < \frac{19~{\rm meV}}{2 \, \sin^2\beta},
\end{equation}
in agreement with Ref.~\cite{Isern:2008nt} who used the SDSS data but a perturbative approach for axion emission. In the end, these results differ only by a factor of 1.4 from the initial estimate \cite{Raffelt:1985nj}.

In 2018, a group \cite{Isern:2018uce} including the authors of Ref.~\cite{Isern:2008nt}, reconsidered this subject and included luminosity functions also for the galactic halo, finding a signature of extra cooling that could be caused by axions and was compatible with Eq.~\eqref{Eq:BoundMillerBertolamiWDLF}. However, the authors admit that their conclusions should be regarded as rather tentative, given the large uncertainties plaguing the determination of both observed and theoretical luminosity functions.

\subsection{Pulsating white dwarfs}

\label{sec:ZZCeti}

Like many other types of stars, WDs can be pulsationally unstable when they fall on the instability strip in the Hertzsprung-Russell diagram, with the star ZZ Ceti the first pulsating WD to be discovered~\cite{Landolt68} and namesake for this entire class. The pulsation represents ``the most stable optical clock known,'' which in the case of the star G117-B15A has been observed since 1974 with a main pulsation period of 215.19738823(63)~s that increases by $(5.12\pm 0.82)\times 10^{-15}~{\rm s/s}$ and shows no glitches, as pulsars do \cite{Kepler:2020qzt}. The slowly increasing pulsation period corresponds to a 26~s phase shift in 45~years. Based on its standard cooling rate, one expects a period decrease of only $1.25\times 10^{-15}~{\rm s/s}$, much smaller than the observed value. After not finding any plausible standard-physics explanation, one titillating explanation could be a new cooling channel \cite{Isern:1992gia}, which in the case of axion bremsstrahlung would require $g_{aee}=(5.66\pm0.57)\times10^{-13}$. Similar results derive from other variable WDs as listed in Table~\ref{tab:WD-Cooling}. 

\begin{table}[t]
 \caption{Constraints and measurements of the axion-electron coupling from WDs if nonstandard behavior is interpreted as axion cooling, cited after Table~5 of Ref.~\cite{Isern:2022vdx}.}
\label{tab:WD-Cooling}
    \begin{tabular*}{\columnwidth}{@{\extracolsep{\fill}}llll}
    \hline\hline
     White Dwarfs       &$g_{aee}$ [$10^{-13}$]&Authors&Ref.\\
     \hline
     WDLF              &${}<2.1$ & Miller Bertolami et al.\ (2014)&\cite{MillerBertolami:2014rka}\\
     Pulsating WDs\\
     \quad G117-B15A   & $5.66\pm0.57$ & Kepler et al.\    (2021) & \cite{Kepler:2020qzt} \\
     \quad R548        & $4.8 \pm 1.6$ & C\'orsico et al.\ (2012) & \cite{Corsico:2012sh} \\
     \quad L19-2 (113) & $4.2 \pm 2.8$ & C\'orsico et al.\ (2016) & \cite{Corsico:2016okh}\\
     \quad L19-2 (192) & ${}<5$        & C\'orsico et al.\ (2016) & \cite{Corsico:2016okh}\\
     \quad PG1351+489  & ${}<5.5$      & Battich et al.\   (2016) & \cite{Battich:2016htm}\\
     \hline
     \end{tabular*}
\end{table}

However, this suggested $g_{aee}$ value is a large number and would require axions to play a prominent role in WD cooling that should show up in the WDLF, in contrast to the limit of Eq.~\eqref{Eq:BoundMillerBertolamiWDLF}, and is in even stronger conflict with the TRGB bound Eq.~\eqref{eq:TRGB-gaee} discussed in Sect.~\ref{sec:TRGB}. The axion luminosity scales with $g_{aee}^2$, so the value implied by G117-B15A would require $L_a$ twenty times larger at the TRGB than the limit---a huge effect. As both the WDLF and TRGB limits involve far less complicated physics than pulsating stars, at this stage we would dismiss the cooling hint from ZZ Ceti stars as a systematic effect in the pulsation physics that needs to be understood. Another question is if one could cook up a particle-physics explanation for the period drift that would not violate the other bounds. Irrespective of the TRGB limit, it would be an urgent exercise to reconsider the WDLF arguments with more recent data and a fresh analysis.

\section{Neutron-star cooling}
\label{sec:NS}

Axions derive from QCD so that nuclear interactions are particularly relevant. Shortly after invisible axions had been proposed, Iwamoto recognized that axion emission by nucleon bremsstrahlung $NN\to NNa$ can dominate neutron-star (NS) cooling~\cite{Iwamoto:1984ir}. In Sect.~\ref{sec:SN} we will see that the cooling speed of a nascent neutron star, during the first few seconds after core collapse, can be constrained by the duration of the neutrino signal of SN~1987A. Shortly after this initial burst, a NS becomes transparent to neutrinos, but continues to cool by neutrino volume emission~\cite{Yakovlev:2000jp}. For the first 30--100~years, the main contribution comes from the crust, leading to rapid thermal relaxation. Thereafter, the stellar interior is isothermal and core cooling dominates until it is so cold that surface photon emission takes over, approximately after $10^5$--$10^6$~years. There exist only a few cases with age and surface temperature determinations so reliable that one can test NS thermal evolution and inner properties and thus constrain new cooling channels. We specifically consider two age classes: young NSs, a few hundred years old, when neutrino emission strongly dominates, and old ones, age $10^5$--$10^6$~years, which are at the verge of photon cooling. 

\subsection{Young neutron star in the SN remnant Cassiopeia A}

A young non-pulsar NS was discovered in 1999 by the X-ray telescope \textit{Chandra}~\cite{CasAChandra} in the SN remnant Cassiopeia A (Cas~A). It was associated with SN~1680~\cite{SNCasA}, making it roughly 340~years old, as confirmed by kinematic data~\cite{Fesen:2006zma}. Analyzing archival data spanning a decade, Heinke and Ho (2010) reported a rapidly decreasing Cas~A surface temperature~\cite{Ho:2009mm, Heinke:2010cr}, much faster than expected from Modified Urca (MU)~\cite{Yakovlev:2000jp} or medium-modified Urca~\cite{Grigorian:2005fn}, the main neutrino emission processes for non-superfluid conditions. Therefore, many authors interpreted the fast Cas~A cooling as evidence of core superfluidity, with neutrino emission enhanced by the \textit{recent} onset of the breaking and formation of neutron Cooper pairs in the  $\prescript{3}{}{P}_2$ channel \cite{Page:2010aw, Shternin:2010qi, Leinson:2014cja}. In this state, Cooper pairs have spin one, nuclear spin fluctuations are possible, and neutrinos are very efficiently produced. Protons, on the other hand, were likely in a superconducting $\prescript{1}{}{S}_0$ state, suppressing MU processes~\cite{Page:2010aw}. Interestingly, even with these assumptions, neutrino cooling was not enough to explain the rapid temperature drop and this stimulated works in which axions were added to obtain extra cooling \cite{Leinson:2014ioa, Hamaguchi:2018oqw}. (For a critique of Ref.~\cite{Hamaguchi:2018oqw} see Ref.~\cite{Leinson:2021ety}.) Unfortunately, recently it was found that the rapid temperature drop was an artifact of a systematic drift of the detector energy calibration~\cite{Posselt:2018xaf, Wijngaarden:2019tht, Posselt:2022pch}, and therefore it appears that any need for axion cooling has disappeared. Nevertheless, Cas A can still be used to \textit{constrain} the QCD axion. 

The latest analysis \cite{Leinson:2021ety} uses the Cas~A temperature data from Ref.~\cite{Wijngaarden:2019tht} and numerical cooling sequences with the publicly available NS Cooling Code {\tt NSCool} \cite{NSCool_Code}. It was modified to include axion losses, but also to correct the crucial Pair Breaking and Formation (PBF) neutrino emission process in the neutron $\prescript{3}{}{P}_2$ channel. The axion-nucleon couplings were chosen (i) for the KSVZ case as the nominal values $C_{ann}= -0.02$ and $C_{app}=-0.47$ and a bound of $f_a>0.3\times10^8$~GeV was found. (ii)~For the DFSZ case, $\cot\beta=10$ was used (in our convention for $\beta$, see footnote~\ref{fn:beta}), implying $C_{ann}=-0.16$ and $C_{app}=-0.19$ and a bound $f_a>4.5\times10^8$~GeV was stated.

On the other hand, within errors or for a specific choice of $\beta$, the axion-neutron coupling can vanish for both KSVZ and DFSZ axions so that none of these limits is truly generic. Therefore, we try to frame them as limits on the Yukawa couplings. The proton processes contribute far less for comparable coupling, so the DFSZ limit is easily backward-engineered to be $|g_{ann}|=|C_{ann}|\,m_N/f_a<3.4\times10^{-10}$. Even in the KSVZ case, according to Fig.~2 of Ref.~\cite{Leinson:2021ety}, protons contribute little at the age of Cas~A, so again we assume neutron dominance, leading to a less restrictive constraint of $|g_{ann}|<6\times10^{-10}$. If protons were to contribute significantly after all, the constraint on the neutron coupling would improve, not diminish, and so something is inconsistent about these results. Given the uncertainties about the Cas~A cooling data and these inconsistencies, we do not know how to interpret these results except in a broad order-of-magnitude sense.

\subsection{Young neutron star HESS J1731--347}

Another young NS that was recently used to constrain QCD axions~\cite{Beznogov:2018fda} is HESS J1731--347, which was discovered by Suzaku, XMM--Newton and Chandra as an X-ray point source in the SN remnant G353.60.7~\cite{Acero:2009yz, Tian:2009yb}. The age was estimated to be around $30~{\rm kyr}$~\cite{Tian:2008tr}, older than Cas~A, but young enough to be dominated by neutrino cooling. Considering its age, its surface is quite hot and requires slow neutrino cooling. Data can be fit only if one avoids triplet pairing of neutrons, while having protons in singlet superconducting states~\cite{Beznogov:2018fda}. The logic is to avoid triplet neutron superfluidity that would lead to very efficient PBF emission, but \textit{strong} proton superconductivity, which has a much smaller PBF rate, but has the main effect of reducing standard processes such as $np$ and $pp$ bremsstrahlung. An additional requirement was an envelope with light elements to increase the thermal conductivity in the external layers, matching a hotter surface to a given core temperature \cite{Beznogov:2018fda}.    

For these inner NS conditions, axions are mostly emitted by bremsstrahlung $n n \rightarrow n n a$ \cite{Beznogov:2018fda}. For the cooling simulation, once more the Code {\tt NSCool} \cite{NSCool_Code} was used with unspecified updates, the APR Equation of State \cite{Akmal:1998cf}, and a carbon atmosphere for the envelope, which fits the soft X-ray spectrum well. To avoid excessive axion cooling, a bound $g_{ann}<2.8\times10^{-10}$ at 90\% C.L.\ was found. However, according to the more recent literature (see for example Ref.~\cite{Maxted:2017xfs}), J1731-347 could be much younger, with an age of 2--4 kyr instead of 30~kyr. In this case, probably no reduced cooling would be needed and the analysis should be reconsidered (see also related comments in Sec.~III-B in the Supplemental Material of Ref.~\cite{Buschmann:2021juv}).

\subsection{Old neutrons stars: Magnificent seven and millisecond pulsars}

\label{sec:OldNS}

The neutrino luminosity in a NS core scales rapidly with core temperature $\Tc$. In particular, $L_\nu \propto \Tc^8$ for non-superfluid cores, and $L_\nu \propto\Tc^6$ for the superfluid case. This means that as time goes by, neutrino emission becomes inefficient and surface photon emission, $L_\gamma = 4\pi\,\Te^4$, takes over, where $\Te$ is the \textit{effective surface} temperature. It must be connected to $\Tc$ with a model for the envelope, which is a thin external layer with a steep temperature gradient. Its composition is rather uncertain, but numerical simulations indicate that as a rule of thumb $\Te \sim 10^{6}\, {\rm K}\, (\Tc/10^8 \,{\rm K})^{1/2}$~\cite{Envelope}. The basic ingredient for this relation is the thermal conductivity in the envelope, where ions are in the liquid phase and the conductivity is dominated by electrons~\cite{Potekhin99}. The presence of light elements and accretion are also important \cite{Chabrier97}.

Old NSs, at the verge of photon domination, can decisively constrain QCD axions, in particular four of the nearby isolated Magnificent Seven NSs along with PSR J0659, for which kinematic age data are available (0.35--0.85~Myr) and the luminosity is well measured \cite{Buschmann:2021juv}. Numerical cooling sequences with axions, once more using the public code {\tt NSCool} \cite{NSCool_Code}, were performed for different EOSs, superfluidity prescriptions, NS masses, and prescriptions for the light elements in the envelope. This set of cooling curves were compared with observations using a joint likelihood for all 5~stars and reveals that, independently of axion cooling, the best-fitting models are the ones without superfluidity. 

\begin{figure}[t]
  \centering
\includegraphics[width=0.65\textwidth]{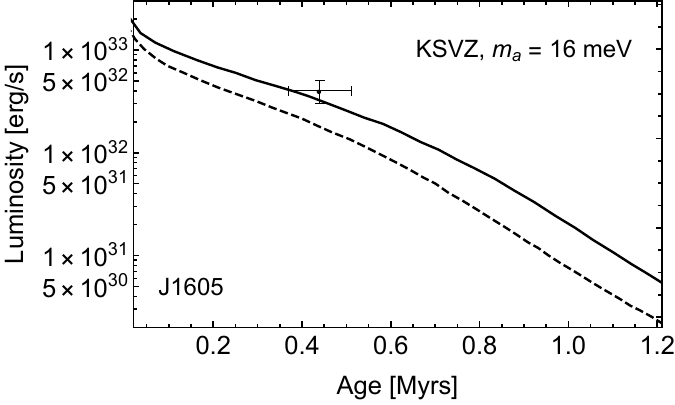}
  \caption{Best-fit luminosity curve for the pulsar J1605 \cite{Buschmann:2021juv}, both under the null-hypothesis (solid curve) and with the addition of axion cooling in the KSVZ model with $m_a=16$~meV (dashed curve), compared with the measurements of Refs.~\cite{Tetzlaff:2012rz, Pires:2019qsk}. This pulsar drives the constraints.
  The choice of $m_a$ corresponds to their original 95\% C.L.\ limit as explained around Eq.~\eqref{Eq:NSOld-KVSZ}.}
  \label{Fig:OldNS}
\end{figure}

More specifically, the constraints are driven by the pulsar J1605 (see Fig.~S4 in their Supplemental Material), using the standard EOS BSk22 \cite{Pearson:2018tkr} \textit{without} superfluidity. For nuclear couplings corresponding to the DFSZ model, $m_a$ is constrained by the thick red solid line in their Fig.~3. We find that it is well approximated (within better than  $\pm0.5$~meV) by
\begin{equation}\label{Eq:NSOld-DFSZ}
    m_{\rm limit}^{\rm DFSZ}= \bigl(33-17\,\sin^2\beta-1.8\sin^4\beta\bigr)\,{\rm meV}
\end{equation}
and varies between 33 and 14~meV for $0<\sin^2\beta<1$. For our reference value $\sin^2=1/2$ ($\tan\beta = 1$), the bound reads $m_a < 24$~meV. In the DFSZ model, the ratio of neutron/proton couplings is the same as in KSVZ for $\sin^2\beta=0.352$, essentially corresponding to a vanishing neutron coupling. The limit on the proton coupling in this case is $g_{app}<1.5\times10^{-9}$, corresponding to $m_a<19$~meV in the KSVZ model, some 20\% larger than the $16$~meV quoted in their abstract~\cite{Buschmann:2021juv}. We were not able to track down the origin of this small difference and for KSVZ adopt 
\begin{equation}\label{Eq:NSOld-KVSZ}
    m_a<19~{\rm meV}
    \quad\hbox{and}\quad f_a>3\times10^8~{\rm GeV}
\end{equation}
for consistency and to err on the conservative side. In practice, of course, such small differences are mostly of cosmetic character in view of all other uncertainties. In Fig.~\ref{Fig:OldNS} we show the best-fit luminosity curve of Ref.~\cite{Buschmann:2021juv} for the crucial pulsar J1605, both under the null hypothesis (solid curve) and with the addition of axion cooling (dashed curve) in the KSVZ model with $m_a=16$~meV, i.e., axions that couple only to protons with $g_{app}\simeq1.3\times10^{-9}$, together with the measurement~\cite{Tetzlaff:2012rz, Pires:2019qsk}. For this plot the authors fixed the axion mass to be $16$~meV, because this corresponds to their official 95\% C.L.\ KSVZ constraint as just explained.

We find it also useful to report the typical axion energy loss rate per unit mass in this case. For a typical NS with nuclear densities $\rho = 3 \times 10^{14} \, \rm g/cm^3$, core temperature $\Tc = 10^8\,{\rm K}$ and typical Fermi momenta $p_n = 360\,{\rm MeV}$ and $p_p = 120\,{\rm MeV}$, the corresponding axion energy loss rate per unit mass on the curve in Eq.~\eqref{Eq:NSOld-DFSZ} is $\epsilon_a \simeq 1~{\rm erg}~{\rm g}^{-1}~{\rm s}^{-1}$. Converted to a volumetric luminosity, this is of the same order as photon surface emission for a NS with typical radius $10 \, \rm km$ and $\Te$ related to $\Tc$ by standard relations~\cite{Gudmundsson1983, Envelope}, with $\Tc = 10^8 \, {\rm K}$ corresponding to $\Te\simeq 10^6 \, {\rm K}$. These old NS are at the verge of photon domination and so the axion luminosity near the limit should indeed be comparable to the photon one.

Summing up, the old NSs analysed in Ref.~\cite{Buschmann:2021juv} seem to provide the strongest and most reliable bounds. However, all of them lack a proper treatment of matter effects which can drastically modify $C_{ann}$ and $C_{app}$ at nuclear densities, an issue that will be studied in a forthcoming paper~\cite{Caputo-inpress-b}. The axion sensitivity of NS cooling may improve by ongoing and future surveys, notably by the recently launched eROSITA X-ray satellite \cite{Kurpas:2023tsk}.

\clearpage

\section{Supernova neutrinos}
\label{sec:SN}

We primarily discuss the traditional SN~1987A cooling bound based on the neutrino signal duration and highlight recent improvements (the axion emission rates), problems (lack of a direct comparison of self-consistent SN models with the data), and overall critique (have we really seen proto-neutron star cooling?).

\subsection{Traditional cooling bound from SN~1987A signal duration}

\begin{figure}[t]
    \centering
    \includegraphics[width=\textwidth]{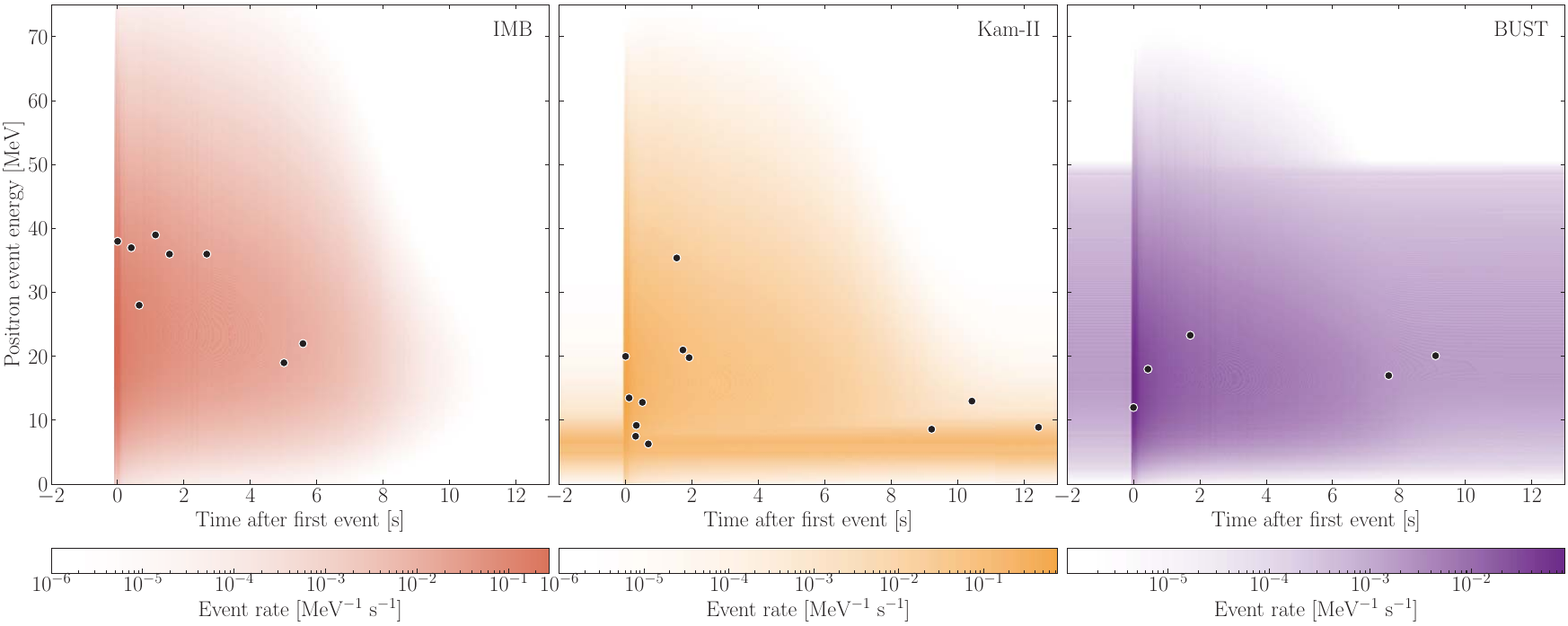}
    \caption{Neutrino signal from SN~1987A, where the energy is that of the detected positron from IBD $\overline\nu_e p\to n e^+$ \cite{Fiorillo:2023frv}. The expected event rates in the detectors, shown as shaded contours, are from the SN model 1.44-SFHo in Ref.~\cite{Fiorillo:2023frv} without flavor swap. In each detector, $t=0$ coincides with the first event except for a minimal offset chosen as the best-fit value for the overall signal. Detector background is also shown except for IMB that was essentially background free, but had a high detection threshold.}
    \label{fig:SN1987A-Signal}
\end{figure}

Supernova (SN) neutrinos were observed for the first and only time on 23~February~1987 from the historical SN~1987A, a few hours before the optical brightening, in the IMB \cite{Bionta:1987qt, IMB:1988suc} and Kamiokande-II (Kam-II) \cite{Kamiokande-II:1987idp, Hirata:1988ad} water Cherenkov detectors and the Baksan Scintillator Underground Telescope (BUST) \cite{Alekseev:1987ej, Alekseev:1988gp}. The main detection channel is inverse beta decay (IBD) $\overline\nu_ep\to ne^+$; we show in Fig.~\ref{fig:SN1987A-Signal} the measured positron energies vs.\ time. The much larger IMB detector (6800~t) was more sparsely instrumented and thus had a much larger detection threshold than Kam-II (2180~t), yet they registered comparable numbers of events. The water equivalent of BUST was only 280~t, yet it registered 6~events, of which the first (not shown) was attributed to background. For many details about these historical data see a recent review~\cite{Fiorillo:2023frv}.

Despite various ``anomalies'' in the sparse data \cite{Fiorillo:2023frv}, the total number of events, their energies, and the distribution over several seconds correspond reasonably well to theoretical expectations. In the standard paradigm \cite{Bethe+1985, Janka:2012wk, Janka2017Handbooka, Burrows+2021}, the core collapse of a massive star leads to a proto neutron star (PNS), a solar-mass object at nuclear density and a temperature of roughly 30~MeV, where even neutrinos are trapped. The long emission time scale is usually attributed to neutrino transport being diffusive. Volume emission of more feebly interacting particles, not trapped in the SN core, can be more efficient for losing energy, resulting in a reduced neutrino burst duration. The late-time signal is most sensitive to this effect because early neutrino emission is powered by accretion and thus not sensitive to volume losses.

\begin{figure}
  \centering
  \includegraphics[width=0.45\textwidth]{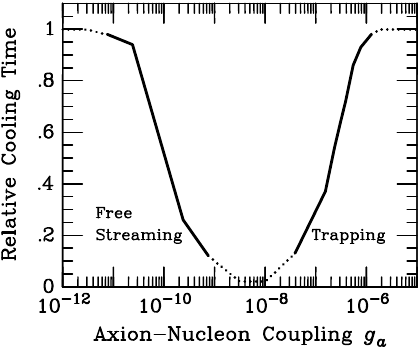}
  \caption{Relative duration of a SN neutrino burst as a function of the axion-nucleon coupling~\cite{Raffelt:1996wa}. Freely streaming axions are emitted from the entire core volume, trapped ones from an ``axion sphere.'' The solid line is from numerical calculations~\cite{Burrows:1988ah, Burrows:1990pk}.  The dotted line is an arbitrary continuation to guide the eye.\label{fig:SN-Axions}}
  \vskip-6pt
\end{figure}

This argument has been applied to many cases, from right-handed neutrinos to Kaluza-Klein gravitons, but axions are the earliest and most widely discussed example~\cite{Raffelt:1987yt, Ellis:1987pk, Turner:1987by, Mayle:1987as, Mayle:1989yx, Brinkmann:1988vi, Burrows:1988ah, Burrows:1990pk, Janka:1995ir, Keil:1996ju, Hanhart:2000ae}. They are emitted by nucleon bremsstrahlung $N+N\to N+N+a$ that depends on the axion-nucleon Yukawa coupling $g_{aNN}$, taken to be some average of neutrons and protons. Figure~\ref{fig:SN-Axions} illustrates that axion emission leaves the signal duration unchanged when $g_{aNN}$ is very small.  For larger couplings, the signal shortens until it reaches a minimum, roughly when the axion mean free path corresponds to the geometric size of the SN core. For yet larger $g_{aNN}$, axions are trapped and emitted from an ``axion sphere'' \cite{Burrows:1990pk, Caputo:2022rca}.\footnote{Some authors have questioned that in the trapping regime, axion emission asymptotes to effective surface emission \cite{Chang:2016ntp, Chang:2018rso, Lucente:2020whw}. However, this conclusion is an artifact of an erroneous axion transmittance, considering only radial propagation. With a proper angular average, the emission in the trapping regime is well described by black-body surface emission even though fundamentally, of course, it emerges from a layer with significant thickness \cite{Caputo:2022rca}. We mention in passing that, beginning with Table~1 in Ref.~\cite{Chang:2016ntp}, repeatedly the gain radius was invoked as an upper limit in the integral defining the optical depth for escaping particles during the cooling phase \cite{Chang:2016ntp, Chang:2018rso, Croon:2020lrf, Chauhan:2024nfa}. However, the concept of gain radius makes only sense during the accretion phase when the explosion has not yet happened, not during the cooling phase. Extended discussions about how to define the trapping regime are provided in Refs.~\cite{Caputo:2021rux, Caputo:2022rca}.}
 When it moves beyond the neutrino sphere, the impact on the neutrino signal becomes ever smaller. Of course, such ``strongly'' interacting axions are not necessarily harmless. First, they may play an important role during the SN collapse phase. Second, in the water Cherenkov detectors that registered the SN~1987A neutrinos, these axions would have interacted with oxygen nuclei, leading to the release of $\gamma$ rays and causing too many events~\cite{Engel:1990zd, Lella:2023bfb}. 

However, for axions and other particles, the trapping regime is often excluded by other arguments so that the free-streaming regime is of greater interest. An approximate analytic constraint on the energy-loss rate per unit mass is~\cite{Raffelt:1990yz}
\begin{equation}\label{eq:raffelt-SN1987A}
  \epsilon_a\lesssim1\times10^{19}\,{\rm erg\,g^{-1}\,s^{-1}},
\end{equation}
to be calculated at $\rho=3\times10^{14}\,{\rm g}\,{\rm cm}^{-3}$ and $T=30\,{\rm MeV}$. If we take the SN core to have a mass of about $1\,M_\odot=2\times10^{33}\,{\rm g}$, this corresponds to an axion luminosity $L_{a}=\epsilon_a M_\odot=2\times10^{52}\,{\rm erg\,s^{-1}}$. The gravitational binding energy of the neutron star is about $3\times10^{53}\,{\rm erg}$ and the emission lasts up to 10~s, and so axion losses would compete significantly with neutrino emission. The criterion Eq.~\eqref{eq:raffelt-SN1987A} was distilled from several numerical simulations that consistently showed that the burst duration was roughly halved when Eq.~\eqref{eq:raffelt-SN1987A} was saturated~\cite{Raffelt:1990yz}. Self-consistent cooling calculations for Kaluza-Klein gravitons \cite{Hanhart:2001fx}, together with a statistical comparison with the data, also confirm Eq.~\eqref{eq:raffelt-SN1987A}. Still, one cannot expect such a simple criteria to provide a precision limit on all cases.

\subsection{Axion bound from nucleon bremsstrahlung}

Applying the nominal criterion of Eq.~\eqref{eq:raffelt-SN1987A} requires a calculation of the energy loss rate in a hot nuclear medium. Using axion-nucleon interactions, this is a formidable nuclear-physics problem, similar to the analogous neutrino processes. Axions couple essentially to the nucleon spins, and also for neutrinos, the axial-vector interaction is far more important than the vector interaction. In this sense, one needs essentially the same medium response functions to calculate neutrino transport or axion emission and absorption. The dominant emission process is nucleon bremsstrahlung $NN\to NNa$, and in the early papers it was calculated using the one-pion exchange (OPE) potential to model the nucleon-nucleon interaction \cite{Raffelt:1987yt, Ellis:1987pk, Turner:1987by, Mayle:1987as, Mayle:1989yx, Brinkmann:1988vi}, as traditionally had been done for neutrino pair emission \cite{Friman:1979ecl}. Later it was argued that this approach led to an inconsistently large emission rate. What emits axions (or neutrino pairs) are the fluctuating nucleon spins, but if the fluctuations are too fast, individual emission events overlap destructively, in analogy to the Landau-Pomeranchuck-Migdal (LPM) effect, or in other words, the nucleon spin evolution is in the hydrodynamic, not kinetic, regime. Phenomenological arguments suggested a significant reduction of the axion emission rate \cite{Raffelt:1991pw, Raffelt:1993ix, Janka:1995ir, Raffelt:1998pa, Sedrakian:2000kc, vanDalen:2003zw, Lykasov:2008yz}. Later it was proposed to use nucleon-nucleon scattering data to extract the low-energy bremsstrahlung rate, leading to a factor of four suppression relative to the OPE-based rate \cite{Hanhart:2000ae}, which then however means that the LPM suppression is less relevant. The logic of the argument was that the axion emission rate, as a function of the nucleon spin fluctuation rate $\Gamma_\sigma$, was near its conceivable maximum and therefore not very sensitive to the exact value of~$\Gamma_\sigma$. Assembling these arguments, the axion energy-loss rate per unit mass from a single-species nuclear medium was estimated to be $\epsilon_a=g_{aNN}^2 (T^4/4\pi^2 m_N^3)\,F$, where $F\simeq1$ was an uncertain numerical factor \cite{Raffelt:2006cw}. For KSVZ axions, which couple only to protons, this reasoning led to an estimated bound $f_{a}\gtrsim4\times10^{8}\,{\rm GeV}$, corresponding to $m_{a}\lesssim16\,{\rm meV}$.

More recently, this estimate was critiqued in an often-cited paper that set out to improve the bound by applying various correction factors \cite{Chang:2018rso}. In particular, these ``results incorporate three classes of corrections to the tree-level, massless pion calculation: a cutoff for scattering at arbitrarily low energies, a factor for the nucleon phase space that accounts for the finite pion mass, and a factor that introduces higher orders in the nucleon scattering.'' Finally a much weaker fiducial constraint was found (see their Fig.~11) of $f_{a}\gtrsim1\times10^{8}\,{\rm GeV}$, corresponding to $m_{a}\lesssim64\,{\rm meV}$. Notice that a factor of 4 in coupling strength corresponds to a factor of 16 in reduced axion luminosity. We seem to track at least part of this unrealistically large reduction to the fact that they treated the different corrections as independent, multiplicative fudge factors (see their Eq.~4.2), which they call $\gamma_{\rm f}$ (cut off for the low energy divergences), $\gamma_{\rm p}$ (finite pion mass and nucleon degeneracy effects), and $\gamma_{\rm h}$ (higher order corrections to the dynamical spin structure function in chiral perturbation theory). However, these corrections are not independent. For example, the factor $\gamma_{\rm f}$ gets significantly closer to one if the other corrections are included consistently~\cite{Carenza:2019pxu}.

\begin{figure}[ht]
  \centering
  \includegraphics[height=5.8cm]{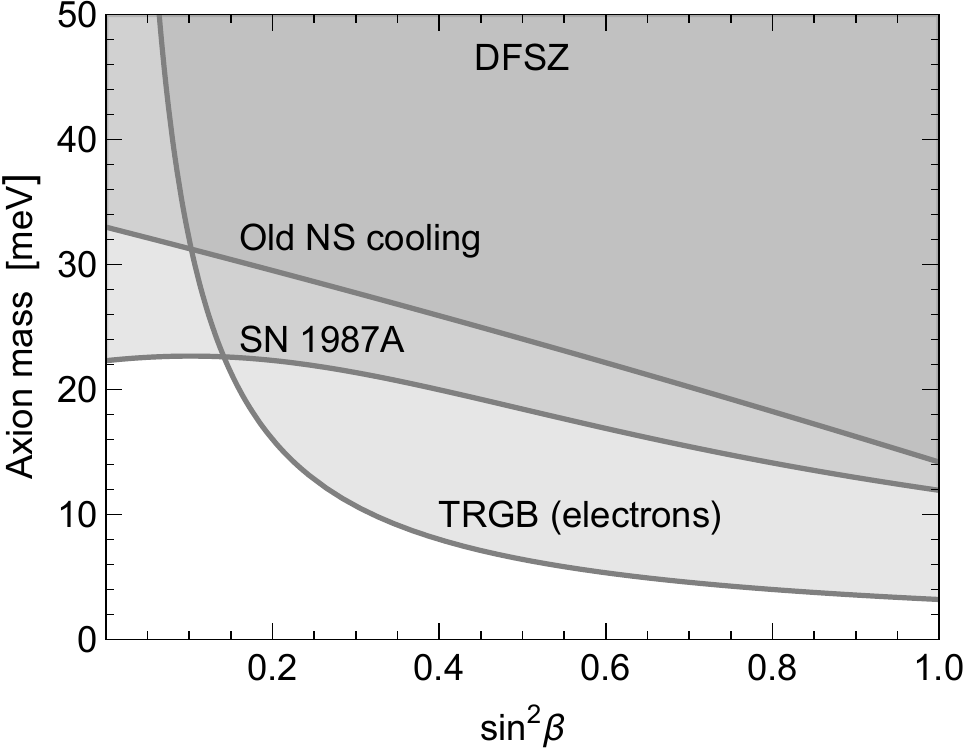}\includegraphics[height=5.8cm]{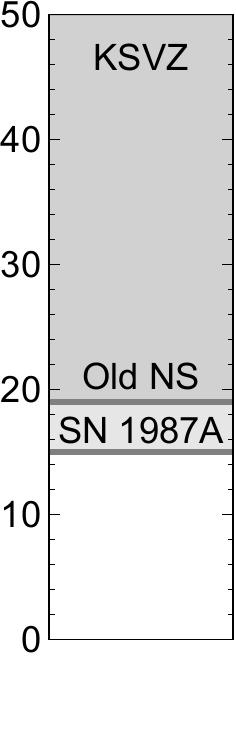}
  \caption{Our adopted bounds on DFSZ axions as a function of $\sin^2\beta$ and on KSVZ axions from SN~1987A and old NS cooling (nucleon couplings), and from the electron coupling constrained by the TRGB.
  \label{fig:DFSZ-Bounds}}
  \vskip-6pt
\end{figure}

Most recently, a fresh {\em ab initio} calculation of the emission rate went beyond the OPE approximation by including two-pion exchange, modeled as $\rho$ exchange, an effect that reduces the OPE rates by a large factor \cite{Carenza:2019pxu}. The results were shown to be consistent with a $T$-matrix approach in chiral perturbation theory \cite{Guo:2019cvs, Entem:2017gor, Bartl:2016vwk}. The $\rho$ exchange corrections were also found to have similar behavior and strength to the factor $\gamma_{\rm h}$ of Ref.~\cite{Chang:2018rso}. Multiple-scattering corrections were also included, although these are now small because the spin fluctuation rate is much smaller than in~OPE. To obtain an axion constraint, the criterion $L_a<L_\nu$ at 1~s after core bounce was applied, based on an unperturbed numerical SN model. The final constraint can be expressed as \cite{Carenza:2019pxu}
\begin{equation}\label{eq:SN1987A-coupling-bound}
    g_{app}^2+1.64\,g_{ann}^2+0.87\,g_{ann}g_{app}<1.35\times10^{-18}
\end{equation}
and equivalently as
\begin{equation}
    m_a<7.0~{\rm meV}\Big/\sqrt{C_{app}^2+1.64\,C_{ann}^2+0.87\,C_{ann}C_{app}}.
\end{equation}
For the KSVZ model with $C_{ann}=-0.02$ and $C_{app}=-0.47$ one finds
\begin{equation}\label{eq:SN1987A-KSVZ-bound}
    m_a<15~{\rm meV}
    \quad\hbox{and}\quad f_a>4\times10^8~{\rm GeV},
\end{equation}
identical to the traditional bound from the original Lecture Notes \cite{Raffelt:2006cw}. While this exact agreement must be fortuitous, it points to the stability of these constraints in terms of detailed approach. We recall that the earlier results were not based on a naive OPE calculation, but took into account the LPM suppression and measured nucleon-nucleon scattering data. For the DFSZ model, Eq.~\eqref{eq:SN1987A-coupling-bound} can be approximately expressed in the form
\begin{equation}\label{eq:SN1987A-DFSZ-bound}
m_{\rm limit}^{\rm DFSZ} = \bigl(23 + 6 \sin^2\beta - 38 \sin^4\beta + 22 \sin^6\beta \bigr)~{\rm meV}.
\end{equation}
For our reference value $\sin^2\beta=1/2$ (meaning $\tan\beta=1$), the DFSZ bound is $m_a<19$~meV. We show these bounds in Fig.~\ref{fig:DFSZ-Bounds}, together with those from old NS cooling and from the bound on the electron coupling from the TRGB.

\subsection{Pion-nucleon scattering}

Another potentially important axion emission process is $\pi^-p\to n a$ \cite{Turner:1991ax, Raffelt:1993ix, Keil:1996ju} that has received renewed attention recently \cite{Carenza:2020cis, Fischer:2021jfm}. With a vacuum mass of $m_{\pi^\pm}=139.6$~MeV and inner SN temperature of, say, 30~MeV, the abundance of one species of thermal pions is $n_{\pi}=3\times10^{-5}~{\rm fm}^{-3}$, to be compared with nuclear saturation density of $0.16~{\rm fm}^{-3}$ (corresponding to $2.6\times10^{14}~{\rm g}~{\rm cm}^{-3}$) and so there are only $Y_{\pi}=2\times10^{-4}$ pions per nucleon. However, they participate in the equilibrium between nucleons so that the chemical potential of charged pions is $\mu_\pi=\mu_n-\mu_p=\mu_e-\mu_{\nu_e}=\mu_\mu-\mu_{\nu_\mu}$, enhancing $\pi^-$ over $\pi^+$ and balancing some of the positive charges carried by protons in the medium. The abundance is enhanced roughly by $e^{\mu_\pi/T}$ and taking, e.g., $\mu_\pi=120$~MeV implies $Y_{\pi}=0.013$ and a substantial rate for $\pi^-p\to n a$. If $\mu_\pi>m_\pi$, a Bose-Einstein condensate of $\pi^-$ forms. Therefore, the abundance and character of the $\pi^-$ distribution strongly depends on their properties in a nuclear medium (effective mass and refractive potential) as well as the nuclear EOS---see the introductions of Refs.~\cite{Fore:2019wib, Fore:2023gwv} for discussions and references. In the EOSs used in typical present-day SN simulations, pions are ignored, just like muons (mass 105.7~MeV) were ignored until recently. However, adding muons simply mends a previous simplification, whereas adding pions requires new assumptions and ultimately parametric studies---for a recent case study see Ref.~\cite{Vijayan:2023qrt}. For practical SN physics, the question of pions is usually taken as yet another aspect of the nuclear~EOS uncertainties.

The renewed interest in pionic axion emission was triggered by a study that found that the $\pi^-$ abundance could be reliably calculated in hot nuclear matter based on the virial expansion~\cite{Fore:2019wib}. It~applies for small particle fugacities, defined as $z_i=e^{(\mu_i-m_i)/T}$, where $m_i$ is the particle mass and $\mu_i$ its relativistic chemical potential (that includes the mass). 
The spin-averaged squared matrix element is $|\mathcal{M}|^2=4 \overline{g}_{aNN}^2 (g_A/2f_\pi)^2\,{\bf p}_\pi^2$, where $g_A=1.27$ is the axial coupling, $f_\pi=92.4$~MeV the pion decay constant, and ${\bf p}_\pi$ the pion momentum. The effective axion-nucleon coupling for this process is ${\overline g}_{aNN}^2=\frac{1}{2}(g_{app}^2+g_{ann}^2)+\frac{1}{3}g_{app}g_{ann}$, where the sign of the interference term corrects an earlier expression \cite{Turner:1991ax}; we have checked that the new sign is correct. The pion dispersion relation in the medium is $E_\pi=(p_\pi^2+m_\pi^2)^{1/2}+\Sigma_{\pi}(p_\pi)$ with $\Sigma_{\pi}(p_\pi)$ varying from 0 at $p_\pi=0$ to around $-100$~MeV for $p_\pi\simeq300$~MeV at its minimum~\cite{Fore:2019wib}, so probably a nontrivial vertex renormalization factor should appear in the matrix element. In the limit of $E_\pi=p_\pi$, the volume energy loss rate is \cite{Carenza:2020cis, Fischer:2021jfm}
\begin{equation}
 Q_a\simeq\frac{\overline{g}_{aNN}^2}{4m_N^2}\,\left(\frac{g_A}{f_\pi}\right)^2 \frac{n_p}{\pi^3} \int dp_\pi p_\pi^5 f(p_\pi)
 \simeq \frac{30}{\pi^3}\,\frac{\overline{g}_{aNN}^2}{m_N^2}\,\left(\frac{g_A}{f_\pi}\right)^2 n_p z_\pi T^6.
\end{equation}
In the second expression we used $f(p_\pi)=z_\pi/[e^{(E_\pi - m_\pi)/T}- z_\pi]\simeq z_\pi e^{-p_\pi/T}$, where we assumed the pion fugacity to be small~\cite{Fore:2019wib} and neglected $m_{\pi}$ everywhere except in the fugacity. 

Neutron Pauli blocking was expressed by a factor $(1+z_n)^{-1}$, which corresponds to neglecting the neutron kinetic energy, i.e., it is the largest possible Pauli-blocking factor taken as representative for all neutrons. On the other hand, in the limit where we neglect nucleon recoil, the proton phase-space integral, including neutron Pauli blocking, is the first integral in Eq.~(3.6) of Ref.~\cite{Choi:2021ign}, where $1-f_n=e^{x_p^2}/(e^{x_p^2}+z_n)$ if $x_n=x_p$ and $x_p^2/(e^{x_p^2}+z_p)$ is the proton phase-space factor. If we ignore the small proton fugacity, this corresponds to an average Pauli-blocking factor of 
\begin{equation}
    \langle 1-f_n\rangle\simeq    
    \int_0^\infty dx_p \frac{e^{x_p^2}}{e^{x_p^2}+z_n}\,\frac{x_p^2}{e^{x_p^2}}\Bigg/ \int_0^\infty dx_p \frac{x_p^2}{e^{x_p^2}}
    \simeq \frac{\pi}{\pi+z_n},
\end{equation}
where the empirical approximation is good to better than 1\% for $0<z_n<1$.

In another calculation of the pionic emission rate~\cite{Choi:2021ign}, the direct axion-pion-nucleon vertex Eq.~\eqref{Eq:AxionNucleonPion} was included. To lowest order in a $p_\pi/m_N$ expansion, it has the effect of modifying the effective axion-nucleon coupling in the form
\begin{equation}
    \overline{g}_{aNN}^2\to\frac{1}{2}\left(g_{app}^2+g_{ann}^2\right)
    \left(1+g_A^{-4}\right)   
    +\frac{1}{3}g_{app}g_{ann}\left(1-3g_A^{-4}\right),
\end{equation}
which returns the original expression when deleting the terms proportional to $g_A^{-4}$. Numerically, the enhancement is roughly $1+g_A^{-4}=1.40$ and thus not a huge effect.

In principle, the process is reduced if spin fluctuations of the nucleons are too large; they are accounted for by a factor $\gamma_{\rm sf}$, which however is not a big effect \cite{Carenza:2020cis}. Isospin fluctuation are caused, for example, by $\pi^- p\to n \pi^0$, where we estimate the rate to be around $\Gamma\simeq T$ and thus also not a large effect relative to typical emitted axion energies of perhaps 5--$7\,T$.

Therefore, in the limit of vanishing pion mass and ignoring all the other tens-of-percent effects, the energy-loss rate per unit mass $Q_a/\rho$ is
\begin{equation}\label{Eq:OurEquationPions}
    \epsilon_a =\overline{g}_{aNN}^2~  2.6 \times 10^{37}~{\rm erg}~{\rm g}^{-1}~{\rm s}^{-1}~
     \left(\frac{T}{30~\rm MeV}\right)^6\,
     \left(\frac{Y_p}{0.3}\right)\biggl(\frac{z_\pi}{0.4}\biggr).
\end{equation}
Comparing with the criterion of Eq.~\eqref{eq:raffelt-SN1987A} implies a constraint
$\overline{g}_{aNN}<0.6\times10^{-9}$. For vanishing neutron coupling, this implies $g_{app}<0.9\times10^{-9}$. For comparison, the bremsstrahlung bound of Eq.~\eqref{eq:SN1987A-coupling-bound} implies $g_{app}<1.2\times10^{-9}$. Including both processes improves the bound roughly by a factor of 2, in approximate agreement with Ref.~\cite{Carenza:2020cis}.

However, fresh doubts about the pion process were recently raised by an overlapping cast of authors who computed $m_{\pi^\pm}$ in neutron-rich matter based on heavy-baryon chiral perturbation theory~\cite{Fore:2023gwv}. At nuclear saturation density, they found $m_{\pi^-}=200$--260~MeV that could strongly reduce the fugacity and thus axion emission. A more recent study of ALP emission from SNe~\cite{Lella:2023bfb}, including authors from Ref.~\cite{Carenza:2020cis}, also warned about using the pion-related bounds and show results with or without the pion process. In other words, while one cannot exclude that the emission rate is enhanced by pions, one also cannot rely on them for a trustworthy bound.

\subsection{Second-generation matter}

The microphysics and nuclear physics in SN simulations typically includes many simplifications and traditionally only uses first-generation hadronic and leptonic matter, except for the inclusion of heavy-lepton neutrinos usually referred to as $\nu_x$. It is only recently that six-species neutrino transport is becoming the norm and that muons are included as discussed earlier. On the hadronic side, even pions are usually neglected, but may or may not be quite abundant as discussed in the previous section, depending on their in-medium properties.

In addition, one could expect that strange matter (baryons including an $s$ quark) or kaons might be quite abundant~\cite{Sedrakian:2022ata, Sagert:2010yg, Raduta:2022elz, Oertel:2016xsn, daSilvaSchneider:2020ddu, Banik:2014qja} (recall that the lightest hyperon is the neutral $\Lambda$ with a mass 1115.68~MeV, not much larger than $m_n=939.57$~MeV). If this is the case, axion emission would be modified~\cite{Cavan-Piton:2024ayu, MartinCamalich:2020dfe, Camalich:2020wac} and the bound on $m_a$  would typically improve. On the other hand, a large hyperon population would soften the nuclear equation of state, in conflict with observed NS masses of up to $2\,M_\odot$, a problem called the hyperon puzzle \cite{Chatterjee:2015pua, Bombaci:2016xzl, Bombaci:2021ffs, Vidana:2022tlx}. The inclusion of three-nucleon interactions seems to solve this problem~\cite{Logoteta:2019utx, Gerstung:2020ktv, Tolos:2020aln}, but for now it appears difficult to provide robust predictions that could be implemented in practical SN simulations. Once the role of hyperons and/or kaons becomes better understood, it could modify SN physics and PNS cooling and provide new channels for axion emission. The exact overall impact of such modifications is impossible to predict, but specifically for axions it appears that the emission rates would probably get larger. As in the case of pions discussed earlier, one cannot exclude that in the nominally allowed parameter range, axions could still play a huge role in SN physics.

%\textcolor{blue}{Pions aside, the equation of state (EoS) of nuclear matter in SNe is in general poorly known. It may well be, for example, that $\Delta$ baryons, hyperons (\textit{i.e} baryons with strangeness), as well as kaons are also present in non-negligible amount in the SN core~\cite{Sedrakian:2022ata, Sagert:2010yg, Raduta:2022elz, Oertel:2016xsn, daSilvaSchneider:2020ddu, Banik:2014qja}. If this is the case, axion emission would be modified~\cite{Cavan-Piton:2024ayu, MartinCamalich:2020dfe, Camalich:2020wac} and the bound on the axion mass would typically improve. The presence of hyperons in the interior of neutron stars is puzzling per se~\cite{Chatterjee:2015pua, Bombaci:2016xzl}, because if only two-nucleons forces are included in the EoS calculation then this latter will be too soft and difficult to reconcile with neutron stars mass measurements. The inclusion of three-nucleons interactions seems to solve this problem~\cite{Logoteta:2019utx,Gerstung:2020ktv,Tolos:2020aln}, but it appears difficult to have robust predictions. Therefore, as for pions, while the possibility to have strange baryons (as well as other particles, such as kaons) in SNe cannot be excluded, it also cannot be used to place reliable and conservative limits.}

%EOS with strangeness vs.\ three-body forces: \cite{Logoteta:2019utx,Gerstung:2020ktv,Tolos:2020aln}

\subsection{For how long lasted SN~1987A neutrino emission?}
\label{sec:convection}

In its traditional form, the SN~1987A cooling argument is somewhat schematic in that the impact of axion losses on the neutrino signal was not directly compared with data. In the early simulations represented by Fig.~\ref{fig:SN-Axions}, the signal duration was defined as the time when 95\% of the expected events would have been recorded, and Eq.~\eqref{eq:raffelt-SN1987A} corresponds to this signal duration being roughly halved. In principle, of course, one should simulate self-consistent SN models that include axion losses and directly compare with data. Before such an exercise, however, one should first study if the SN~1987A signal is consistent with theoretical expectations beyond the general impression that the event energies and time structure roughly correspond to expectations. 

At the time of SN~1987A, the now-standard neutrino-driven explosion mechanism had only been proposed a few years earlier by Bethe and Wilson \cite{Bethe+1985} and detailed modeling was still in its infancy. One physically unavoidable effect is Ledoux convection driven by entropy and lepton-number gradients \cite{Epstein1979, Burrows+1988, Keil+1996, Janka+2001proc, Buras+2006a, Dessart+2006, Roberts+2012a, Roberts+2017Handbook, Nagakura+2020, Pascal+2022}, which actually determines the overall time scale of deleptonization and energy loss. Recently, SN~1987A data were confronted with a suite of spherically symmetric models with different equations of state and different final NS masses \cite{Fiorillo:2023frv}. PNS convection was included in a mixing-length approximation. The signal prediction of a typical model is shown in Fig.~\ref{fig:SN1987A-Signal}, overlaid with the actual SN~1987A data. 

The cooling time scale and expected signal duration in different detectors, once more defined as the period of 95\% of the energy emitted or signal received, of this class of convective models is too short to explain the late event triplet in Kam-II at 10--12~s post bounce or the last two events in BUST. In principle, these late events can represent an unlikely background fluctuation, but not PNS cooling of these models. Including axions would further reduce the signal duration. Even if one were to rely only on the last IMB events (as done in one of the early papers \cite{Raffelt:1987yt}) begs the question of interpreting the late Kam-II and BUST events. One speculation considers the fallback of material on the NS, causing a second flash of neutrino emission, although it is hard to obtain such a large signal \cite{Janka1996, Fiorillo:2023frv}. However, the sparse data prevent one from reaching strong conclusions.

\subsection{Have we really seen PNS cooling?}

The overall logic of the traditional SN~1987A cooling bound was questioned on the grounds that the late-time neutrino signal need not signify PNS cooling altogether \cite{Bar:2019ifz}. One key argument is that no compact remnant has yet been observed at the location of SN~1987A and especially no pulsating signal. Of course, the remnant could be a non-pulsar neutron star similar to Cas~A and the hot blob recently reported by the ALMA radio telescope \cite{Cigan:2019shp} can be interpreted as first evidence for such a source \cite{Page:2020gsx} and a very recent James Webb Space Telescope observation points in the same direction \cite{Fransson:2024}. Of course, direct evidence for a non-BH compact remnant would provide more confidence that the neutrino signal is indeed a good proxy for the PNS cooling speed.

In Ref.~\cite{Bar:2019ifz} a non-standard scenario is envisaged, where neutrino emission for the first 2--3~s from the collapsed SN core causes the initial signal in the detectors shown in Fig.~\ref{fig:SN1987A-Signal}, followed by BH formation. The actual explosion would be caused by the Collapse-Induced Thermonuclear Explosion (CITE) mechanism \cite{Kushnir:2014oca, Blum:2016afe}, whereas the late neutrino events, after 5~s or so, would be attributed to an accretion disk. This scenario has its own difficulties. First, it requires a mixture of helium into the carbon-oxygen core, replacing the carbon layer typically found in stellar evolution studies for pre-supernova stars~\cite{Hirschi:2004ks, Yusof_2013}. The carbon-oxygen layer being much hotter than the ignition temperature of helium, it is not obvious how often and for how long such mixed shells can occur. Second, the formation of a disk around the new-born BH requires a large amount of angular momentum deep inside the stellar core. This is expected to happen only in very rare events, namely gamma-ray-burst and hypernova cases \cite{1999ApJ...524..262M} as acknowledged in Ref.~\cite{Kushnir:2014oca}, but SN~1987A was not of that type. In fact, a preliminary study showed---using the MESA stellar evolution code---that the formation of mixed oxygen-helium shells is extremely rare and in the successful cases the angular momentum of the stellar core is never as high as required by CITE~\cite{Gofman:2018ldp}. 

Moreover, progress in 3D modeling over the past few years has lent strong support to the neutrino-driven delayed explosion mechanism, in particular also for SN~1987A. Specifically, 3D models of neutrino-driven explosions can explain many of its observational properties, employing pre-collapse stellar models compatible with its progenitor. The most important observables in this context are (i)~the explosion energy, (ii)~the ejected masses of chemical elements such as oxygen and radioactive isotopes like ${}^{56}$Ni and ${}^{44}$Ti, (iii)~observed ejecta asymmetries and radial mixing of chemical elements, in particular also of (radioactive) iron-group species, and (iv)~the measured light curve. Additional support for the neutrino-driven mechanism is provided by recent work that has linked 3D explosion models to observations~\cite{Sieverding2023ApJ, Bollig2021, Jerkstrand2020, Utrobin2021}. Based on the CITE-required progenitor structure, such a comprehensive picture is missing. Even the original authors do not seem to have pursued it beyond initial studies. But of course, it would be interesting to investigate all of these observables within any alternative mechanism for SN~1987A.

In any event, these doubts as well as the difficulty of explaining the late events in the presence of PNS convection (Sect.~\ref{sec:convection}) motivate a new study of PNS cooling with axions to understand better the impact of axion cooling together with convection and to understand better the impact on the early signal during the first few seconds. In Ref.~\cite{Bar:2019ifz} it was argued that axion emission could not strongly affect the first few seconds, but we are not convinced that such a conclusion is strongly supported by their simulation for the first 0.2~s post bounce. A better quantitative understanding of a possible fallback signal in 3D simulations is also on our wish list.

\subsection{Axion cooling and supernova explosions}

Supernova bounds based on the SN 1987A neutrino signal duration apply in the free-streaming regime, where axion losses are a local energy sink and do not transfer energy, for example to the stalling shock wave. On the other hand, they could still affect explosion physics because the collapsed SN core could become more compact more quickly and in this way indirectly heat the neutrino-sphere region. This effect was studied with 1D and 2D SN models, confirming the expectation that axions that obey the energy-loss bound of Eq.~\eqref{eq:SN1987A-KSVZ-bound} would not modify explosion physics \cite{Betranhandy:2022bvr}. However, for a coupling strength around twice larger (energy loss rate a factor of 4 larger), the effect begins to become noticeable. Given the systematic uncertainties of the SN~1987A and NS cooling bounds, the possibility of such effects cannot be completely dismissed. Beyond QCD axions, ALPs or FIPs with suitable parameters could transfer energy to the outer SN regions and strongly modify the explosion physics (see Sec.~\ref{sec:SN-ALPs} below).

\subsection{Diffuse supernova axion background (DSAB)}

Based on current evidence, axion emission from a core-collapse SN could be a large effect, the total energy loss could be comparable to that in neutrinos. The diffuse SN neutrino background (DSNB) is the largest cosmic neutrino radiation \cite{Beacom:2010kk, Vitagliano:2019yzm}, given that primordial neutrinos form a hot dark matter contribution today, not radiation. The energy density of the DSNB is comparable to the extra-galactic background light (EBL), the electromagnetic radiation emitted from all stars over the age of the universe. Therefore, axions emitted by all stellar core collapses likewise could contribute a comparable energy density \cite{Raffelt:2011ft}. While the DSNB is at the verge of detection with Super-Kamiokande Gd and JUNO \cite{Li:2022myd}, there is no immediate idea or prospect to detect the much more feebly interacting DSAB. If pion production is indeed the dominant emission process, the spectrum would be much harder than originally foreseen. Beyond QCD axions, the diffuse SN particle flux provides interesting constraints, for example on ALPs that would subsequently decay \cite{Caputo:2021rux} or the DSNALPB could be converted to photons in astrophysical magnetic fields \cite{Calore:2020tjw, Calore:2021hhn, Calore:2021lih, Eby:2024mhd}. Of course, a diffuse SN background is not peculiar to axions or ALPs---any FIP produced in core-collapse SNe can form a diffuse background (DSFB) and could lead to interesting signals. This is certainly the case for sterile neutrinos \cite{Carenza:2023old}, dark photons \cite{Kazanas:2014mca}, or CP-even scalars \cite{Caputo:2021rux}.

\subsection{Supernova constraints on ALPs and other FIPs}

\label{sec:SN-ALPs}

Besides recent revisions of the traditional SN~1987A axion bound that we have discussed in this chapter, numerous other SN-related arguments about a variety of ALPs and FIPs have been advanced recently. The richest phenomenology arises when FIPs decay into photons after being produced in the SN. Many different signatures arise depending on their mean free path:
\begin{itemize}
    \item If the mean free path $\lambda$ is between the radius of the new-born NS and $R_*$ of the SN progenitor, then the explosion energy will be enhanced~\cite{Falk:1978kf}. Effectively, FIPs would dump energy from the SN core into the mantle, showing up as explosion energy or radiation, that is typically less than 1\% of the energy carried by neutrinos. To avoid overly energetic SN explosions imposes severe constraints that are even stronger when compared with Low-Energy SNe \cite{Yang:2021fka, Spiro_2014}. This class presents particularly low explosion energies and thus is perfect to constrain exotic energy deposition~\cite{Caputo:2022mah}. The \textit{explosion criterion} applies to particles decaying into photons~\cite{Caputo:2021rux, Caputo:2022mah}, but also to other interacting final states such as $e^+e^-$ \cite{Sung:2019xie, Calore:2021lih}.
    
    \item If $\lambda$ is large enough for the particle to make it out of the progenitor, radiative decays en route to Earth can become observable. If $R_* \lesssim \lambda \lesssim 50\,{\rm kpc}$ (distance to SN~1987A), then the strongest limits come from FIPs produced in SN~1987A, where the decay $\gamma$-rays would have been picked up by the Gamma-Ray Spectrometer on board of the Solar Maximum Mission (SMM) \cite{Caputo:2021rux, Jaeckel:2017tud}. For smaller FIP masses, or smaller couplings, $\lambda$ increases to cosmological distances and the strongest limits come from FIP emission of all past SNe (the DSFB discussed in the previous section). When the FIPs later decay, they contribute to the diffuse cosmic $\gamma$-ray background~\cite{Fermi-LAT:2014ryh}, leading to competitive bounds~\cite{PhysRevLett.39.784, Caputo:2021rux, Calore:2021klc}.
    
    \item There can be another twist when $\lambda \gtrsim R_*$. The produced photons may be unable to escape because they are so dense that they form a fireball~\cite{Diamond:2023scc}, creating a QED plasma originating from pair production. This plasma quickly evolves to much lower temperatures mainly via pair bremsstrahlung, and the emerging photons have much lower energies of 0.1--1~MeV. Most of them would have escaped detection by SMM, but could have been seen by the Pioneer Venus Orbiter (PVO) Satellite \cite{evans1979gamma}, which had an energy window 0.2--2~MeV. These ideas also apply to FIPs decaying directly to $e^+e^-$.
\end{itemize}

These signatures are important for large FIP masses, typically at the MeV scale or above, whereas otherwise different probes are more relevant. In this mass window, we mention in passing the recent use of other spectacular astrophysical transients, such as hypernovae~\cite{Caputo:2021kcv} and NS mergers~\cite{Diamond:2023cto,Diamond:2021ekg}, where decaying FIPs produce striking signatures. For much smaller masses, the decay rate is strongly phase-space suppressed. However, in this case FIPs produced in SN explosions (or also in other astrophysical objects, such as Wolf-Rayet (WR) stars~\cite{Dessert:2020lil}) may convert into photons in the Milky Way magnetic field, also producing $\gamma$-ray signals (X-ray for WR stars). For ALPs these constraints are competitive for $m_a \lesssim 10^{-8} \, \rm eV$~\cite{Calore:2021hhn, Payez:2014xsa}.

Another case worth mention is a boson $\phi$ coupled to neutrinos, such as majorons \cite{Chikashige:1980ui, Chang:1988aa}. Besides SN~1987A cooling \cite{PhysRevD.39.985, Choi:1989hi,  Heurtier:2016otg, Farzan:2002wx, Grifols:1988fg, Aharonov:1988ee}, another powerful constraint derives from the non-observation of 100-MeV-range neutrinos in the IMB and Kam-II signals of SN~1987A that could arise from majoron decays en route to Earth \cite{Fiorillo:2022cdq}. This argument is particularly relevant for $m_\phi \gtrsim 100 \, \rm eV$ when neutrino coalescence is the main production channel in the SN core. A future galactic SN offers the opportunity to detect such signatures~\cite{Akita:2022etk}.

\section{Black-hole superradiance}
\label{sec:BH}

Our final class of stellar axion factories consists of astrophysical black holes (BHs), the most compact of all stars. This may seem like a self-contradiction because BHs are thought to be ideal black bodies with very low temperature $T_{\rm BH}=(8\pi G_{\rm N}M_{\rm BH})^{-1}= 0.61 {\rm nK}~(M_\odot/M_{\rm BH})$, much colder than $T_{\rm CMB}=2.73$~K. Astrophysical BHs live in environments of baryonic matter, dark matter, the CMB, and cosmic background neutrinos, absorbing all of these, and so BHs do not evaporate. Assuming a massless scalar boson $a$ without cosmic background flux, the BH luminosity in this channel would be $L_a=\bigl(2^{11} 15\pi G_{\rm N}^2M_{\rm BH}^2\bigr)^{-1}=4.45\times10^{-29}~{\rm Watt}\,(M_\odot/M_{\rm BH})^2$ and thus extremely small. However, unrelated to Hawking radiation, rotating (Kerr) BHs can efficiently flush out low-mass bosons at the expense of rotational energy, so-called superradiance.

Following a recent review \cite{Brito:2015oca}, we note that the concept of superradiance was introduced by Dicke (1954) for the then-hypothetical idea of
radiation amplification by coherence of emitters~\cite{Dicke:1954zz}. In 1971, Zeldovich showed that radiation that scatters off rotating absorbing surfaces can result in amplification \cite{Zeldovich:1971, Zeldovich:1972}, an effect today widely called (rotational) superradiance. A wave with frequency $\omega$ in a spherical-harmonic distribution with azimuthal number $m$ is amplified if
\begin{equation}\label{Eq:SuperradianceCondition}
    \omega < m \, \Omega,
\end{equation}
where $\Omega$ is the angular frequency of the rotating body. Around the same time, 
following an earlier paper by Penrose (1969) \cite{Penrose:1969pc}, Penrose and Floyd (1971) showed that one can extract energy and angular momentum from a rotating BH in analogous fashion \cite{Penrose71}, in what today we call {\em black hole superradiance}. $\Omega$ is now to be interpreted as the angular velocity of the BH horizon defined as
\begin{equation}
    \Omega_{\rm H}=\frac{1}{2r_{\rm g}}\,
    \frac{a/r_{\rm g}}{1+\sqrt{1-(a/r_{\rm g})^2}},
\end{equation}
where $r_{\rm g}=G_{\rm N}M_{\rm BH}$ is the BH gravitational radius and $0<a<r_{\rm g}$ its spin-to-mass ratio.

The case of \textit{massive} bosonic waves (mass $m_a$) is special because they can form bound states around the BH, allowing for exponential growth by steady superradiant amplification~\cite{Zouros:1979iw, Detweiler:1980uk}. The bound states are similar to those of hydrogen atoms, with a ``gravitational fine structure constant'' $\alpha_{\rm g}=G_{\rm N}M_{\rm BH} m_a$. The orbitals around the BH are characterized by the principal, orbital, and magnetic quantum numbers $\{n, l, m\}$ and their energy is~\cite{Arvanitaki:2010sy, Arvanitaki:2014wva}
\begin{equation}\label{Eq:EnergyOrbitals}
    \omega = m_a \bigg(1 - \frac{\alpha^2_g}{2 \, n^2}\biggr).
\end{equation}
The orbital velocity is roughly $\alpha_{\rm g}/l$ and the size of the cloud $r_{\rm c} \sim (n/\alpha_{\rm g})^2\,r_{\rm g}$. If one approximates the bound-state energy as $\omega\simeq m_a$, the superradiance condition of Eq.~\eqref{Eq:SuperradianceCondition} reads $m_a < m \, \Omega_{\rm H}$. In this limit one can also derive a simple parametric form for the \textit{imaginary part} $\Gamma$ of the frequency, the growth rate of the superradiant cloud,
\begin{equation}\label{Eq:ImaginaryPartSuperradiance}
\Gamma \propto \alpha_{\rm g}^{4 \, l + 5} \bigl(m \, \Omega_{\rm H} - m_a\bigr) \propto m_a^{4 \,l + 5}.
\end{equation}
This expression reveals a crucial aspect of superradiance: on the one hand, $m_a$ should be small enough or the BH spin large enough to satisfy Eq.~\eqref{Eq:SuperradianceCondition}; on the other hand, $m_a$ cannot be too small or else $\Gamma$ becomes too small for any interesting phenomenological consequences. Therefore, the BH mass determines the $m_a$ range that one can probe via superradiance. 

Axion clouds grow at the expense of the BH spin. Therefore, observing a sufficiently fast-spinning old BH can be enough to constrain the existence of bosons with $m_a\simeq 1/r_{\rm g}$~\cite{Arvanitaki:2014wva, Baryakhtar:2022hbu, Cardoso:2018tly, Stott:2018opm}. The leading method to measure BH spin is continuum fitting and X-ray relativistic reflection, which have been used to measure BH spins in X-ray binary systems~\cite{Zhou:2019fcg}. Both methods measure the innermost stable circular orbit (ISCO) of the accretion disk, which in turns provides the BH spin. Of course, the spins of stellar-mass BHs have also been measured in the gravitational-wave (GW) detectors LIGO and VIRGO, but the uncertainties are much larger than those in X-ray binaries.

For example, Ref.~\cite{ Baryakhtar:2022hbu} used the five stellar-mass BH X-ray binaries M33 X-7, LMC X-1, GRO J1655-40, Cyg X-1, and GRS1915+105 to exclude the mass range
\begin{equation}\label{Eq:SuperradianceBounds}
0.29 \times 10^{-12} \lesssim m_a \lesssim 7 \times 10^{-12} \, {\rm eV}.
\end{equation}
In principle, this constraint depends only on the boson mass, but other interactions than gravity can modify superradiance. For example, if we give up the canonical relation Eq.~\eqref{eq:axmass} between $m_a$ and $f_a$ and the latter is too small for given $m_a$, then axion self-interactions will decrease the occupation number in the equilibrium state and the axion cloud will collapse \textit{before} extracting much BH spin~\cite{Baryakhtar:2022hbu}. The bounds in Eq.~\eqref{Eq:SuperradianceBounds} do not apply if $f_a$ is at least some five orders of magnitude smaller than implied by the standard relation.

Similar ideas apply also to supermassive BH spins, probing much smaller $m_a$ \cite{Stott:2018opm, Davoudiasl:2019nlo, Stott:2020gjj}, which would typically imply trans-Planckian values for $f_a$ in the QCD-axion case.

Besides BH spin down, there can be other signatures of superradiance. Transitions between different states of the gravitational atom formed by the boson field can emit continuous-wave GWs~\cite{Arvanitaki:2014wva}, in contrast to the chirps produced by the usual compact-star mergers. Searches with LIGO O2 data \cite{Palomba:2019vxe} and LIGO O3 data \cite{LIGOScientific:2021rnv, Yuan:2022bem} exclude some domain in the $m_a$--$M_{\rm BH}$ plane, but no directly tangible QCD axion parameters. Still, this channel offers a future detection opportunity for very low-mass axions. Binary mergers of BHs surrounded by bosonic gravitational atoms would provide modified GW signatures \cite{Baumann:2018vus, Baumann:2019eav, Baumann:2021fkf, Baumann:2022pkl}, offering yet another future detection opportunity.

Finally, one may wonder if rotational superradiance can occur in other rotating astrophysical objects, for example pulsars? Superradiance requires \textit{dissipation} which in rotating BHs is provided by their ergoregion,\footnote{We stress that rotational superradiance does \textit{not} rely on the BH horizon. This follows from a simple causality argument: waves would take infinite time to reach the horizon and therefore the superradiance dynamics cannot depend on boundary conditions. Moreover, working in the time domain, it has been shown explicitly that superradiance happens for horizonless geometries~\cite{Vicente:2018mxl}.} but does not exist in neutron stars. Recent works have proposed that enough dissipation may come from finite conductivity either in the stellar magnetosphere~\cite{Day:2019bbh, Chadha-Day:2022inf} or within the star itself~\cite{Cardoso:2015zqa, Cardoso:2017kgn}, but one may wonder if the required conditions can be met in \textit{realistic} pulsars. More work may be needed to substantiate this exciting possibility.

\section{Modified equation of state in compact stars}
\label{sec:EoS}

As discussed earlier, the derivative axion-nucleon interactions are modified in the dense nuclear matter of neutron stars and SN cores \cite{Balkin:2020dsr, Stelzl:2023}. However, dense matter may spawn another intriguing effect through a coherent interplay between the nucleon density and the axion field, potentially leading to entirely different structures for compact stars~\cite{Hook:2017psm, Springmann:2023}. A nonvanishing nucleon density $n_N$ modifies the usual axion potential in the form
\begin{equation}\label{Eq:AxionPotentialMatter}
    V(a) \simeq \frac{m_\pi^2 f_\pi^2}{4}\left(\epsilon - \frac{\sigma_{N}n_N}{m_\pi^2 f_\pi^2}\right)\left(1-\cos\frac{a}{f_a}\right),
\end{equation}
where $\sigma_N\simeq50$~MeV is the pion-nucleon sigma term and $\epsilon$ is a numerical coefficient. This formula applies in the approximation of $m_u=m_d$, with  Eq.~\eqref{eq:axmass} implying an axion vacuum mass $m_a=m_\pi f_\pi/2$ and thus $\epsilon\simeq1$. Numerically, for nuclear saturation density $n_0=0.15~{\rm fm}^{-3}$, the second term is $\sigma_N n_0/m_\pi^2 f_\pi^2\simeq0.4$. Therefore, if the axion is only minimally lighter than standard, the axion potential switches sign and the minimum is not at $a=0$ but at $a=\pi f_a$.

Conversely, a nonvanishing expectation value for the axion field produces a potential for nucleons and shifts their mass according to
\begin{equation}
    m_N(a) \simeq m_N   
    \left[1+\frac{\sigma_N}{2m_N}\left(\cos\frac{a}{f_a}-1\right)\right]
\end{equation}
and thus for a flipped axion minimum, a mass shift of $\delta m_N\simeq -\sigma_N$, but using a more precise potential, $\delta m_N\simeq-30$~MeV \cite{Balkin:2022qer}.

For $\epsilon\ll 1$, as a function of radius in a neutron star, the axion field would be in the wrong minimum deeply inside, and in the usual minimum outside, with an axion-field brane at some transition radius. In an ordinary nucleus, this effect will not occur because the axion gradient energy would prevent such a solution, but in a neutron star, a modification of the nuclear equation of state would obtain \cite{Balkin:2023xtr}. In the context of binary neutron star mergers and concomitant gravitational wave emission, large effects can arise \cite{Hook:2017psm}. Shifting the nucleon mass in white dwarfs modifies their equation of state and their mass-radius relationship and requires $\epsilon \gtrsim 10^{-8}$ \cite{Balkin:2022qer}.  Similar arguments apply to the Sun and Earth~\cite{Hook:2017psm}, as well as other astrophysical objects. As a rule of thumb, denser objects will probe larger value of $\epsilon$, closer to the standard QCD axion. For the latest status, see Konstantin Springmann's recent PhD Thesis \cite{Springmann:2023}.

\section{Hot dark matter axions and telescope searches}
\label{sec:HDM}

We next turn to a possible axion hot dark matter (HDM) component that provides constraints directly complementary to stars and thus warrants mention here. Of course, the main cosmological interest in axions derives from their possible role as cold dark matter in the form of low-momentum classical field oscillations initiated in the early universe, see Ciaran O'Hare's Lectures at this School and Ben Safdi's recent TASI Lectures~\cite{Safdi:2022xkm}. In addition, thermal axions are produced by interactions with the quark-gluon plasma before QCD confinement \cite{Turner:1986tb, Masso:2002np} and afterwards by $\pi N\leftrightarrow N a$ \cite{Turner:1986tb, Berezhiani:1992rk} and mostly $\pi\pi\leftrightarrow\pi a$ \cite{Chang:1993gm}, spawning a radiation density comparable with that of one neutrino. With $m_a$ in the eV-range, axions would provide HDM analogous to neutrinos. After initially postulating some axion HDM \cite{Moroi:1998qs}, the steady advance of $\Lambda$CDM cosmology invariably diminished possible HDM components. A historical selection over 20 years is $m_a<3.0$~eV \cite{Hannestad:2003ye}, 1.05~eV \cite{Hannestad:2005df}, 0.67~eV \cite{Archidiacono:2013cha}, 0.529~eV \cite{DiValentino:2015wba}, and 0.83--0.89~eV \cite{DiValentino:2015zta}, where here and henceforth the pion couplings of the KSVZ model are used. For an explicit study of the DFSZ case in this context see \cite{Ferreira:2020bpb}. 

However, these results were called into question in that pion-axion thermalization in the 100~MeV temperature range puts the interactions outside the validity of lowest-order chiral perturbation theory \cite{DiLuzio:2021vjd}. Recently, however, these difficulties were elegantly overcome by scaling instead to measured $\pi \pi$ scattering data \cite{Notari:2022ffe}. These authors also consider the changing number of thermal degrees of freedom during the freeze-out process and find $m_a < 0.24~{\rm eV}$ at 95\%~C.L., as well as $\sum m_\nu < 0.14~{\rm eV}$. Meanwhile, the authors of \cite{DiLuzio:2021vjd} have issued a new paper \cite{DiLuzio:2022gsc}, where they use unitarized NLO chiral perturbation theory and correct an earlier error in the loop function of the NLO scattering amplitude. For the phenomenological outcome they seem to largely agree with \cite{Notari:2022ffe}, although this seems to be just a numerical coincidence. Another contemporaneous paper found $m_a < 0.18~{\rm eV}$ and $\sum m_\nu<0.16~{\rm eV}$ \cite{DiValentino:2022edq}, using an interpolation of the axion production rate across the QCD phase transition between standard results far above and far below the confinement scale~\cite{DEramo:2021psx, DEramo:2021lgb}. In addition, the authors of Ref.~\cite{DiValentino:2022edq} have marginalized over an ensemble of cosmological models and in this case found 0.21~eV for both quantities, slightly stronger than their previous limit of $m_a < 0.28~{\rm eV}$ at 95\%~C.L.~\cite{DEramo:2022nvb}. Very recently, yet another analysis adopted unitarized NLO chiral perturbation theory and with a detailed cosmological analysis found $m_a < 0.18~{\rm eV}$ at 95\% C.L.~\cite{bianchini2023qcd}, including constraints from BBN, which is the main factor driving the improvement with respect to \cite{Notari:2022ffe}. Their bound improves further to $m_a < 0.16~{\rm eV}$ at 95\% C.L.\ after including the ground-based Atacama Cosmology Telescope (ACT) and the 2018 South Pole Telescope (SPT) CMB data that were also included in Ref.~\cite{DiValentino:2022edq}. Unsurprisingly, the overall situation is similar to HDM constraints on neutrinos masses, where recently published bounds are in the range $\sum m_\nu<0.082$--0.54~eV, depending on data sets and analysis \cite{Lesgourgues:2023}. A detailed assessment of the different published $m_a$ bounds is beyond our scope, but we rather interpret the range 0.16--0.24~eV between different groups and assumptions as impressively concordant, especially considering that this entire approach had been questioned only a few years ago. As a representative value, we adopt a nominal limit
\begin{equation}\label{eq:HDMlimit}
    m_a < 0.20~{\rm eV}
\end{equation}
as a number to put on our summary plot Fig.~\ref{fig:Summary}.

It is striking that a forecast with state-of-the-art likelihood tools for DESI and CMB-S4 suggest that the cosmological sensitivity may soon become competitive with the astrophysical bounds~\cite{bianchini2023qcd}. Moreover,
future cosmological probes should finally turn up the minimal HDM fraction implied by neutrino oscillations and concomitant minimal neutrino masses of $\sum m_\nu\gtrsim 60$~meV. Finding rather than constraining this unavoidable HDM component will take precision cosmology to yet another level and beyond this feat, to distinguish between neutrino and axion contributions.

When $m_a\gtrsim 24$~eV, axions decay within the age of the universe---see the discussion below Eq.~\eqref{eq:ALP-decay}---producing too much cosmic background radiation. If the decay is very fast, one needs to appeal to effects in the early universe, notably big-bang nucleosynthesis (BBN). One sensitive probe is the cosmic deuterium abundance and accordingly, axions with $m_a\lesssim 300$~keV are excluded \cite{Cadamuro:2010cz}. More recent studies include Refs.~\cite{Depta:2020wmr, Langhoff:2022bij}. Keeping our focus on QCD axions, it is probably fair to assume that they are cosmologically excluded down to the 100~keV range shown in Fig.~\ref{fig:Summary}.

When the lifetime exceeds the age of the universe, axion decays need not be harmless because the resulting photons can be observed, e.g.~coming from galaxy clusters, where HDM axions would have somewhat accumulated. Early ``telescope searches'' targeting the clusters Abell 1413, 2218 and 2256 excluded $m_a=3.2$--7.8~eV \cite{Ressell:1991zv, Bershady:1990sw}. Later observations of Abell 2667 and 2390 with the VIMOS integral field unit at the Very Large Telescope (VLT) excluded $m_a=4.5$--7.7~eV \cite{Grin:2006aw}. Recently, the MUSE instrument at the VLT looked at five galaxies (classical and ultra-faint) and excluded the range 2.7--5.3~eV \cite{Todarello:2023hdk}, assuming axions are all of the DM in these small systems. While this could not be true for HDM, these bounds apply even for strongly suppressed $G_{a\gamma\gamma}$ and therefore also to a reduced DM density. In any case, the compound mass range 2.7--7.8~eV excluded by telescope searches is already ruled out by the HDM bounds of Eq.~\eqref{eq:HDMlimit}. We also briefly mention that in the 1--10~eV mass range, other interesting probes include the use of optical depth measurements of distant blazars~\cite{Bernal:2022xyi}, emerging line intensity mapping techniques~\cite{Bernal:2020lkd}, the angular power spectrum of the anisotropies of the extragalactic background light~\cite{Caputo:2020msf, Carenza:2023qxh, Gong:2015hke, Kohri:2017oqn, Kalashev:2018bra, Nakayama:2022jza}, as well as direct line searches with the James Webb Space Telescope~\cite{Bessho:2022yyu, Janish:2023kvi, Roy:2023omw} or the Subaru Telescope~\cite{Yin:2023uwf}.

\section{Conclusions}
\label{sec:Conclusions}

Over the past few years, many of the traditional astrophysical axion bounds have been extended to broader classes of ALPs, WISPs and FIPs, and new arguments have been advanced, assuming independent interactions to different classes of standard-model particles and/or masses unrelated to the interaction strength. Summarizing this recent explosion of activities was too ambitious within the limits of these Lecture Notes, but would be urgently needed in the literature. A fantastic overview of the excluded parameter domains for different couplings, and the corresponding references, is found in the GitHub pages of Ciaran O'Hare \cite{cajohare}. 

\begin{figure}[b]
    \centering
    \includegraphics[width=1\textwidth]{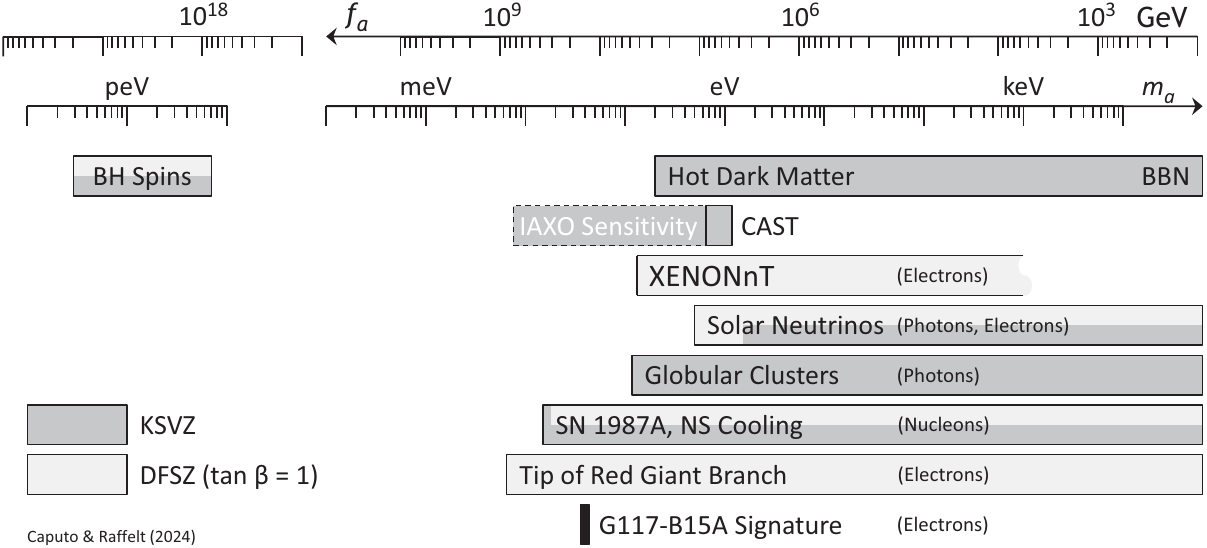}
    \caption{Summary of astrophysical axion bounds. For DFSZ axions, $\tan\beta=1$ or $\sin^2\beta=1/2$ is assumed. We also show the signal if the period drift of the variable white dwarf G117-B15A in interpreted as axion cooling. The high-mass end of the XENONnT exclusion range is not exactly known because the experimental analysis was for massless solar axions. Of course, this panoramic plot must be interpreted with care because it combines information from very heterogeneous sources with different levels of confidence and reliability.}
    \label{fig:Summary}
\end{figure}

With a more limited scope, we have focussed on the quasi one-dimensional parameter space of QCD axions, where all properties derive from $f_a$ or equivalently $m_a$. Figure~\ref{fig:Summary} and Table~\ref{tab:Summary} provide summaries, distinguishing only between hadronic models, represented by the KSVZ case, and nonhadronic ones which also include couplings to charged leptons, represented by the DFSZ model. In this case, we assume an axion-electron coupling of $g_{aee}=m_e/6 f_a$, corresponding in the DFSZ model to $\tan\beta=1$ or $\sin^2\beta=1/2$, and more general bounds scale from here. (See footnote~\ref{fn:beta} for our $\beta$ convention.)

\begin{table}[t]
 \caption{Summary of the astrophysical axion bounds shown in Fig.~\ref{fig:Summary}. The mass bounds are upper limits, where for the DFSZ case, $\tan\beta=1$ or $\sin^2\beta=1/2$ is assumed, corresponding to $C_e=1/6$.
 (See footnote~\ref{fn:beta} for our $\beta$ convention.)}
 \vskip-3pt
\label{tab:Summary}
    \begin{tabular*}{\columnwidth}{@{\extracolsep{\fill}}Sllllll}
    \hline\hline
     Coupling$^{a,b,c}$       &$m_a^{\rm KSVZ}$&$m_a^{\rm DFSZ}$&Argument&Eqs.&References\\
                        &[meV]&[meV]\\
     \hline
     ---&\multicolumn{2}{Sl}{0.29--7~peV excluded}
     &BH superradiance&\eqref{Eq:SuperradianceBounds}&\cite{Baryakhtar:2022hbu}\\[3pt]
     %%%%%%
     ---&200&similar
     &Hot dark matter&\eqref{eq:HDMlimit}&\cite{Notari:2022ffe, DiValentino:2022edq, bianchini2023qcd}\\[3pt]
     %%%%%%
     $G_{a\gamma\gamma}$ (Fig.~\ref{fig:CAST})&\multicolumn{2}{l}{0.64--1.17~eV exclud.}
     &CAST solar axions&\eqref{eq:CAST-mass}&\cite{CAST:2013bqn}\\
     %%%%%%
     $G_{a\gamma\gamma}^{(10)}<6$ &1500 &--- 
     &Solar neutrinos&\eqref{eq:SolarLimit-Photon} &\cite{Vinyoles:2015aba}, Sect.~\ref{sec:SolarPrimakoff} \\
     %%%%%%
     $g_{aee}^{(13)}<75$ &--- &500 
     &Solar neutrinos&\eqref{eq:SolarLimit-electroncoupling} &Sect.~\ref{sec:solar-axion-electron} \\
     %%%%%%
     $G^{(10)}_{a\gamma\gamma}<4.8$ &1200 &--- 
     &XENONnT solar ax. &\eqref{eq:XENON-photon} &\cite{XENON:2022ltv} \\
     %%%%%%
     $g_{aee}^{(13)}<19$ &--- &130 
     &XENONnT solar ax. &\eqref{eq:XENONnT-electroncoupling} &\cite{XENON:2022ltv} \\
     %%%%%%
     $G^{(10)}_{a\gamma\gamma}<0.65$&170&---&HB/RGB stars in GCs
     &(\ref{Eq:Bound_R_parameter}, \ref{Eq:HB-MassBound})&\cite{Ayala:2014pea, Straniero:2015nvc}\\
     %%%%%%
     $G^{(10)}_{a\gamma\gamma}<0.47$&120&---&AGB/RGB stars in GCs
     &(\ref{Eq:Bound_R2_parameter}, \ref{eq:R2-KSVZ})&\cite{Dolan:2022kul}\\
     %%%%%%
     $g_{aee}^{(13)}=5.7\pm0.6$ &--- &$38\pm4$& Period G117-B15A
     &Table~\ref{tab:WD-Cooling}&\cite{Isern:2022vdx, Kepler:2020qzt} \\
     %%%%%%
     $g_{aee}^{(13)}<2.1$ &--- &14 &WD luminos.\ function
     &Table~\ref{tab:WD-Cooling}&\cite{Isern:2022vdx, MillerBertolami:2014rka} \\
     %%%%%%
     $g_{aee}^{(13)}<1.6$ &--- &11 &TRGB, gal.\ NGC~4258
     &Sect.~\ref{sec:TRGB}&\cite{Capozzi:2020cbu} \\
     %%%%%%
     $g_{aee}^{(13)}<0.96$ &--- &6.4 &TRGB, 21~GCs 
     &(\ref{eq:TRGB-gaee}, \ref{eq:TRGB-mass-bound}) &\cite{Straniero:2020iyi, Straniero:2023} \\
     %%%%%%
     $g_{app}^{(9)}<1.2$ &15 & 19  &SN 1987A neutrinos
     &(\ref{eq:SN1987A-coupling-bound}--\ref{eq:SN1987A-DFSZ-bound})&\cite{Carenza:2019pxu}\\
     %%%%%%
     $g_{app}^{(9)}<1.5$ &19 & 24  &Old NS cooling 
     &(\ref{Eq:NSOld-DFSZ}, \ref{Eq:NSOld-KVSZ})&\cite{Buschmann:2021juv}\\[2pt]
     \hline
    \multicolumn{6}{Sl}{$^a$ $G^{(10)}_{a\gamma\gamma}=G_{a\gamma\gamma}/10^{-10}\,{\rm GeV}^{-1}$%}\\
     %\multicolumn{6}{Sl}{
     \qquad 
     $^b$ $g^{(13)}_{aee}=g_{aee}/10^{-13}$}\\
     \multicolumn{6}{Sl}{$^c$ $g^{(9)}_{app}=g_{app}/10^{-9}$. The limit applies for the KSVZ case of $g_{ann}\simeq0$.}\\
     \multicolumn{6}{Sl}{\kern0.7em Otherwise the exact constrained combination of $g_{app}$ and $g_{ann}$
     depends on environment.}
     \end{tabular*}
\end{table}

The bound from BH superradiance in Eq.~\eqref{Eq:SuperradianceBounds} depends only on the boson mass and not on the specific axion model. Likewise, the hot-dark matter bound of Eq.~\eqref{eq:HDMlimit} depends primarily on $m_a$ with weak dependence on the exact interactions with pions and nucleons at freeze out. We have included this cosmological constraint because it invades the normal jurisdiction of astrophysical arguments and is beginning to compete in earnest. Telescope limits in the $m_a$ range 2.7--7.8~eV (Sect.~\ref{sec:HDM}), based on the nonobservation of decay photons, are then both irrelevant and inconsistent because axions in this mass range are cosmologically excluded and cannot act as a source. 

The traditional bound on the axion-photon coupling from the helium-burning lifetime of horizontal-branch (HB) stars Eq.~\eqref{Eq:Bound_R_parameter} has been supplemented recently by a slightly more restrictive bound Eq.~\eqref{eq:R2-KSVZ} using number counts of asymptotic giants in globular clusters, contradicting the idea of a cooling hint in the HB number counts. In both cases one would worry about systematic issues in the underlying data when it comes to interpreting the results at their literal values and more importantly, the treatment of convection and core-breathing pulses during helium burning.

Axions derive from QCD so that bounds on their interactions with nucleons and pions, already underlying the hot dark matter argument, are of particular interest. After some significant recent critique of the axion bremsstrahlung emission rates, the SN~1987A cooling bound has settled back to its long-standing value Eq.~\eqref{eq:SN1987A-KSVZ-bound}, although with a more explicit dependence on the proton and neutron couplings Eq.~\eqref{eq:SN1987A-coupling-bound}. Such results remain subject to overall doubts about the sparse SN~1987A data (what is the meaning of the late events in the presence of PNS convection?) and the absence of a direct NS remnant observation. The recurring idea of pion scattering as a dominant emission process remains intriguing, but the nuclear physics very uncertain, so we have not used it in Fig.~\ref{fig:Summary}. Observations of certain old NSs provide uncannily similar constraints stated in  Eq.~\eqref{Eq:NSOld-DFSZ} and \eqref{Eq:NSOld-KVSZ}, whereas arguments based on the young NS in Cas~A and HESS J1731–347 now appear questionable in view of doubts about their luminosity or age. Still, many things would have to conspire to completely invalidate the combined SN and NS bounds shown in Fig.~\ref{fig:Summary}.

Taking uncertainties in the opposite direction, even for much smaller $m_a$ than excluded in Fig.~\ref{fig:Summary}, axions could carry away large amounts of SN energy and produce a sizeable cosmic diffuse axion background (DSAB) with a rather hard spectrum if pionic processes were to dominate.

The strongest constraints derive for nonhadronic axions based on their coupling to electrons, although with the strongest axion model dependence. The emission by degenerate electrons is the dominant source, either in white dwarfs (WDs), or the cores of low-mass red giants, effectively helium WDs. The observed period drift of certain pulsating WDs is a significant and intriguing effect, but cannot be attributed to axion cooling without strongly violating the most restrictive limit that arises from the brightness of the tip of the red giant branch (TRGB) in globular clusters. There has been significant progress, in particular on the distance determinations of globular clusters and thus the absolute brightness determination. We do not perceive a credible cooling hint from WD and TRGB observations.

In the end, the main interest is not in constraining QCD axions, but in detecting them. The backreaction on stars is a difficult signature because any putative deviation between standard stellar theory and observations is unlikely to be very specific. The unsolved puzzle of the period drift of variable WDs discussed in Sect.~\ref{sec:ZZCeti} is a case in point. As far as stars are concerned, the best bet for a future axion detection probably remains the (Baby)IAXO search for solar axions. IAXO is sensitive to the KSVZ model for $m_a \gtrsim 6.8 \, \rm meV$ as indicated in Fig.~\ref{fig:Summary}. We are also eagerly awaiting the high-statistics neutrino signal from the next galactic SN that would probably solidify and improve what we have learned from SN~1987A, even if direct evidence for QCD axions is hard to imagine. Looking beyond QCD axions, there is of course no telling where a smoking-gun signature for ALPs, WISPs or FIPs might eventually show up, in astrophysics or laboratory experiments, and perhaps in a not yet foreseen channel related to stars or otherwise.

\acknowledgments

We thank many colleagues for recent discussions, clarifications, or comments on the manuscript, notably Salvatore Bottaro, Malte Buschmann, Enrico Cannizzaro, Pierluca Carenza, Alexander Derbin, Chris Dessert, Johannes Diehl, Miguel Escudero, Damiano Fiorillo, Diego Guadagnoli, Ciaran O'Hare, Nick Houston, Joerg Jaeckel, Thomas Janka, Alessandro Lella, Daniele Lombardi, Giuseppe Lucente, Luca Di Luzio, Jamie McDonald, Alessandro Mirizzi, Maxim Pospelov, Andreas Ringwald, Fabrizio Rompineve, Ben Safdi, Aldo Serenelli, Andrea Shindler, Konstantin Springmann, Oscar Straniero, Mauro Valli, Edoardo Vitagliano, Jordy de Vries, and Andreas Weiler. This article is based upon work from COST Action COSMIC WISPers CA21106, supported by COST (European Cooperation in Science and Technology). GR acknowledges partial support by the German Research Foundation (DFG) through the Collaborative Research Centre ``Neutrinos and Dark Matter in Astro- and Particle Physics (NDM),'' Grant SFB-1258-283604770, and under Germany’s Excellence Strategy through the Cluster of Excellence ORIGINS EXC-2094-390783311.

\bibliographystyle{JHEP}
\bibliography{References}

\end{document}